\documentclass[aj_pt4]{aastex}

\slugcomment{Not to appear in Nonlearned J., 45.}

\shorttitle{CO, HCN, and  \hcosi\  Lines of  Bubbles }
\shortauthors{Q. Z. Yan et al.}

\usepackage{hyperref}
\usepackage{float}
\usepackage{natbib}
 
\usepackage{tikz}
\usepackage{blindtext}
\usepackage[flushleft]{threeparttable}
\usepackage{pst-node,graphicx}
\usepackage{multirow,bigdelim}
\graphicspath{{pic/}} 

\def\meth    {CH$_3$OH}   
\def\hho     {H$_2$O}
\def\hcosi {HCO$^+$}

\def\kms     {km~s$^{-1}$}

\def\Msun    {$M_{\odot}$}

\def\cof {$^{12}$CO (J=1$\rightarrow$0)}
\def\cos {$^{13}$CO (J=1$\rightarrow$0)}
\def\cot {C$^{18}$O (J=1$\rightarrow$0)}
 \def\intunit{K km s$^{-1}$}

\def\hcn {HCN (J=1$\rightarrow$0)}
\def\hco {HCO$^+$ (J=1$\rightarrow$0)}
\def\cofs {$^{12}$CO}
\def\coss {$^{13}$CO}
\def\cots{C$^{18}$O}
\def\hcns {HCN}
\def\hcos{HCO$^+$}

\def\mum {$\mu \rm m$}

\def\h       {\ifmmode{^{\rm h}}\else{$^{\rm h}$}\fi}
\def\m       {\ifmmode{^{\rm m}}\else{$^{\rm m}$}\fi}
\def\s       {\ifmmode{^{\rm s}}\else{$^{\rm s}$}\fi}
\def\deg     {\ifmmode{^{\circ}}\else{$^{\circ}$}\fi}
\def\decdeg  {\ifmmode{{\rlap.}^{\circ}} \else ${\rlap.}^{\circ}$\fi}
\def\decs    {\ifmmode{{\rlap.}^{\rm s}} \else ${\rlap.}^{\rm s}$\fi}
\def\decas   {\ifmmode{{\rlap.}{''}}\else{${\rlap.}{''}$}\fi}

\newcommand{\HII}{\mbox{H\,\textsc{ii}}}%

\begin{document}


\title{MOLECULAR LINES OF 13 GALACTIC INFRARED  BUBBLE REGIONS}

\author{Qing-zeng Yan  }
\affil{Shanghai Astronomical Observatory, Chinese Academy of Sciences,
              Shanghai 200030, China;  }
 \affil{Purple Mountain  Observatory, Chinese Academy of Sciences, Nanjing 210008, China}
  \affil{International Center for Radio Astronomy Research, Curtin University, GPO Box U1987, Perth WA 6845, Australia}

 \email{qzyan@shao.ac.cn}
\author{Ye  Xu}
\affil{Purple Mountain  Observatory, Chinese Academy of Sciences, Nanjing 210008, China}
 \email{xuye@pmo.ac.cn}
\author{  Bo Zhang  }
\affil{Shanghai Astronomical Observatory, Chinese Academy of Sciences, Shanghai 200030, China}
   
\author{  Deng-rong Lu  }
\affil{Purple Mountain  Observatory, Chinese Academy of Sciences, Nanjing 210008, China}
\author{  Xi Chen }
\affil{Shanghai Astronomical Observatory, Chinese Academy of Sciences, Shanghai 200030, China}

\author{Zheng-hong Tang    }
\affil{Shanghai Astronomical Observatory, Chinese Academy of Sciences,
              Shanghai 200030, China;  }



\begin{abstract}

We investigated the physical properties of molecular clouds and star formation processes around infrared bubbles which are essentially expanding \HII\ regions. We performed observations of 13 galactic infrared bubble fields containing 18 bubbles. Five molecular lines, \cof, \cos, \cot, \hcn, and \hco, were observed, and several publicly available surveys, GLIMPSE, MIPSGAL, ATLASGAL, BGPS, VGPS, MAGPIS, and NVSS, were used for comparison.  We find that these bubbles are generally connected with molecular clouds, most of which are giant. Several bubble regions display velocity gradients and broad shifted profiles, which could be due to the expansion of bubbles. The masses of molecular clouds within bubbles range from 100 to 19,000 $M_\odot$, and their dynamic ages are about 0.3-3.7 Myr, which takes into account the internal turbulence pressure of surrounding molecular clouds. Clumps are found in the vicinity of all 18 bubbles, and molecular clouds near four of these bubbles with larger angular sizes  show shell-like morphologies, indicating that either collect-and-collapse or radiation-driven implosion processes may have occurred. Due to the contamination of adjacent molecular clouds, only six bubble regions are appropriate to search for outflows, and we find that four of them have outflow activities. Three bubbles display ultra-compact \HII\  regions at their borders, and one of them is probably responsible for its outflow. In total, only six bubbles show star formation activities in the vicinity, and we suggest that  star formation processes might have been triggered.


\end{abstract}






\section{Introduction}           
\label{sect:intro}
 
  The \emph{spitzer} Galactic Legacy Infrared Mid-Plane Survey Extraordinaire (GLIMPSE; \citet{2003PASP..115..953B,2009PASP..121..213C}) identifies almost 600 bubbles~\citep{2006ApJ...649..759C,2007ApJ...670..428C}, extended to 5106 by the Milky Way Project (MWP)~\citep{2012MNRAS.424.2442S}. Bubbles were defined by the 8.0 micron emission~\citep{1984A&A...137L...5L}, which contains 7.7 micron and 8.6 micron polycyclic aromatic hydrocarbon (PAH) features, including the continuum~\citep{2010ApJ...713..592E}.  \citet{2006ApJ...649..759C} claimed that 25\% of 322 bubbles outside 10\deg\ of the Galactic center coincide with known radio \HII\ regions, and~\citet{2010A&A...523A...6D} extended this proportion to 86\% based on a sample of 102  bubbles selected from the catalog provided by \citet{2006ApJ...649..759C}.

Most bubbles are Photodissociation Regions (PDRs) ~\citep{1997ARA&A..35..179H,1999ApJ...527..795K} produced by high-mass stars ionizing  atomic or molecular hydrogen. \citet{2010A&A...523A...6D} proposed a simple model for \HII\ region evolution involving two main phases, rapid ionization of the neutral medium followed by a long expansion. During the second phase, shock and ionization fronts form, and neutral material and cold dust collected between them. There are some studies supporting that the expanding of \HII\  regions are three-dimensional. \citet{2010ApJ...709..791B} find three-dimensional structures throughout a sample of 43 bubbles, using the observations of  CO (J=3-2) and \hcosi\ (J=4-3)  observed by the James Clerk Maxwell Telescope (JCMT). With more sensitive \emph{Herschel} data, \citet{2012A&A...542A..10A} detected emission from "near-side" and "far-side" of bubbles along the line of sight, which suggesting three-dimensional structures for bubbles.


 Molecular lines, usually characterized with particular critical densities, excitation temperatures, and kinematics information, are superb tools for  studying molecular clouds including those around infrared bubbles. Observations of molecular lines contain physical, chemical, and dynamic information which is essential to the study of molecular clouds. CO, a good tracer of molecular clouds due to its low excitation temperature and low critical density~\citep{2001ApJ...547..792D,2015ARA&A..53..583H}, has been widely used to study infrared bubbles.  For example,  the Galactic Ring Survey (GRS) $^{13}$CO data~\citep{2006ApJS..163..145J} has been present in many bubble papers~\citep{2014A&A...569A..36X,2014A&A...565A...6S,2010A&A...513A..44P,2014MNRAS.438..426H}. Other molecular lines, such as \hcn\ and \hco\  which are  probes of dense clumps and cores~\citep{2010ApJS..188..313W,2005ApJ...622..346C} can also been found throughout bubble studies~\citep{2012ApJ...755...71K,2010ApJ...709..791B}.  Moreover, \hcos\ and \cof\ are good tracers of bipolar outflows, which are ubiquitous phenomena in star forming regions~\citep{2001ApJ...552L.167Z,2004MNRAS.351.1054R}, and \cofs\ has also been used to identify outflows around infrared bubbles~\citep{2010ApJ...709..791B}.

A number of papers argue that star formation processes can be triggered by expanding \HII\ regions, and mainly two mechanisms for the triggering are  proposed: collect-and-collapse~\citep{1977ApJ...214..725E}, and radiation-driven implosion (RDI)~\citep{1989ApJ...346..735B}. \citet{1977ApJ...214..725E} proposed that star formation can be triggered by the propagation of ionization and shock fronts through a molecular cloud complex. \citet{2005ApJ...623..917H} analyzed the dynamical expansion of \HII\ regions and the outer PDRs around a high-mass star by solving the UV and FUV radiation transfer and the thermal and chemical processes, using time dependent hydrodynamics. They find that a molecular gas shell with a mass in the order of $10^4$ $M_\odot$ can be shaped in $\sim$1 Myr, and this triggering process is of great importance for star formations of next generation. Seventeen candidate regions for the collect-and-collapse process were identified by  \citet{2005A&A...433..565D}, and a large number of young stellar objects (YSOs) were found in the vicinity of  bubbles  \citep{2008ApJ...681.1341W,2009ApJ...694..546W,2010ApJ...716.1478W}. \citet{2012ApJ...755...71K} found a strong positional correlation between high-mass YSOs (HMYSOs) and \HII\ regions with MWP bubbles at separations of $\textless$2 bubble radii. A statistical study of YSOs around 322 \emph{Spitzer} mid-infrared bubbles has been done by \citet{2012MNRAS.421..408T}, and they found a significant overdensity of Red MSX-Source Survey (RMS) YSOs around the bubbles.  These results  support that expanding \HII\ regions can provide an effective mechanism to form next generation stars.  


However, this scenario is not universal.  \citet{2012A&A...542A..10A} found the cold gas lies in a ring instead of a sphere, indicating a flattened molecular clouds which could be not greatly compressed by expanding shock fronts. In this case, the formation of new stars could be hindered.  \citet{2015MNRAS.450.1199D} investigated the most commonly used signposts and found none of them significantly improved the chances of correctly identifying a given star as triggered. They argued that we should be cautious in interpreting observations of star formation near feedback driven structures in terms of triggering. 
 
If bubbles can truly trigger star formation, then the characteristics of star formation, i.e., collapse, outflows, or  masers, should be found around them.  Although these phenomena are not specific for triggered star formation, the more star formation process we find around bubbles, the safer we can say they are triggered or accelerated by expanding \HII\ regions. 

Identifying YSOs is the  most commonly used method to study  triggered star formation. However, it is difficult to determine the distances, ages, and masses of YSOs. If  we could find outflows or specific masers, such as OH, H$_2$O, and CLASS \textsc{ii} CH$_3$OH  \citep{1995MNRAS.272...96C},  which are direct signposts of star formation, we can at least confirm that star formation processes are indeed present around bubbles, although some of these processes are probably spontaneous. \citet{2009ApJ...702.1615C} made a class \textsc{i} and \textsc{ii}  \meth\ masers survey of approximately  20 HMYSO outflow candidates,  referred to as Extend Green Objects (EGOs) which are candidates of massive outflows,  and three of them  are associated with bubbles. \citet{2010ApJ...709..791B} proposed 12 eye-based outflow candidates, and however, stronger evidence is required to confirm them because moments maps are inadequate to confirm outflows due to the complex environments around bubbles. 

We studied 13  infrared bubble  regions selected from the bubble catalog of~\citet{2006ApJ...649..759C}, including 18 bubbles. For each bubble, five molecular lines were observed, revealing the dynamic and physical features of molecular clouds around   bubbles. Among the five lines, C$^{18}$O (J=1-0), HCN (J=1-0), and HCO$^+$(J=1-0) are not widely present in previous observations. Several publicly available surveys are also involved, such as  the GLIMPSE~\citep{2003PASP..115..953B,2009PASP..121..213C},  the Multiband
Imaging Photometer for \emph{Spitzer} (MIPS)  Galactic Plane
Survey (MIPSGAL)~\citep{2009PASP..121...76C,2015AJ....149...64G}, the APEX Telescope Large Area Survey of the Galaxy (ATLASGAL)~\citep{2009AA...504..415S}, the Bolocam Galactic Plane Survey (BGPS)~\citep{2011ApJS..192....4A,2010ApJS..188..123R}, the VLA Galactic Plane Survey (VGPS)~\citep{2006AJ....132.1158S},  the Multi-Array Galactic Plane Imaging Survey (MAGPIS)~\citep{2006AJ....131.2525H}, and the NRAO VLA Sky Survey (NVSS)~\citep{1998AJ....115.1693C}. Details of these surveys are described in Table \ref{Tab:surveys}. Combing these surveys, we  performed  a multi-wavelength  analysis of  the 18 bubbles, focusing on their physical properties  and  star formation processes around them. 
 
\begin{table}[H]
 \footnotesize
     
  \begin{center}
 
   \caption{ Observation  parameters of surveys. }
 \label{Tab:surveys}
 \begin{tabular}{lccccccccc}
   \tableline\tableline
 
Survey  &Wavelengths &Resolutions    &Facilities   &     References  \\

   \tableline
  
   GLIMPSE& 3.6, 4.5, 5.8, 8.0 \mum & $\sim2''$ &\emph{Spitzer}   & \citet{2003PASP..115..953B,2009PASP..121..213C}   \\

   MIPSGAL& 24 \mum &6$''$&\emph{Spitzer}  &\citet{2009PASP..121...76C,2015AJ....149...64G} \\
 ATLASGAL  &  0.87 mm& $19.2''$ &APEX &    \citet{2009AA...504..415S} \\
BGPS   &1.1 mm &33$''$ &Bolocam &\citet{2011ApJS..192....4A,2010ApJS..188..123R}\\
 VGPS &20 cm&60$''$&VLA& \citet{2006AJ....132.1158S} \\
 MAGPIS&20 cm&6$''$&VLA& \citet{2006AJ....131.2525H} \\
NVSS &20 cm&45$''$&VLA&\citet{1998AJ....115.1693C}\\
\tableline
\end{tabular}
 \end{center}

\end{table}

\section{ OBSERVATIONS AND DATA REDUCTIONS}

We selected 13 bubble regions from the catalog provided by \citet{2006ApJ...649..759C} by checking the 8 \mum\ band image  of GLIMPSE surveys visually,  and 18 bubbles in total were identified. Three criteria were applied to identify a bubble: (1) Located in the northern sky; (2) With outer major axis larger than 1.5 arcmin;  (3)   With approximate circular or elliptical shape. For small size bubbles, the dynamic expansion of \HII\ regions and the bipolar motion of  outflows  are mixed together, which cannot be resolved by the limited resolution (about $52''$) of the Purple Mountain Observatory Delingha  (PMODLH) 13.7 m  millimeter telescope. We assume that regular shapes of bubbles indicating relatively simple environments, which make it possible to do outflow and bubble structure analysis.  Unfortunately, only 3 of the 12 outflow candidates identified by \citet{2010ApJ...709..791B} are included in our samples.   
 
In Table~\ref{Tab:bubbles}, we list the parameters of observed bubbles.  From left to right, the columns are the bubble identity,  galactic longitude, galactic latitude, inner X diameter, inner Y diameter, outer X diameter, ellipse position angle (from the Y axis), velocity of ionized gas, flux density at 1.4 GHz, and the distance.  Bubbles in the same region are bracketed together. The morphological parameters are provided by \citet{2012MNRAS.424.2442S}. The integrated flux density of \HII\ regions at 1.4 GHz  are offered by~\citet{1998AJ....115.1693C}. Distances of these bubbles are generally provided by~\citet{2010A&A...523A...6D}, and the near distance, which is more statistically realistic~\citet{2005IAUS..227..174S}, is adopted when the kinematic distance is ambiguous.

Observations were performed with the PMODLH 13.7 m  millimeter telescope at Qinghai station, latitude 37\deg22$'$.4, from 30 May to 27 December, 2013, employing a superconducting array receiver (SSAR) with 3$\times$3 beam array  running in  sideband separation mode~\citep{shan2012development}. Spectral analysis was performed using the fast Fourier transform spectrometer (FFTS) with velocity resolution 0.16 \kms. Details about the telescope are described in the PMODLH status report\footnote{\url{http://www.radioast.nsdc.cn/zhuangtaibaogao.php}}.

Five molecular lines were observed: \cof, \cos, \cot, \hcn, and \hco. Observations of these three CO lines were performed simultaneously, separated from observations of  \hcos\ and \hcns. The half power beamwidth (HPBW) of the telescope is approximately 52$''$ at 110.2 GHz, with pointing and tracking uncertainty approximately 5$''$ and 1.4$''$, respectively. The beam efficiency is approximately 46\% for \cofs, 51\% for \coss, and 55\%  for  \cots, \hcns, and \hcos. The system temperature was approximately 280 K for  \cofs, 185 K for \coss\  and \cots, and 140 K for \hcns\ and \hcos. The on the fly (OTF) scanning model was used for the observations, with scanning rate of 50$''$ s$^{-1}$, and spectra were recorded every 0.3 s. For each bubble, the total observation time for CO was approximately 2 hours,  and 5 hours for \hcns\ and \hcos.  Further observations were taken for bubble regions with weak emissions.

\begin{table}[H]
 \scriptsize
  \centering

\setlength{\tabcolsep}{3.5pt}

 \caption{Parameters of the observed bubbles.\label{Tab:bubbles}}
  \begin{center}

 \begin{tabular}{clcccccccccccc}
   \tableline\tableline
    
 &Name\tablenotemark{a}   &  l$\rm ^b$ &  b$\rm ^b$ &   iXdiam$\rm ^b$&   iYdiam$\rm ^b$  &oXdiam$\rm ^b$    &PA$\rm ^b$ &   V$_{\rm ion}$$\rm ^h$  &  S$_{\rm 1.4 GHz}$  &Distance$\rm ^{c}$   \\
 
 & &(deg)&(deg)&(arcmin)& (arcmin)& (arcmin)  &$^\circ$ &  (\kms)&(mJy)&(kpc)  \\
   \tableline

&N4   &      11.895   &  +0.751 & 3.5 & 3.3 & 5.8 &   7  &       25.1 & 2109.9&       3.2  \\
      
&N14      &  13.992	  & -0.129  & 2.7 & 2.6 & 4.3 &      38 &    36   &1462.6&       3.7   \\
      
&N37   &   25.291   &  +0.294     & 2.7 & 3.4 & 4.0 &  	13    &  39.6  &222.4   &    12.6 \\

&N44  &  26.822	  & +0.383 & 1.8 & 1.9 & 2.8 &  48   &   82.0  &58.4   &  5.0/10.1\\
      
&N49 &   28.828  & 	-0.228 & 2.3 & 2.3 & 3.6 &  30  &   90.6   &642.7    &    5.5\\

\ldelim \{ {3}{1 mm}&MWP1G032057+000783$\rm ^b$  &   32.057  & 	+0.078 & 1.2 & 1.1 & 1.9 &   8 &     96.3 &100.0 & 8.4 \\
      
&N55  &   32.099  & 	+0.092 & 1.0 & 1.1 & 1.5 &  	59 &     93.0 &  43.4  & 8.4 \\
      
&MWP1G032158+001306$\rm ^b$  &  32.158  & 	+0.131 & 1.1 & 1.2 & 1.8 &  	16 &     95.0  &  653.3 & 8.4 \\

\ldelim \{ {2}{1 mm}&N74&   38.907  & 	-0.439  & 2.4 & 2.1  & 3.9  &  49   &  42.1 &7.3&  2.8/10.4\\
 
&N75 &   38.928  & 	-0.386  & 1.2 & 1.3 & 2.0 &  25  &  42.1 & 6.2   &   2.8/10.4\\
      
&N82 &   42.104   & 	-0.623  & 2.7 & 2.6 & 4.7 &  10  &  66.0   & 786.4  &   4.3$\rm ^d$ \\
      
 \ldelim \{ {2}{1 mm}&N89  &  43.734  & 	+0.117 & 1.8 & 2.0 & 2.8 &  15	 &   73.1   &14.1    & 6.1\\
      
&N90  &  43.775  & 	+0.061 & 2.5 & 2.5 & 3.9 &  9	 &  70.5  &168.3    &  6.1\\
      
&N95 &   45.387  & 	-0.715   & 2.8 & 2.8 & 4.0 &  	27 &      52.5 &   315.6  & 8.0\\
      
&N105 &   50.078  & 	+0.570 & 1.3 & 1.3 & 2.4 &  8 &   -1.1$\rm ^g$  &  101.2 & 11.2 $\rm ^i$\\
      
&N123 &   57.544  & 	-0.282	 & 2.1 & 2.0 & 3.3 &   	37 &  2.0$\rm ^e$ &  787.6    &8.6$\rm ^e$\\
      
\ldelim \{ {2}{1 mm}&N132 &   63.121	  & +0.386 & ... & ...	 & 0.3 &  ...  &    16.4	 & 9.6  &  2.1$\rm ^f$\\
      
&N133 &   63.163	  & +0.441  & 3.9 & 3.7 & 6.8 &  22 &   16.4 &  2900.4 &  2.1$\rm ^f$\\

\tableline
\end{tabular}   
 \flushleft
\tablecomments{0.9\textwidth}{\item[] (1) The columns are, from left to right, the bubble identity,  galactic longitude, galactic latitude, inner X diameter, inner Y diameter, outer X diameter, ellipse position angle (from the Y axis), velocity of \HII\ region, flux at 1.4 GHz, and distance. Bubbles in the same region are marked with brackets.  \item[] (2) For the distance, we adopted 5.0 kpc for N44 and 2.8 kpc for N74 and N75. \item[] (3) References: a~\citet{2006ApJ...649..759C}; b~\citet{2012MNRAS.424.2442S}; c~\citet{2010A&A...523A...6D}; d~\citet{2010MNRAS.407..923S}; e~\citet{2003ApJ...587..714W}; f~\citet{2010ApJ...716.1478W}; g~\citet{2011ApJS..194...32A}; h~\citet{2014ApJS..212....1A}; i~\citet{2012ApJ...754...62A}.} 
\end{center}
 
\end{table}

The typical  observed size for bubbles was approximately $10' \times 10'$, which is large enough to include the molecular clouds surrounding them.  Data reduction was achieved with the Grenoble Image and Line Data Analysis Software (GILDAS) package. After replacing bad channels with adjacent channels in the spectra, the data was regridded to 30$''$$\times$30$''$ pixels and mosaicked to a FITS cube file after calibrating. The root mean square (rms) noise is approximately 0.45 K for \cofs\ at 0.16 \kms\ resolution, 0.20 K for  \coss\ and  \cots\ at 0.17 \kms\ resolution,  and 0.07 K for \hcns\ and \hcos\ at a resolution of 0.18 \kms.

 In Table~\ref{Tab:lineParameters}, from left to right, the columns are molecular line, rest frequency, critical density, half-power beam width (HPBW), system temperature, main beam efficiency, velocity resolution, and the root mean square (rms) of noise. These critical densities are calculated assuming a kinematic temperature of 10 K. The critical densities of three CO lines are given by~\citet{2010ApJ...718.1019Y}, while the values for \hcns\ and \hcos\ are calculated by~\citet{2015PASP..127..299S}.

\begin{table}[H]
 \footnotesize
     
  \begin{center}
 
   \caption{Observational parameters of the molecular lines.\label{Tab:lineParameters}}

      \begin{threeparttable}
 \begin{tabular}{lccccccccc}
   \tableline\tableline
 
 Molecular line  &Rest frequency & Critical density\tnote{a}&HPBW  & $T$$_{\rm sys}$  & $\eta_{\rm mb}$&  $\delta v$  & rms noise   \\
 
  (J=(1-0)) &(GHz)&($10^3$ cm$^{-3}$)& ($''$) & (K) &  &(\kms) &   (k)  \\
   \tableline

\cofs  &   115.271204&  0.0067-0.64\tnote{b}& 49 &220-300  &45.9\%  & 0.16 & 0.5 \\
\coss &  110.201353  &0.38-1.9\tnote{b}& 51    &140 -190  &51.1\%  & 0.17&  0.2 \\
\cots &  109.782183  &1.6-1.9\tnote{b}&   50    &140 -190  &   54.6\% & 0.17& 0.2\\
\hcns &  88.6318473 & 68\tnote{c}&   56   &140 -190    &   55.0\%& 0.18&  0.07\\
\hcos & 89.1885260 & 470\tnote{c}& 58    &140 -190   &  55.7\%& 0.18&  0.07\\

\tableline
\end{tabular}
\begin{tablenotes}
\item[a]  For the kinematic temperature of 10 K. 
\item[b]  \citet{2010ApJ...718.1019Y}  
\item[c]  \citet{2015PASP..127..299S} 
\end{tablenotes}
 
      \end{threeparttable}
\end{center}

\end{table}

\section{RESULTS}


Except N89 and N90  where no \cots\ and  only weak  \hcos\ emission was  detected,  all the molecular lines were detected for the other bubble regions. Two bubbles, N44 and N123,  show relatively low signal to noise ratio (SNR).  For all bubbles, the SNR of \cots\ are generally lower than the other four lines,  because of its equal integration  time with \cofs\ and \coss, but comparable  low antenna temperature  with \hcns\ and \hcos.  

The physical properties for  molecular clouds and clumps associated with these bubbles were calculated using three CO isotopic lines. The velocities were determined according to  ionized gas  velocities~\citep{1989ApJS...71..469L,2014ApJS..212....1A}, CO velocities~\citep{2010ApJ...709..791B}, and were confirmed by the spatial coherence between the \coss\ integrated intensity and bubbles. We also searched for outflows around bubbles using the profiles of \cofs\ and \hcos.  Details of these results are presented in this section.

\subsection{Physical properties of molecular clouds around bubbles}

\begin{figure} [H]
\center
\includegraphics[width=0.99\textwidth ]{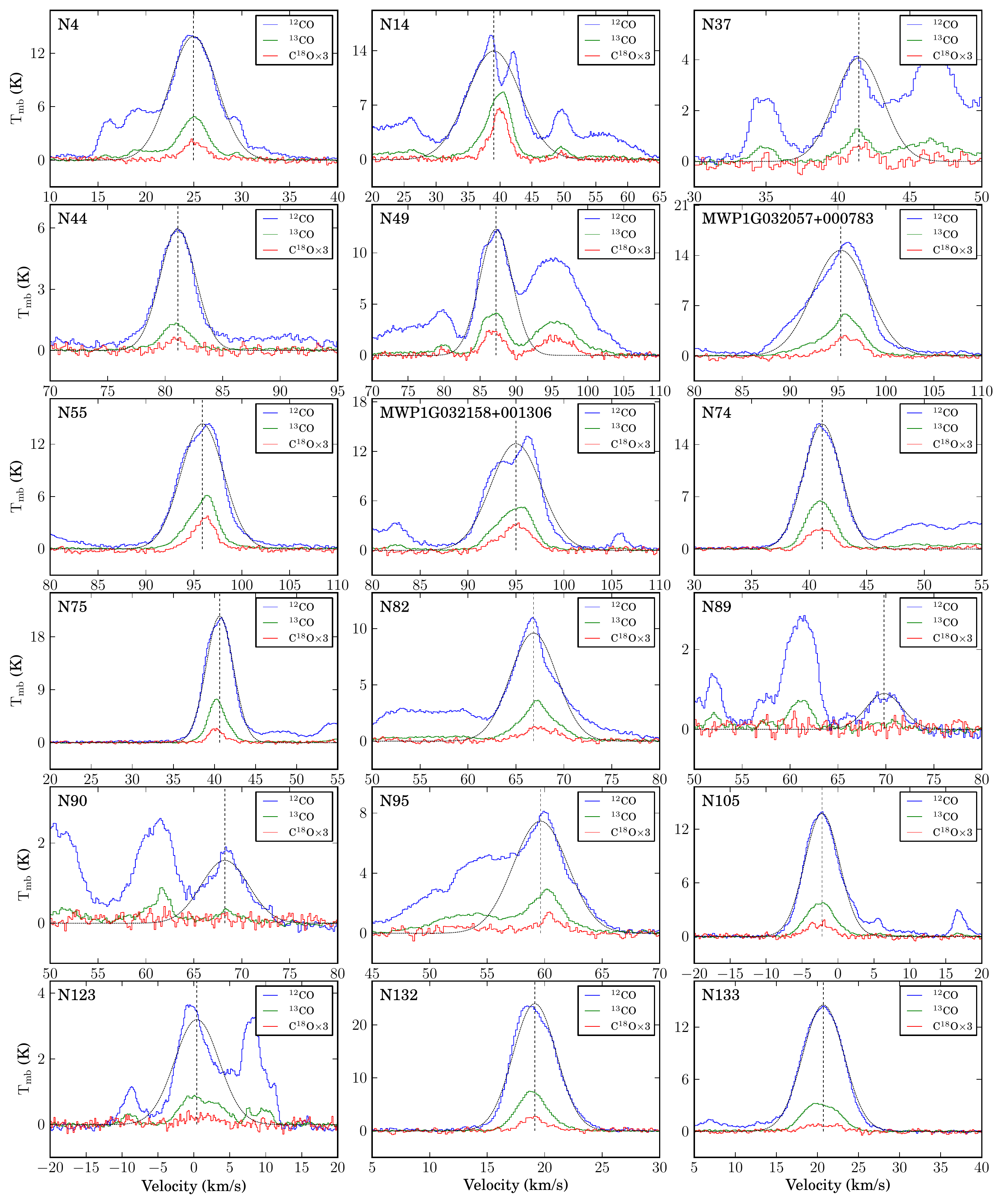}
\caption{Average spectra for all observed bubbles. The \cots\ lines are amplified by a factor of 3 for clarity.  The black dotted lines delineate the fitted Gaussian curves of \cofs, and the dashed vertical lines mark the center velocities of these Gaussian curves.} 
 \label{Fig:allspectral}
\end{figure}

To demonstrate dynamic characteristics of bubbles, we averaged the \cofs\ and \coss\ spectra over the bubble areas which are squares with the side length equal  to the outer major axis provided by~\citet{2012MNRAS.424.2442S}. The spectral profiles are shown in Figure~\ref{Fig:allspectral}.  The fitted Gaussian curves of \cofs\ are plotted in Figure~\ref{Fig:allspectral}. Some average profiles display significant deviation from Gaussian curves. Generally, this could be due to different components with adjacent velocities along the line of sight, which show multi-peaks in the profile. Since \coss\ lines are more  Gaussian than the \cofs\ lines, another alternative is that the outer region of clouds are perturbed, which are mainly traced by \cofs. 

We estimated the  physical  properties of each bubble, with the assumption of local thermodynamic equilibrium (LTE). \cof\ is assumed to be optically thick, whereas \cos\ is generally optical thin, which means we could estimate the excitation temperature (T$_{\rm ex}$) and optical depth ($\tau$), respectively, \citep{1998AJ....116..336N, 2013RAA....13..921L} 
\begin{equation}
 T_{\mathrm{ex}} = 5.53\{\mathrm{ln}[1+\frac{5.53}{T_R^*(^{12}\mathrm{CO})+0.819}]\}^{ -1},
\label{eq:Tex}
\end{equation}
and
\begin{equation}
 \tau(\mathrm{C^{13}O}) = -\mathrm{ln}(1-\frac{T_R^*(^{13}\mathrm{CO})}{ \frac{5.29}{\mathrm{exp}(5.29/T_{\mathrm{ex}}-1)}-0.89 }),
\end{equation}
where $T_R^*(^{12}\mathrm{CO})$ and $T_R^*(^{13}\mathrm{CO})$ are the main beam temperatures of \cofs\ and  \coss, respectively. The optical depths are  generally below 0.5, which indicates the optical thin assumption is mostly valid.

We determined the molecular cloud  angular sizes, A, associated with bubbles from the ellipse centers  and outer axes, as given in~\citet{2012MNRAS.424.2442S}. The estimated diameters of molecular clouds were obtained after deconvolving the telescope beam, 
\begin{equation}
 l=D\sqrt{ 4A/\pi-\theta_{\rm MB}^2 }, 
\end{equation}
 where D is the distance to these bubbles, and $\theta_{\rm MB}$ is the beam size of the telescope.

The second step to deriving the molecular cloud masses was to calculate the averaged $\rm H_2$ column density,  $N(\rm H_2)$,  using averaged \cofs\ and \coss\ spectra over the bubble area \citep{1998AJ....116..336N}, 
\begin{equation}
 N(\mathrm{H_2}) = 1.49\times10^{20}[1-\mathrm{exp}(\frac{-5.29}{T_{\mathrm{ex}}})]^{-1}\times[\frac{\int T_R^*(^{13}\mathrm{CO})dV}{ 1 \rm K\ km s^{-1} }] \ \rm  cm^{-2},
\end{equation}
where $\int T_R^*(^{13}\mathrm{CO})dV$ is the area of fitted \coss\ Gaussian curve. The abundance ratio, $N(\rm H_2)$/$N$(\coss), was set to $7\times10^5$, following~\citet{1982ApJ...262..590F}.

Finally, we convert $N(\mathrm{H_2})$ to cloud masses,
\begin{equation}
\label{eq:mass}
M = \mu m_{\mathrm{H_2}} N{(\rm H_2)}\pi l^2/4,
\end{equation}
where $m$ is the mass of the hydrogen molecule, and  $\mu$ is the ratio of  gas to  hydrogen by mass, approximately 1.36~\citep{1983QJRAS..24..267H}.  
 
 Following~\citet{2014A&A...568A...4T}, we estimated the age for each bubble using parameters listed in Table~\ref{Tab:bubbles}.  \citet{2014A&A...568A...4T} provided a new method to determine the age and size of  \HII\  regions using  3D simulations. They included the internal turbulence of molecular clouds surrounding the bubbles, which we believe is more accurate. 
  
 Details of the physical parameters of the bubbles are shown in Table~\ref{Tab:bubbleMass}, where we list the parameters of fitted Gaussian curve, center velocity, peak temperature, and  full width at half maximum (FWHM) of \cofs\ and \coss, respectively. We also list the average H$_2$ columns densities, total area of the counted molecular clouds, and  the corresponding mass under the assumption of LTE. It should be emphasized that these masses only count the molecular clouds within bubble  squares mentioned above, instead of the entire molecular clouds shown in the maps. 

From Table~\ref{Tab:bubbleMass}, we can see that the masses of clouds within bubble squares are in the range of 100 - 19000 solar mass, and their ages are in the range of 0.3 - 3.7 Myr. 
 
\begin{table}[H]
 \footnotesize
  \centering
  \begin{center}
 
 \caption[]{Physical parameters of the molecular clouds.\label{Tab:bubbleMass}}
    
 \setlength{\tabcolsep}{1.5pt}
       \begin{threeparttable}
 \begin{tabular}{lccccccccccccc}
   \tableline\tableline

 Name  & \cofs$_v$ &\cofs$_{\rm peak}$    & \cofs $_{\Delta V}$ &\coss $_v$  &\coss $_{\rm peak}$  &\coss $_{\Delta V}$ & $N_{\rm H_2}$  & Area\tnote{a}       &   $M_{\rm LTE}$ & Age\tnote{b}  \\
& (\kms)&(K)& (\kms) & (\kms)&  (K) & (\kms) & ($10^{21} \rm cm^{-2}$) &$pc^2$   &$10^3   M_{\odot}$& (Myr) \\
   \tableline

N4 & 24.9 &  13.9& 5.8& 25.0& 4.6& 3.8 &10.1 &35   & 7.7 & 0.8 \\ 
N14 & 39.0 &  14.0& 9.8& 39.9& 8.4& 4.7 &22.6 &23   & 11.2 & 0.6   \\ 
N37 & 41.5 &  4.1& 3.9& 41.6& 1.2& 2.3 &0.8 &406   & 7.3 & 3.7   \\ 
N44 & 81.1 &  6.0& 3.6& 81.0& 1.3& 2.9 &1.3 &26   & 0.8 & 1.7  \\ 
N49 & 87.2 &  12.3& 5.3& 87.1& 4.3& 4.0 &8.9 &46   & 8.8 & 1.2  \\ 
MWP1G032057+000783 & 95.3 &  14.7& 6.9& 95.7& 5.2& 4.9 &15.0 &37   & 12.1 &  1.2  \\ 
N55 & 95.9 &  14.3& 5.3& 96.1& 5.7& 3.6 &12.0 &30   & 7.7 & 1.7 \\ 
MWP1G032158+001306 & 95.0 &  12.9& 6.2& 95.0& 5.2& 4.5 &12.7 &37   & 10.3 &  0.6 \\ 
N74 & 41.1 &  16.8& 3.9& 41.0& 6.5& 2.8 &11.9 &13   & 3.5 &  0.8 \\ 
N75 & 40.6 &  21.5& 3.9& 40.4& 7.1& 2.7 &14.8 &5   & 1.6 & 0.3  \\ 
N82 & 66.8 &  9.6& 5.7& 67.3& 3.2& 4.5 &6.4 &43   & 6.0 & 0.9  \\ 
N89 & 69.8 &  0.9& 4.4& 70.3& 0.2& 3.4 &0.1 &39   & 0.1 & 1.8  \\ 
N90 & 68.2 &  1.6& 5.8& 68.4& 0.2& 4.6 &0.2 &64   & 0.3 & 2.2  \\ 
N95 & 59.6 &  7.5& 5.6& 60.0& 2.7& 4.0 &4.1 &110   & 9.8 & 2.9  \\ 
N105 & -2.2 &  13.8& 5.7& -2.4& 3.8& 4.6 &9.9 & 96  &  24.4 & 1.6  \\ 
N123 & 0.4 &  3.2& 7.3& 1.0& 0.8& 6.7 &1.5 &107   & 3.4 & 1.4 \\ 
N132 & 19.1 &  24.0& 4.4& 19.0& 7.3& 3.2 &19.6 &1   & 0.2 &  0.3 \\ 
N133 & 20.7 &  14.6& 5.9& 20.5& 3.3& 5.1 &9.9 &30   & 6.4 & 0.4 \\

\tableline
\end{tabular} 

 \begin{tablenotes}
\item[a] The areas are determined using the parameters of bubbles given by~\citet{2012MNRAS.424.2442S}.
\item[b]   The method of estimating ages are following \citet{2014A&A...568A...4T}. 
\end{tablenotes}
 
      \end{threeparttable}
 
\end{center}
\end{table}

\subsection{Molecular Clumps } \label{section:molecularclump}

Following \citet{
2007ARA&A..45..339B} and \citet{2015ARA&A..53..583H}, we refer clumps and cores as substructures of molecular clouds, and their typical sizes are about 0.3-3 and 0.03-0.2 pc, respectively. The PMODLH telescope can only resolve about 0.5 pc for the nearest bubble with a distance of 2.1 kpc, and consequently, the dense parts of the molecular clouds identified by the telescope are generally clumps.

Although \hcos\ and \hcns\ lines possess higher critical densities, they are readily optically thick. \cots, however, is always tracing regions with high column densities. Therefore, we used \cots\ to identify  clumps, which were subsequently confirmed by \coss\ lines whose SNR are relatively high.

In total, 23 molecular clumps are found around 18 bubbles, and each bubble contains at least one molecular clump at its border. Since the \coss\ lines  tend to be optically  thick at the peak of clumps, we estimated the mass of molecular clumps using \cofs\ and \cots\ lines, with the assumption that they are optically thick and thin, respectively. N89 and N90 show low SNR for \cots, and consequently we used \coss\ instead for these two bubbles.   
 
Equations~\citep{1986ApJ...303..416S, 2014MNRAS.443.2264L} we used to estimate  \cots\ column density are  
\begin{equation}
\tau(\mathrm{C^{18}O}) = -\mathrm{ln}(1-\frac{T_R^*(\mathrm{C^{18}O})}{ \frac{5.27}{\mathrm{exp}(5.27/T_{\mathrm{ex}})}-0.89 }),
\end{equation}
and
\begin{equation}
  \frac{N(\rm C^{18}O)}{\rm cm^{-2}}= 4.77\times10^{13}\frac{T_{\mathrm{ex}}+0.88}{ \mathrm{exp}(-5.27/T_{\rm ex})} \frac{\tau(\mathrm{C^{18}O})}{1-\mathrm{exp}({-\tau(\mathrm{C^{18}O})})} \frac{\int T_{\mathrm{mb}}\mathrm{d}v}{\rm K\ km\ s^{-1} },
\end{equation}
where $\int T_{\mathrm{mb}}\mathrm{d}v$ is the  area under the fitted \cots\ Gaussian line, and $T_{\rm ex}$ is estimated using Eqn. \ref{eq:Tex}.  The fractional abundance of $\rm C^{18}O/H_2$ is approximately $1.7\times10^{-7}$ ~\citep{1982ApJ...262..590F}.

The angular extent of each clump was determined by  the contour of the half  integrated intensity peak, $A_{1/2}$, and the diameters of the molecular clumps estimated using
 
\begin{equation}
l_{\mathrm{clump}}=D\sqrt{ 4A_{1/2}/\pi-\theta_{\rm MB}^2 },
\end{equation}
 where $D$ is the distance to clumps, and $\theta_{\rm MB}$ is the beam size of the telescope. The clump mass is then given by Eqn.~\ref{eq:mass}.

The virial masses of  clumps are calculated by 
\begin{equation}
\frac{M}{(M_\sun)} = k_2\frac{l_{\mathrm{clump}}/2}{(\mathrm{pc})}(\frac{\Delta V}{(\rm km\ s^{-1})})^2
\end{equation}
where $k_2$ is 210 for $\rho (r) = \rm constant$, and $\Delta V$ is the full width at half-maximum (FWHM) in unit of \kms~\citep{1988ApJ...333..821M}.  We assume the picture that molecular cores spreads throughout the molecular clumps homogeneously, and consequently, we adopt $\rho (r) = \rm constant$ rather than $\rho (r) = 1/r$ which might be more accurate for single cores.  For comparsion, following~\citet{2010ApJS..188..123R}, we calculated the dust mass of the corresponding BGPS sources of these clumps, which is displayed in Table~\ref{Tab:clumps}. The equation we used to estimate the dust mass is 
$$M=\frac{d^2S}{B_\nu(T) _\nu} = 13.1 M_{\odot} (\frac{d}{1~\mathrm{kpc}})^2(\frac{S_\nu}{1~\mathrm{Jy}})[\frac{\mathrm{exp}(13.0~K/T_{\rm ex})-1}{\mathrm{exp}(13.0/20)-1}]$$
 where $T_{\rm ex}$ is estimated using Eqn. \ref{eq:Tex}, $d$ is the distance, and $S_\nu$ is the integrated flux density in the catalog of \citet{2010ApJS..188..123R}.

 In Table~\ref{Tab:clumps}, we summarize the physical properties of molecular clumps. From left to right, the columns are clumps name, peak position of integrated  intensity, integrated velocity range, center of fitted Gaussian curve, sigma of fitted Gaussian curve, peak value of fitted Gaussian curve, optical depth, column density, clump size, mass with LTE assumption, viral mass of the clump,  dust mass of the clump, the line used, and outflow flag. If a clump is associated with an outflow candidate, it is marked by Y.  In channel maps, the positions of all these clumps are marked with green crosses.

\begin{table}[H]
 \tiny
  \centering

\setlength{\tabcolsep}{2.8 pt}
  \begin{center}
 
 \caption{Physical parameters of the molecular clumps. \label{Tab:clumps}}
    \begin{threeparttable}
 
 \begin{tabular}{lcccccccccccccc}
   \tableline\tableline

Clumps& Peak position  &$V_{\rm range}\tnote{a} $& $V_{\rm center}$    & $\Delta V$ & Tex  & $\tau$ &Column density& Area    & $M_{\rm LTE}$& $M_{\rm vir}$  & $M_{\rm dust}$\tnote{b} &Lines &Outflow\tnote{c}\\
 & (l,b)&(\kms)& (\kms)& (\kms) &(k)&   &10$^{15}$cm$^{-2}$ &$pc^2$   &$10^3M_{\odot}$&$10^3M_{\odot}$ &$10^3M_{\odot}$  \\
   \tableline
  N4A & (11.836, 0.742)&20.1 - 29.3 & 24.7 &2.7 & 25.0 &0.08 &7.4 &4.2   & 4.1 &1.8 &-- &\cots   \\ 
 N4B & (11.903, 0.717)&21.1 - 28.7 & 24.9 &2.2 & 29.5 &0.09 &9.6 &2.7   & 3.5 &1.0   &-- & \cots   \\ 
 N14A & (14.009, -0.179)&35.4 - 45.7 & 40.5 &3.0 & 24.5 &0.20 &20.7 &5.3   & 14.7 &2.5   & 0.22& \cots &Y  \\ 
 N37A & (25.320, 0.276)&39.1 - 44.9 & 42.0 &1.7 & 25.1 &0.06 &3.6 &26.3   & 12.7 &1.8   &1.2 & \cots   \\ 
 N44A & (26.844, 0.371)&77.7 - 83.7 & 80.7 &1.7 & 16.6 &0.08 &2.4 &2.5   & 0.8 &0.6   & 0.15& \cots   \\ 
 N49A & (28.834, -0.253)&82.5 - 91.5 & 87.0 &2.6 & 21.4 &0.19 &13.1 &7.6   & 13.3 &2.3  & 1.8 &\cots   \\ 
 N49B & (28.842, -0.212)&82.2 - 89.9 & 86.0 &2.3 & 22.9 &0.11 &7.3 &5.0   & 4.9 &1.4  & 0.67 &\cots   \\ 
  
 MWP1G032057+000783A  & (32.016, 0.063)&89.5 - 105.6 & 97.5 &4.7 & 21.7 &0.12 &15.6 &26.6   & 55.4 &13.7  & 4.6 &\cots   \\ 
 MWP1G032158+001306A & (32.149, 0.138)&88.4 - 102.4 & 95.4 &4.1 & 19.5 &0.13 &11.4 &11.7   & 17.7 &6.9 & -- &\cots   \\ 
 N55A & (32.116, 0.088)&92.0 - 101.0 & 96.5 &2.6 & 24.3 &0.13 &11.1 &14.7   & 21.8 &3.2 & 1.9 &\cots &Y  \\ 
 N75A & (38.925, -0.353)&34.4 - 43.6 & 39.0 &2.7 & 32.3 &0.06 &9.0 &2.5   & 2.9 &1.4  &0.36&\cots   \\ 
 N74A & (38.950, -0.462)&37.8 - 45.9 & 41.9 &2.4 & 23.7 &0.15 &11.4 &3.3   & 5.0 &1.2&  0.23  &\cots   \\ 
 N74B & (38.925, -0.420)&37.0 - 44.4 & 40.7 &2.2 & 25.5 &0.09 &6.9 &3.1   & 2.9 &1.0 &  0.09 &\cots   \\ 
 N82A & (42.120, -0.595)&63.2 - 72.9 & 68.0 &2.8 & 21.1 &0.09 &6.6 &5.8   & 5.1 &2.3 & --  &\cots   \\ 
 N90A & (43.798, 0.051)&64.6 - 72.2 & 68.4 &2.3 & 12.9 &0.31 &5.1 &6.2   & 0.5 &1.5 & -- &\coss   \\ 
 N89A & (43.723, 0.143)&68.2 - 73.3 & 70.7 &1.5 & 10.9 &0.32 &2.4 &7.7   & 0.3 &0.7 & -- &\coss   \\ 
 N95A & (45.389, -0.747)&55.2 - 65.5 & 60.4 &3.0 & 18.3 &0.07 &4.2 &17.4   & 9.8 &4.5& --  &\cots   \\ 
 N105A & (50.077, 0.561)&-8.9 - 1.6 & -3.6 &3.1 & 21.9 &0.06 &4.9 &24.2   & 16.0 &5.7 & --  &\cots   \\ 
 N105B & (50.077, 0.594)&-8.4 - 4.8 & -1.8 &3.9 & 18.6 &0.04 &3.2 &46.4   & 19.8 &12.1&  --  &\cots  &Y \\ 
 N123A & (57.571, -0.288)&-7.7 - 6.4 & -0.6 &4.1 & 14.4 &0.04 &2.0 &19.8   & 5.3 &9.0  & 1.1 &\cots   \\ 
 N133A & (63.115, 0.406)&15.8 - 22.9 & 19.4 &2.1 & 24.7 &0.09 &6.9 &1.6   & 1.4 &0.7 &  0.06 &\cots   \\ 
 N133B & (63.140, 0.439)&17.1 - 28.0 & 22.5 &3.2 & 35.3 &0.03 &7.1 &2.1   & 2.0 &1.8 &  0.04 &\cots &Y   \\ 
 N133C & (63.248, 0.431)&13.8 - 27.3 & 20.5 &4.0 & 14.7 &0.07 &3.6 &1.0   & 0.5 &1.9 & --  &\cots   \\ 

\tableline
\end{tabular}

\begin{tablenotes}
\item[a] The velocity ranges are the spectral line ranges which are used to fit Gaussian curve and integrate intensities which are used to determine the areas and masses of clumps.

\item[b]  The dust mass was calculated following~\citet{2010ApJS..188..123R} 

\item[c] If a clump is associated with an outflow candidate, it is marke by Y. 
 
\end{tablenotes}

      \end{threeparttable}

 \vspace{0.4cm}
\flushleft
 
\end{center}
\end{table}

\subsection{Expanding H \uppercase\expandafter{\romannumeral2}\  Regions}

\begin{figure} [H]
\center
\includegraphics[width=0.9\textwidth]{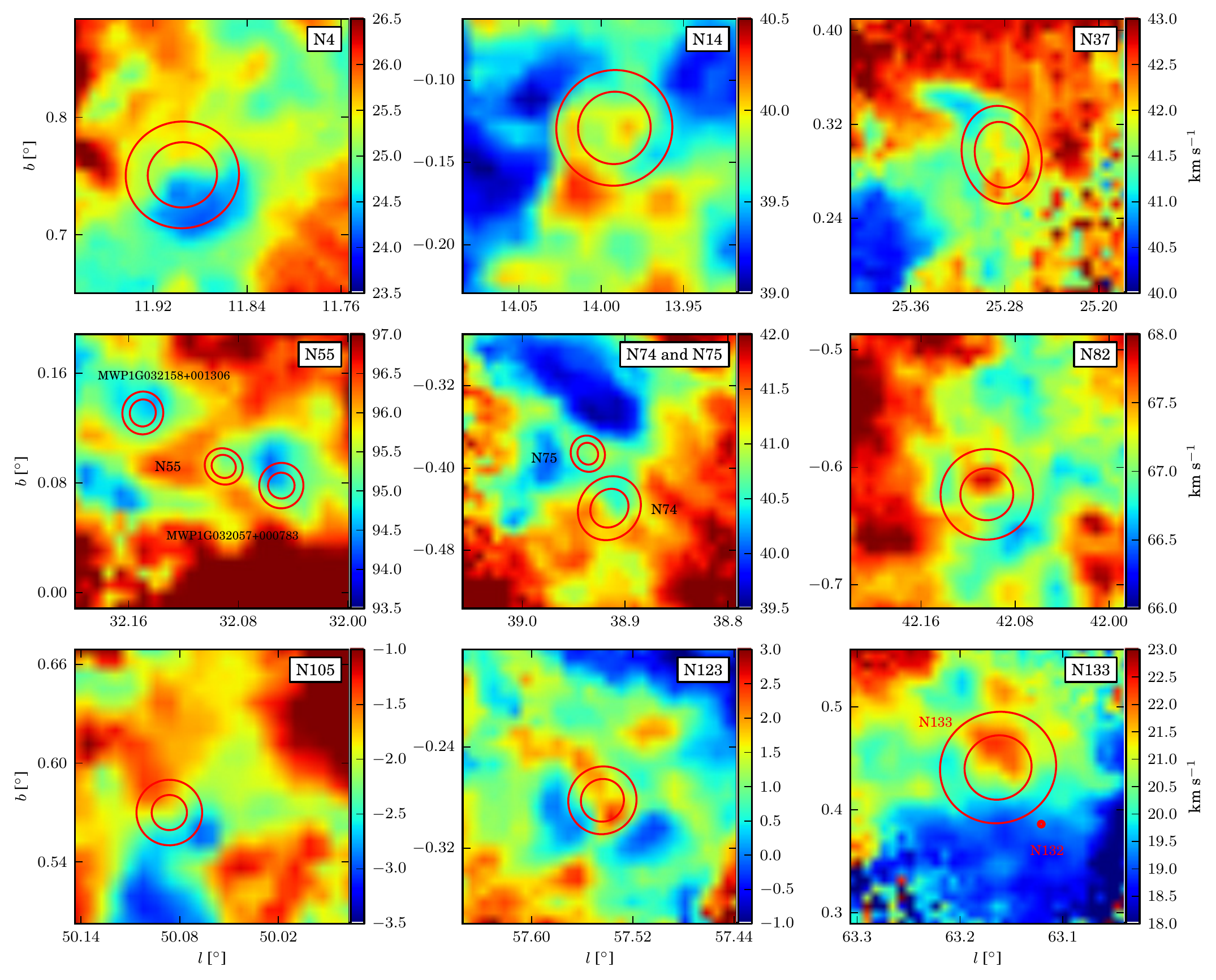} 
\caption[]{Intensity weighted velocity distribution of \cof\ for 9 bubble regions.} 
\label{Fig:velocitygrad}
\end{figure}

 In order to check the velocity field near infrared bubbles, we calculated the moment 1 (the velocity weighted by intensity) of \cof\ for all bubbles. \cof\ possess the highest SNR, and it is more sensitive to the motion of molecular clouds. 9 bubble fields show relatively evident velocity gradients near their borders, as shown in Figure ~\ref{Fig:velocitygrad}. Interestingly, the velocity field around three bubbles, N14, N74 and N133, display  arc-like  structures, which are probably tracing the expanding shell of \HII\ regions.  
 
  These arc-like structures are similar to bubble N6 studied by~\citet{2014ApJ...797...40Y}, although their results show larger velocity dispersions. The magnitudes of velocity dispersions are comparable with those of N22~\citep{2012A&A...544A..39J}. These arc-like structures are probably  produced by the interaction between bubbles and their surrounding molecular clouds. This speculation is supported by the CO profiles shown in Figure~\ref{Fig:allspectral}. Some CO profiles  display significant deviation from Gaussian curves, indicating that the molecular clouds may be interacting with bubbles.

\subsection{Individuals} \label{section:individuals}

 Most of the three color images of bubbles below are composed of 20 cm continuum MAGPIS data shown in red, 1.1 mm continuum BGPS data shown in green, and 8 \mum\ GLIMPSE data shown in blue. For bubble N37, N74, and N75, where 20 cm continuum MAGPIS emissions are faint, we used VGPS data instead due to their high sensitivity to extended structures. For regions where BGPS data are unavailable, we adopted 0.87 mm from ATLASGAL data as an alternative. The positions of identified clumps are marked in all integrated intensity maps of five lines by purple crosses and in all channel maps of \cos\ by green crosses.

\subsubsection{N4}

Bubble N4 shows a superb ring  shape at 8 \mum\ with  a 20 cm continuum disk  enclosed in it. The radio recombination line (RRL) indicates the velocity of the local standard of rest (LSR) is 25.1 \kms\ \citep{1989ApJS...71..469L,2014ApJS..212....1A}, which is in  good agreement with the CO velocity, 24.9 \kms~\citep{2010A&A...523A...6D,2013RAA....13..921L}.  We adopted the near kinematic distance, 3.15 kpc, following ~\citet{2010A&A...523A...6D}.

\citet{2013RAA....13..921L} studied three CO  isotropic lines using the 13.7 m millimeter telescope at  Qinghai Station. They showed that CO emissions were well correlated with 8 \mum\  structure, which is more likely an inclined ring rather than an expanding spherical shell.  They also identified a 15 \Msun\ star with an  age of $\sim$ 1 Myr, which is probably the energy source of bubble N4. In addition, they claimed that they found a possible infall signature  which indicates triggered star formation process may exist  at the border of bubble N4.

\begin{figure} [H]
\center
\includegraphics[width=0.9\textwidth]{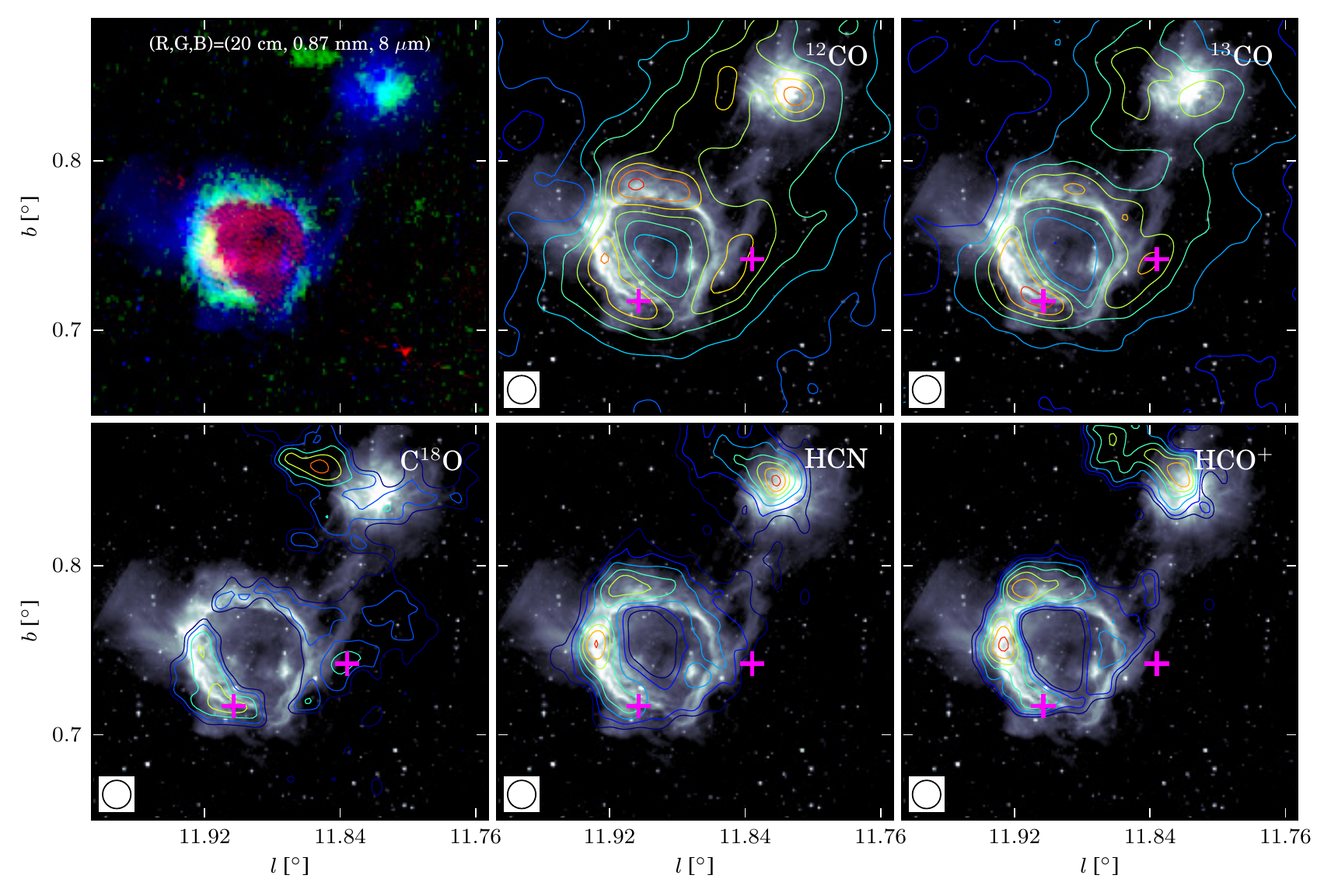}
\caption[]{Images of N4. The upper left panel is a three color image with 20 cm in red, 0.87 mm in green, and 8 \mum\ in blue. The other five panels are contours of five spectral line integrated intensity maps superimposed on the 8 \mum\ image with positions of clumps marked by purple crosses. The contour maps integrate from 21.0 \kms\ to 29.0 \kms. The rms ($\sigma$) of the five line maps, \cofs, \coss, \cots, \hcns, and \hcos, are 0.54, 0.26, 0.26, 0.08, and 0.09 K \kms, respectively, and these contours all begin at the 8$\sigma$ level, spacing with 29, 21, 4, 5, and 3$\sigma$, respectively.}
\label{Fig:N4Fiveline}
\end{figure}

\begin{figure} [H]
\center
\includegraphics[width=0.8\textwidth]{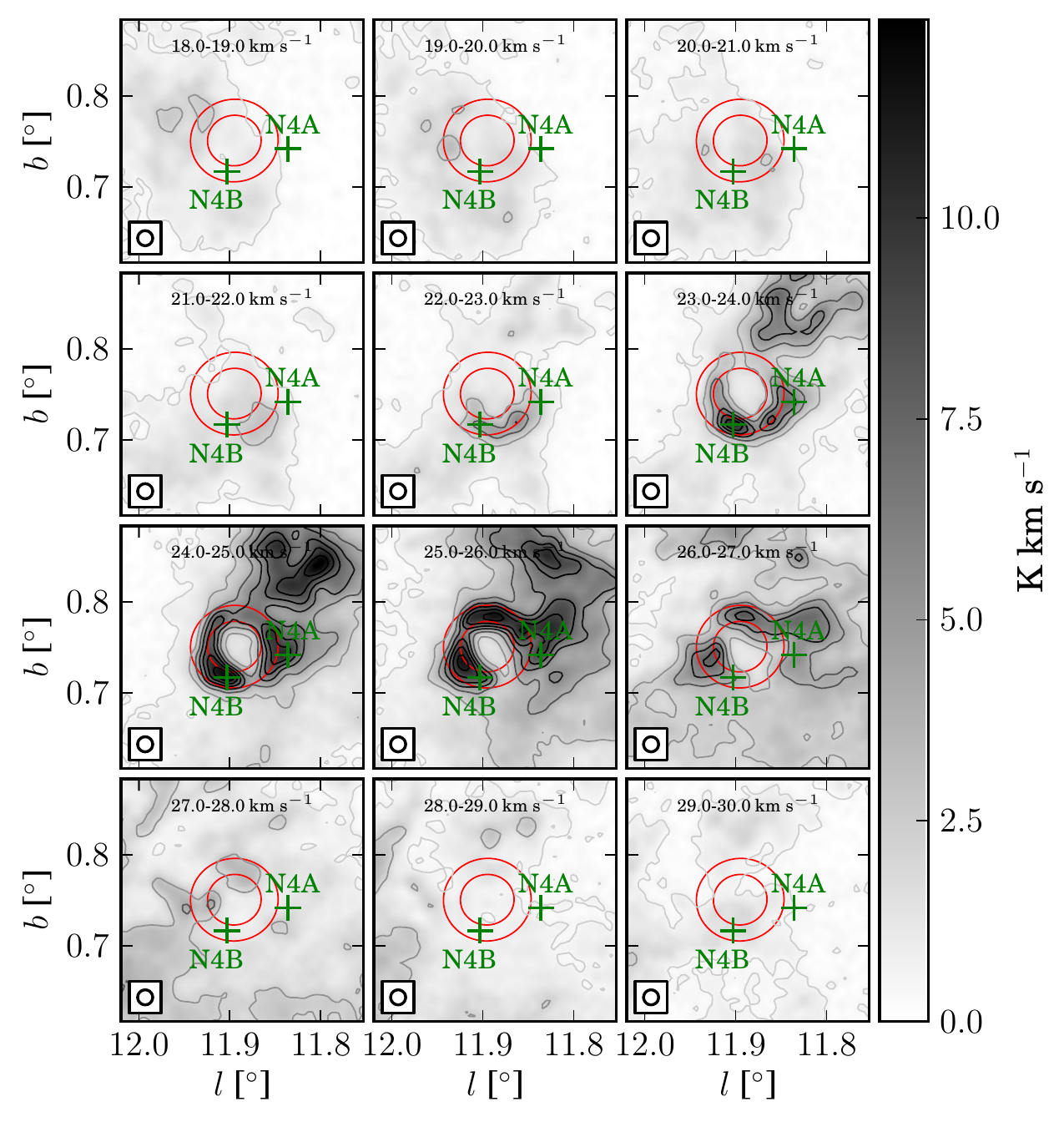}
\caption{Channel map  of \coss\ for bubble N4 from 18.0 \kms\ to 30.0 \kms\ with 1.0 \kms\ intervals. The red circles mark  the position of N4 from~\citet{2012MNRAS.424.2442S}, and the green crosses mark the peak positions of clumps. The rms ($\sigma$) of the image background is about 0.09 \intunit. The contour levels space linearly  from 8.0$\sigma$ to the peak with a step of 18.1$\sigma$. }
\label{Fig:N4chananelmap}
\end{figure}

Our CO observations show  similar results to \citet{2013RAA....13..921L},  and the \cots\ line reveals two molecular clumps, N4A and N4B shown in Figure~\ref{Fig:N4chananelmap}, at the border of N4. The integrated intensity contour maps of five molecular lines are shown in Figure~\ref{Fig:N4Fiveline},  and observation parameters are listed in Table~\ref{Tab:lineParameters} . The molecular cloud mass over the region of N4 is approximately 7.7$\times10^3 M_\sun$, which is comparable to $7\times10^3 M_\sun$, from~\citet{2013RAA....13..921L} using \coss. The morphology shown by \hcns\ and \hcos\ lines is in accordance with 8 \mum, which indicates that collect-and-collapse processes may be occurring in this region. However, we did not find any blue asymmetry at the position of the infall candidate identified by  \citet{2013RAA....13..921L}. Furthermore, both the blue and red ends of line profiles of  CO and \hcos\ are contaminated by components with similar velocities,  which prevents us form identifying outflows.

\subsubsection{N14}

This is a large open bubble with strong 8 \mum\ emissions and diffuse 20 cm continuum enclosed in it. To the south of N14, there is an infrared dark cloud whose position coincides with the 1.1 mm continuum. The RRLs indicate the velocity of N14 is approximately 36.0 \kms\ \citep{1989ApJS...71..469L,2014ApJS..212....1A}, which agrees well with the CO velocity of 40.3 \kms ~\citep{2010ApJ...709..791B}. We prefer the near distance of 3.7 kpc due to the strong infrared emission and H$\alpha$ emission, which is also suggested by \citet{2010A&A...523A...6D}.

\citet{2012ApJ...760...58S} observed seven spectral lines and 3 mm continuum in this field. Among their observations, the N$_2$H$^+$  emission, which is not exposed to the \HII\ region, shows that N14 is expanding into a very inhomogeneous cloud. They identified 10 compact dust sources  in the vicinity of N14 based on 3 mm continuum data, largely correlated with bubble N14, and however, no velocity gradient was detected, which is not consistent with the prediction of the expanding shock.  They proposed that triggered star formation have occurred, although their findings do not conform to a classic collect-and-collapse model.

\citet{2013MNRAS.429.1386D}  performed a  multi-wavelength study of  N14.  They argued that they found observational signatures of the collected molecular and cold dust material  around this bubble. However, they prefer a process of compression of the pre-existing dense clumps by the shock wave and/or small scale Jeans gravitational instabilities in the collected materials,  because of the disagreement between the ages of the \HII\ region and the fragmentation time of the molecular clouds.  A Class \uppercase\expandafter{\romannumeral1} YSO cluster traced by the YSO surface density contours indicates that star formation processes might have been triggered by the expansion of the \HII\  region.

\begin{figure} [H]
\center

\includegraphics[width=0.9\textwidth]{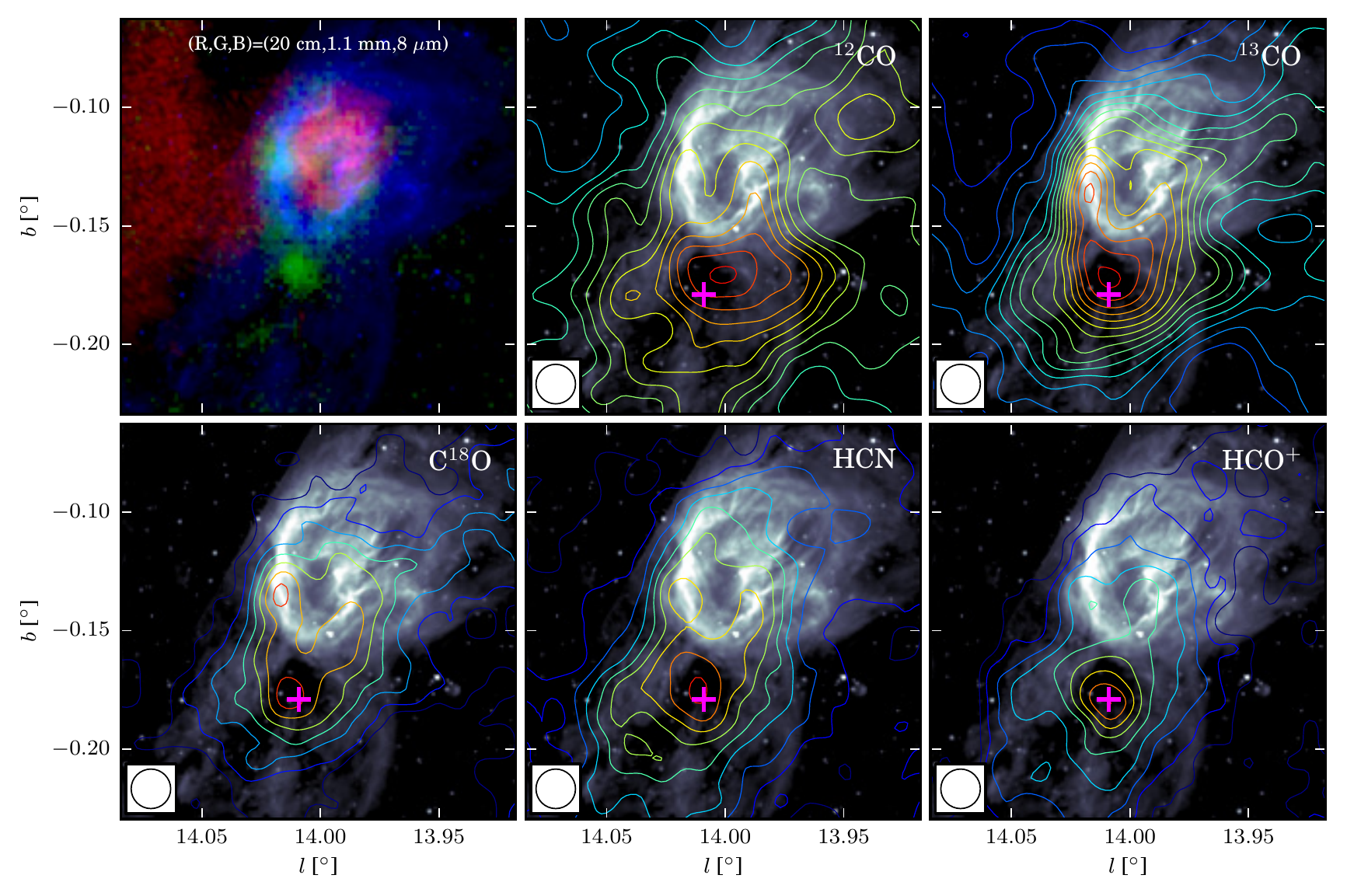}
 
\caption[]{Images of N14. The upper left panel is a three color image with 20 cm in red, 1.1 mm in green, and 8 \mum\ in blue. The other five panels are contours of five spectral line integrated intensity maps superimposed on the 8 \mum\ image with positions of clumps marked by purple crosses. The contour maps integrate from 36.0 \kms\ to 44.0 \kms. The rms ($\sigma$) of the five line maps, \cofs, \coss, \cots, \hcns, and \hcos, are 0.57, 0.32, 0.28, 0.07, and 0.08 K \kms, respectively, and these contours all begin at 8$\sigma$, spacing with 14, 11, 6, 6, and 5$\sigma$, respectively.}
\label{Fig:N14Fiveline}
\end{figure}

\begin{figure} [H]
\center
\includegraphics[width=0.8\textwidth]{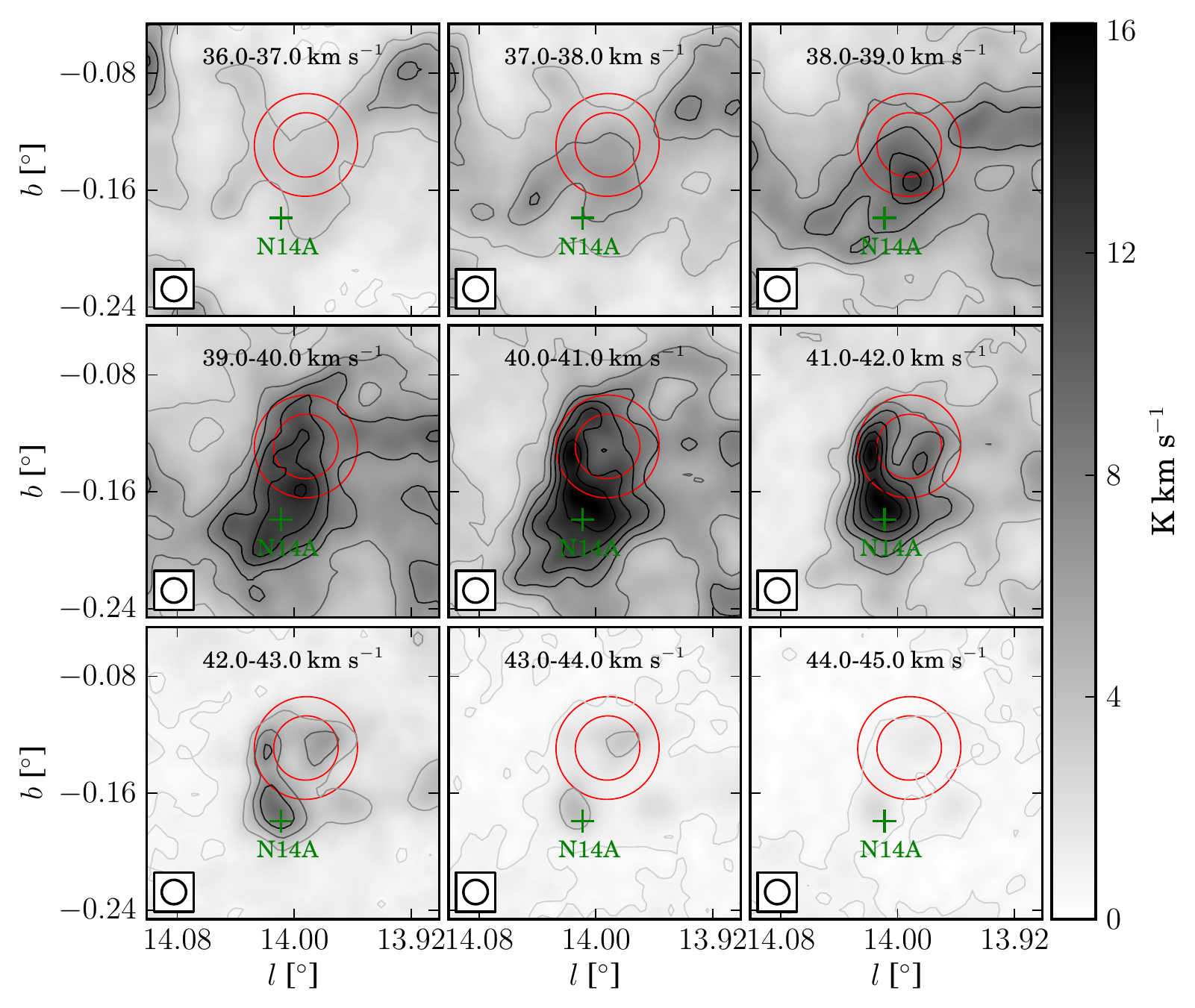}
\caption{Channel map of \coss\ for bubble N14 from 36.0 \kms\ to 45.0 \kms\ with 1.0 \kms\ intervals. The red circles mark the position of N14~\citep{2012MNRAS.424.2442S}, and the green cross marks the peak position of the clump.  The RMS ($\sigma$) of the image background is about 0.11 \intunit. The  contour levels space linearly  from 8$\sigma$ to the peak with step 19.1$\sigma$.}
\label{Fig:N14chananelmap}
\end{figure}

 Our observations show that CO, \hcns, and \hcos\ emissions are in good  agreement with the 8 \mum\ bubble, as shown in Figure~\ref{Fig:N14Fiveline} (see Table~\ref{Tab:lineParameters} for the line parameters). We identified a molecular clump near N14, named N14A, which are marked in Figure~\ref{Fig:N14chananelmap}. This clump overlaps with an extended 1.1 mm continuum emission and an infrared  dark cloud. Although the CO profiles are seriously contaminated by other components, as shown in Figure~\ref{Fig:allspectral}, we still identified an outflow candidate using the \hcos\ line, which shows a clear red lobe profile. Interestingly, we found a compact 24  \mum\ source near this outflow candidate. Detailed discussion concerns this outflow is presented in Section~\ref{secoutflow}.

\subsubsection{N37}

\begin{figure} [H]
\center
\includegraphics[width=0.9\textwidth]{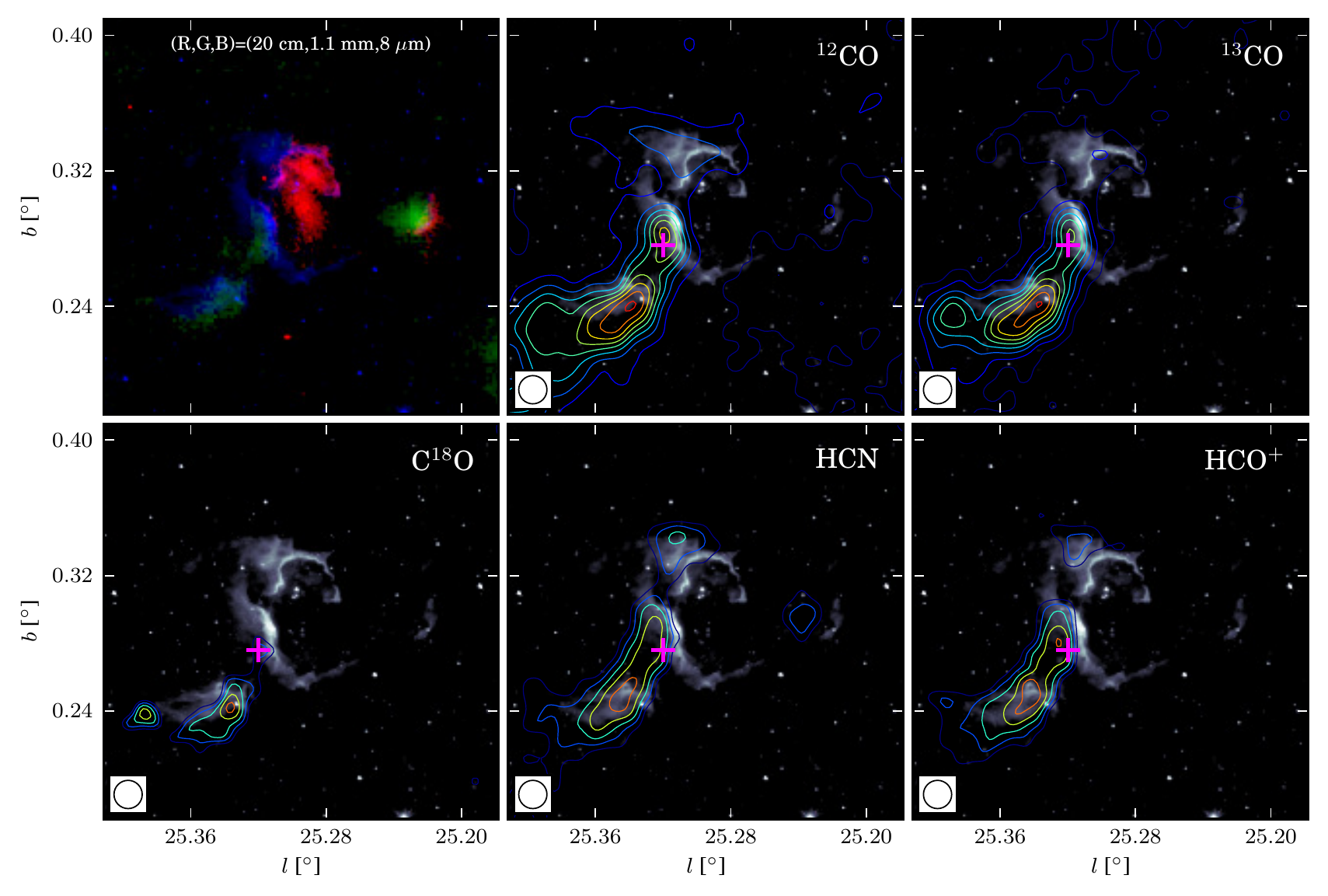}
\caption[]{Images of N37. The upper left panel is a three color image with 20 cm in red, 1.1 mm in green, and 8 \mum\ in blue. The other five panels are contours of five spectral line integrated intensity maps superimposed on the 8 \mum\ image with positions of clumps marked by purple crosses. The contour maps integrate the intensity from 39.0 \kms\ to 44.0 \kms. The rms ($\sigma$) of the five line maps, \cofs, \coss, \cots, \hcns, and \hcos, are 0.36, 0.21, 0.19, 0.05, and 0.05 K \kms, respectively, and these contours all begin at 8$\sigma$, spacing with 24, 12, 3, 7, and 5$\sigma$, respectively. }
 \label{Fig:N37Fiveline}
\end{figure}

N37 is an open bubble with 8 \mum\ emissions encircling  weak 20 cm continuum emissions, which is shown in Figure~\ref{Fig:N37Fiveline} (see Table~\ref{Tab:lineParameters} for the line parameters). The ionized gas velocity of this \HII\ region is 39.6 \kms \citep{1989ApJS...71..469L, 2014ApJS..212....1A}, which is in good accordance with the CO velocity of 40.0 \kms\ given by~\citet{2010ApJ...709..791B}. Although \citet{2010ApJ...716.1478W} used a distance of 3.3 kpc, we prefer the far distance of 12.6 kpc, resolved by \citet{2009ApJ...690..706A} using two methods, H \textsc{i} emission/absorption (H \textsc{i} E/A) and H \textsc{i} self-absorption (H \textsc{i} SA).  This  value is identical with the distance adopted by~\citet{2010A&A...523A...6D}.

\begin{figure} [H]
\center
\includegraphics[width=0.8\textwidth]{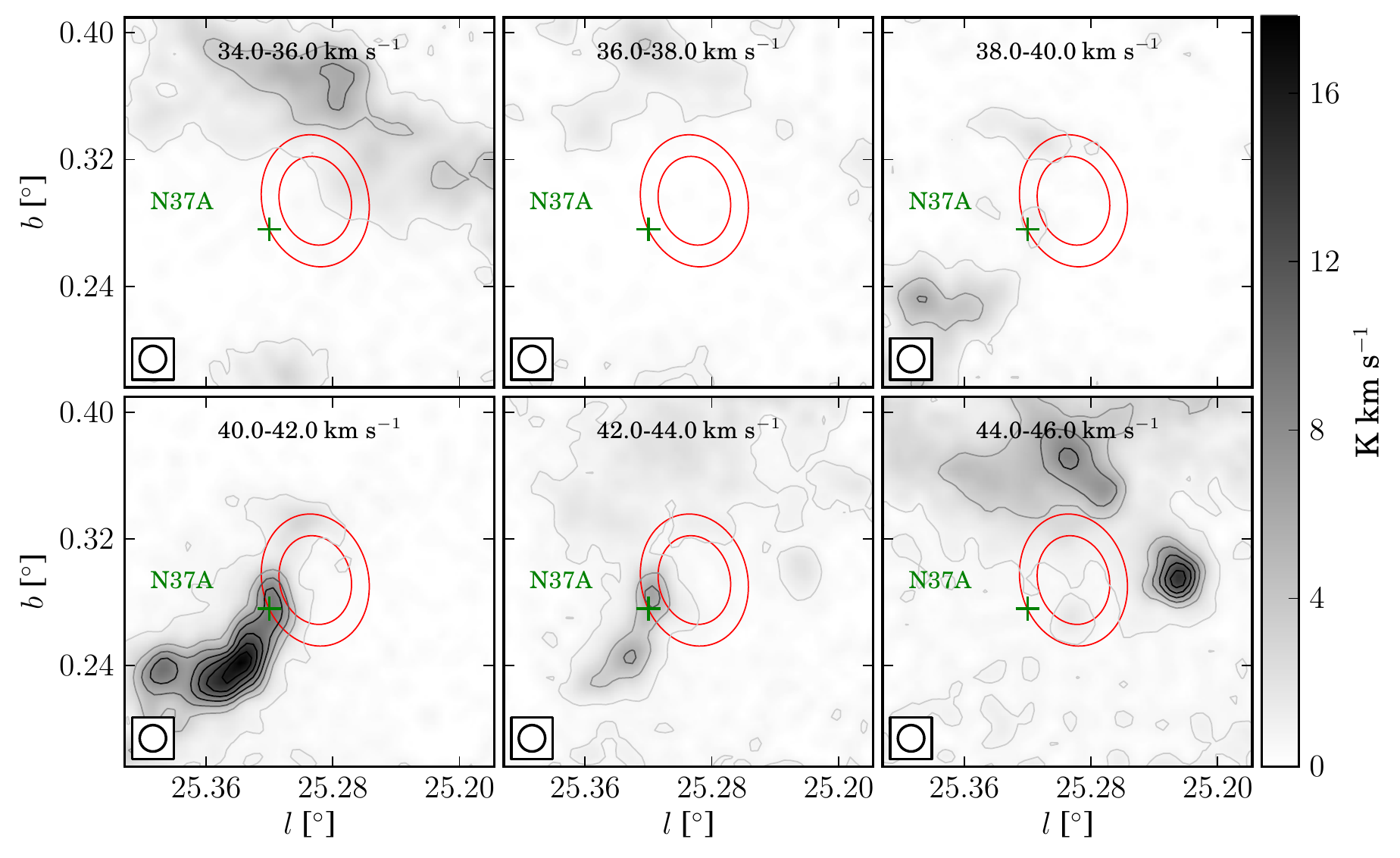}
\caption{Channel map of \coss\ for bubble N37 from 34.0 \kms\ to 46.0 \kms\ with 2.0 \kms\ intervals.  The red circles mark the position of N37~\citep{2012MNRAS.424.2442S}, and the green cross marks the peak position of the clump.  The rms ($\sigma$) of the image background is about 0.13 \intunit, and the contour levels space linearly  from 5$\sigma$ to the peak with step 18.3$\sigma$. }

\label{Fig:N37chananelmap}
\end{figure}

Based on a visual method, \citet{2010ApJ...709..791B} identified an outflow using CO (J=3$\rightarrow$2)  at the position (l,b) = (25.285, 0.266), while we did not find evident outflow activity near this position.  They suggested that the high and low velocities of this outflow were 30 and 40 \kms, respectively. However, the \cofs\ emission is weak and complex in this velocity range, and is not consistent with bubble velocity and morphology.  A molecular clump, N37A, was identified at the border of N37, as shown in Figure~\ref{Fig:N37chananelmap}, while the molecular line emission was rather weak at the open part of the bubble.

Although  \coss\ and \hcos\ display blue wings, no corresponding red wings were found. Furthermore,  no 24 \mum\ source was found near  the peak of the \cots\ intensity.  Therefore, we suggest the blue wings are either produced by expansion of the \HII\ regions, or contaminated by other components. Consequently, we did not detect outflows in this region, but considering the limited spatial resolution of PMODLH and the environment traced by \cof, we cannot rule out the outflow identified by~\citep{2010ApJ...709..791B} using CO (J=3$\rightarrow$2).

\subsubsection{N44}

N44 is an open bubble with a regular elliptical shape with weak, diffuse 20 cm continuum enclosed in it. The ionized gas velocity of this bubble is about 82  \kms~\citep{2014ApJS..212....1A}, which is in good agreement with the CO velocity of 81.1 \kms ~\citep{2010ApJ...709..791B}. The kinematic distance of N44 has not been resolved, and we simply adopted a near kinematic distance of 5 kpc.

\begin{figure} [H]
\center
\includegraphics[width=0.9\textwidth]{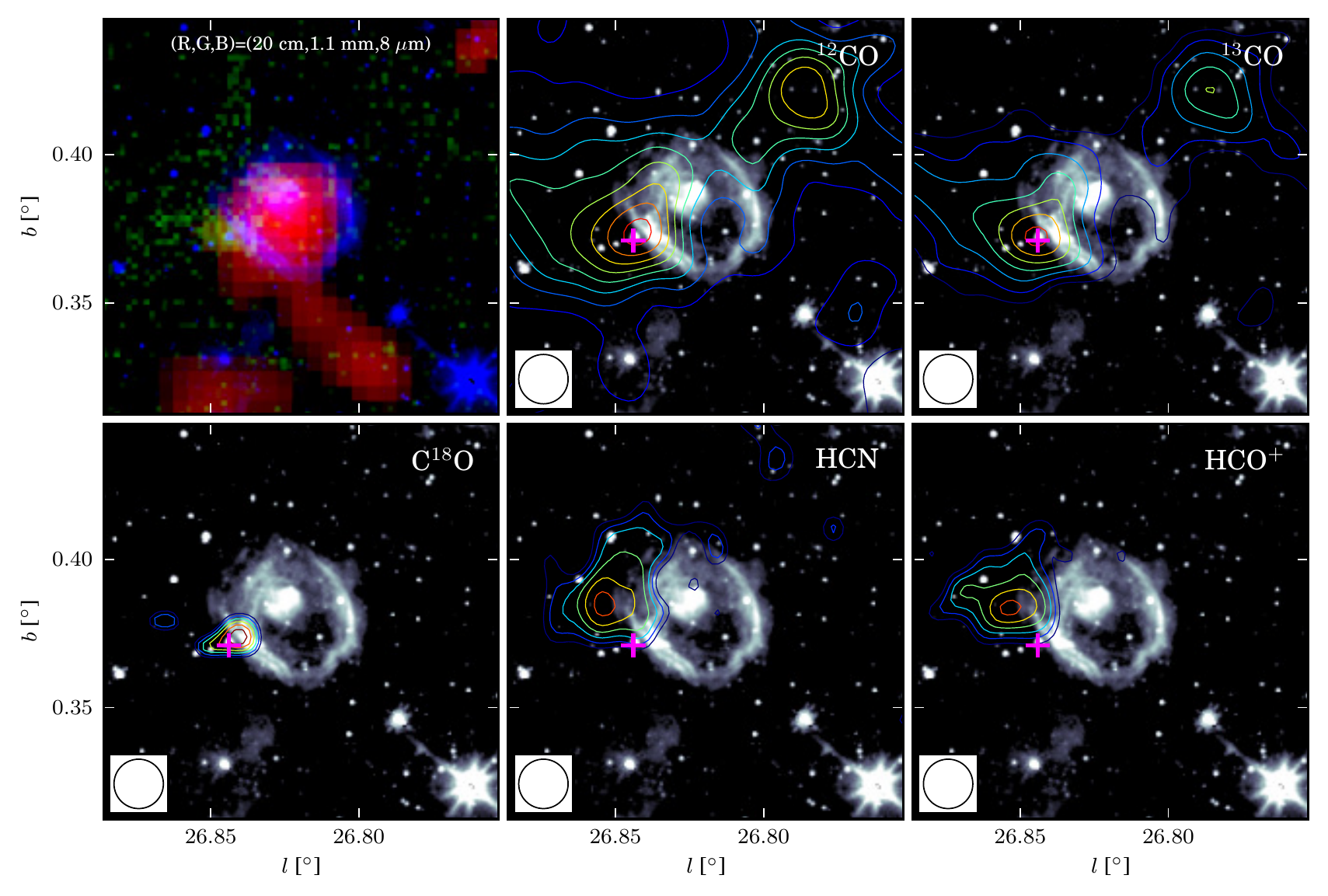}
\caption[]{Images of N44. The upper left panel is a three color image with 20 cm in red, 1.1 mm in green, and 8 \mum\ in blue. The other five panels are contours of five spectral line integrated intensity maps superimposed on the 8 \mum\ image with positions of clumps marked by purple crosses. The contour maps integrate the intensity from 78.0 \kms\ to 84.0 \kms. The background  rms ($\sigma$) of the five line maps, \cofs, \coss, \cots, \hcns, and \hcos, are 0.31, 0.25, 0.16, 0.05, and 0.05 K \kms, respectively, and these contours all begin at 8$\sigma$, with 20, 7, 1, 3, and 3$\sigma$ spacing, respectively.}
\label{Fig:N44Fiveline}
\end{figure}

\begin{figure} [H]
\center
\includegraphics[width=0.8\textwidth]{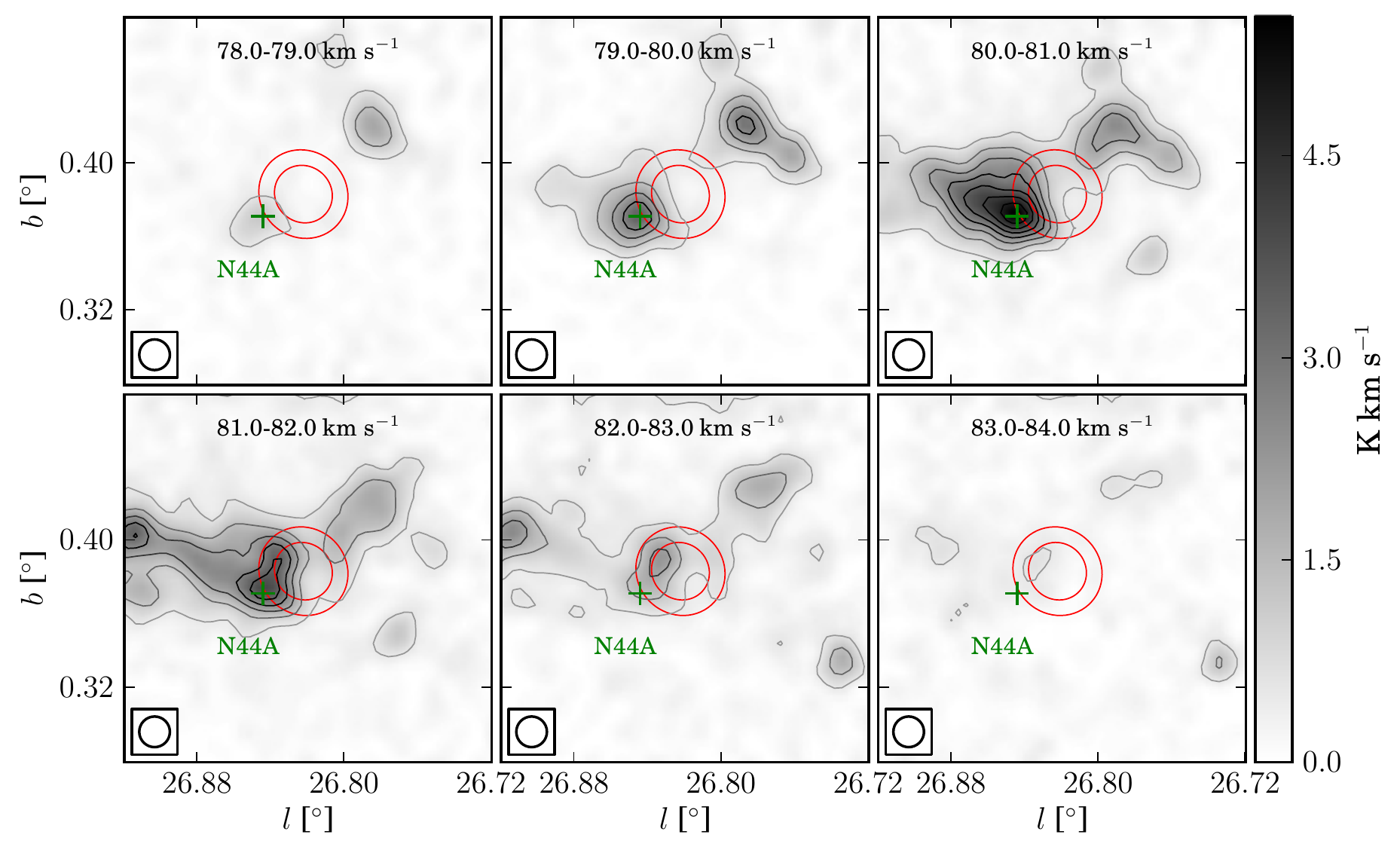}
\caption{Channel map of \coss\ for bubble N44 from 78.0 \kms\ to 84.0 \kms\ with 1.0 \kms\ intervals. The  two red circles mark the position of N44~\citep{2012MNRAS.424.2442S}, and the green cross marks the peak position of  the clump. The rms ($\sigma$) of the image background is about 0.10 \intunit, and the contour levels space linearly from 5$\sigma$ to the peak with step 6.9$\sigma$.}

\label{Fig:N44chananelmap}
\end{figure}

\citet{2010A&A...523A...6D} found two condensations in this region, and both of them were observed by our CO observations, as shown in Figure~\ref{Fig:N44Fiveline} (see Table~\ref{Tab:lineParameters} for the line parameters). However, we only consider the one at the edge of N44, which is N44A, marked by a green cross in Figure~\ref{Fig:N44chananelmap}. The CO line wings of this clump are contaminated by other adjacent components, which hinders us from searching for outflows.

\subsubsection{N49}

\begin{figure} [H]
\center
\includegraphics[width=0.9\textwidth]{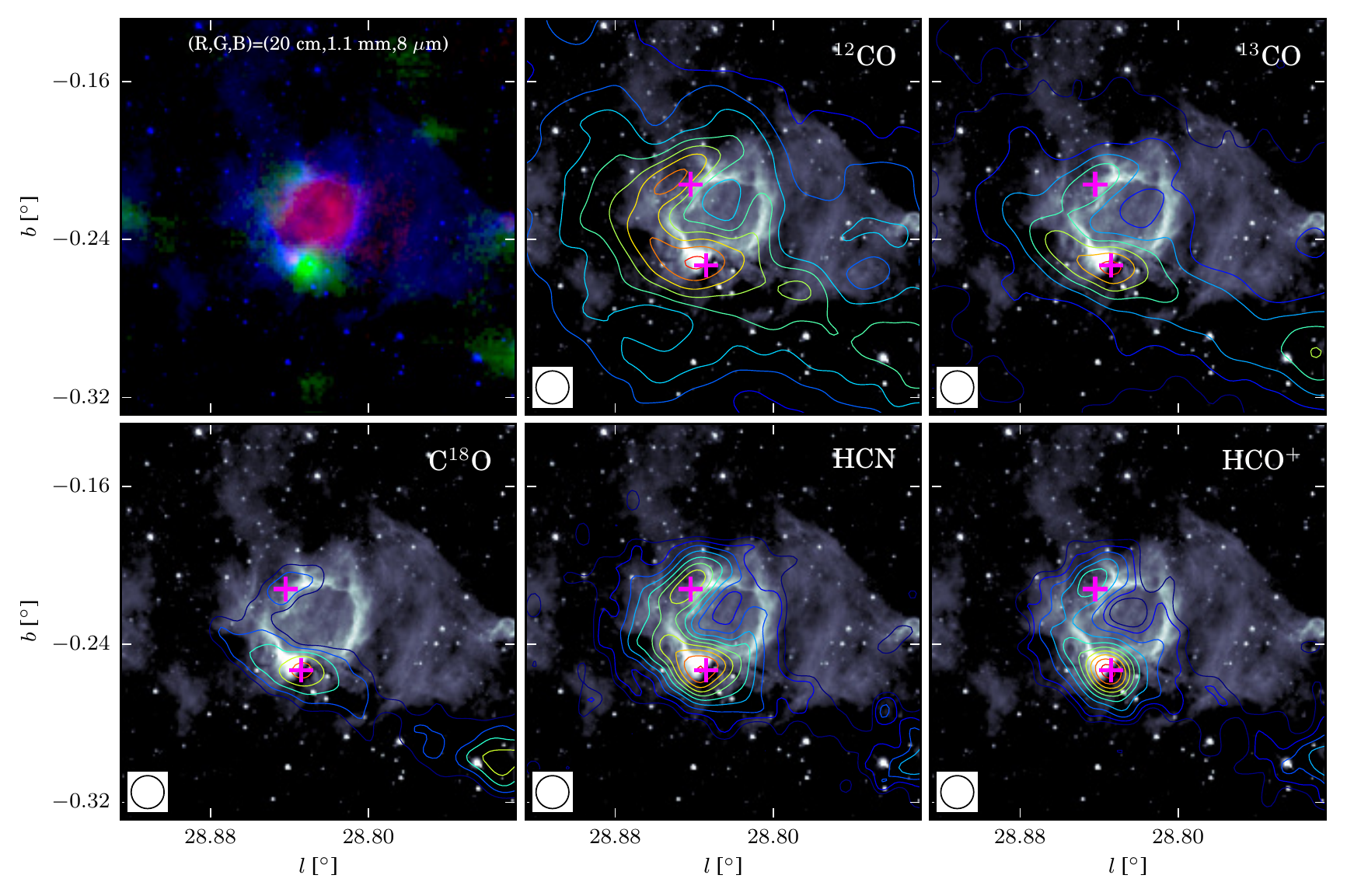}
\caption[]{Images of N49. The upper left panel is a three color image with 20 cm in red, 1.1 mm in green, and 8 \mum\ in blue. The other five panels are contours of five spectral line integrated intensity maps superimposed on the 8 \mum\ image with positions of clumps marked by purple crosses. The contour maps integrate the intensity from 83.0 \kms\ to 90.0 \kms. The rms ($\sigma$) of five line maps, \cofs, \coss, \cots, \hcns, and \hcos, are 0.44, 0.31, 0.24, 0.06, and 0.07 K \kms, respectively, and these contours all begin at 8$\sigma$, spacing with 25, 19, 7, 4, and 5 $\sigma$, respectively.}
\label{Fig:N49Fiveline}
\end{figure}

  N49 is a remarkable bubble, and has been investigated by several studies~\citep{2008ApJ...681.1341W,2009ApJ...702.1615C,2010A&A...518L.101Z}. As shown in Figure~\ref{Fig:N49Fiveline} (see Table~\ref{Tab:lineParameters} for observation parameters), N49 display a clear typical structure of the bubbles, a ring of  8 \mum\ emission surrounding a disc of 20 cm continuum, and interestingly,  there is  an ultra-compact (UC)  \HII\ region at the border. The velocity of ionized gas in this field is approximately 90.6 \kms\ given by~\citet{1989ApJS...71..469L,2014ApJS..212....1A}, and the distance is approximately 5.5 kpc \citep{2010A&A...523A...6D}.

\citet{2008ApJ...681.1341W} identified seven YSO candidates using the Spectral Energy Distribution (SED) fitting method in this region, and two of them which are located at the edge of N49 possess the largest mass. They proposed that those YSOs were triggered by the expanding shell, since this bubble is being blown by an O5 V star with a dynamical age of $\geq$ 10$^5$ yr.

\citet{2009ApJ...702.1615C} surveyed class I and II CH$_3$OH masers  towards EGOs using the Very Large Array (VLA). One of their samples is situated at the border  of  N49, and is likely associated with these two YSOs identified by~\citet{2008ApJ...681.1341W}.  Both class I and II masers were detected by~\citet{2009ApJ...702.1615C} in this region.  The presence of 6.7 GHz  CH$_3$OH masers indicates that at least one high-mass star is forming here. They suggest the class I 44 GHz CH$_3$OH masers are tracing an outflow, while the configuration of class II 6.7 GHz CH$_3$OH masers represents a  Keplerian rotation disk.  \hcos(3-2), which was also observed by \citet{2009ApJ...702.1615C} using the James Clerk Maxwell Telescope (JCMT), shows a similar velocity range to 6.7 GHz methanol masers. However, no \hho\ maser was found by~\citet{2013ApJ...764...61C} in this region using  the Nobeyama Radio Observatory 45 m telescope.
 
\citet{2010A&A...518L.101Z} studied star formation  triggered by the  expansion of \HII\ regions using $Herschel$ Photoconductor Array Camera and Spectrometer (PACS) and  Spectral and Photometric Imaging Receiver (SPRIRE) images from the  $Herschel$ infrared survey of the Galactic plane (Hi-GAL).  The Hi-GAL PACS and SPIRE images reveal a new population of embedded young stars, and five high-mass stars are forming here. They conclude that the high star formation efficiency in this region may be caused by the expanding shell.

\begin{figure} [H]
\center
\includegraphics[width=0.8\textwidth]{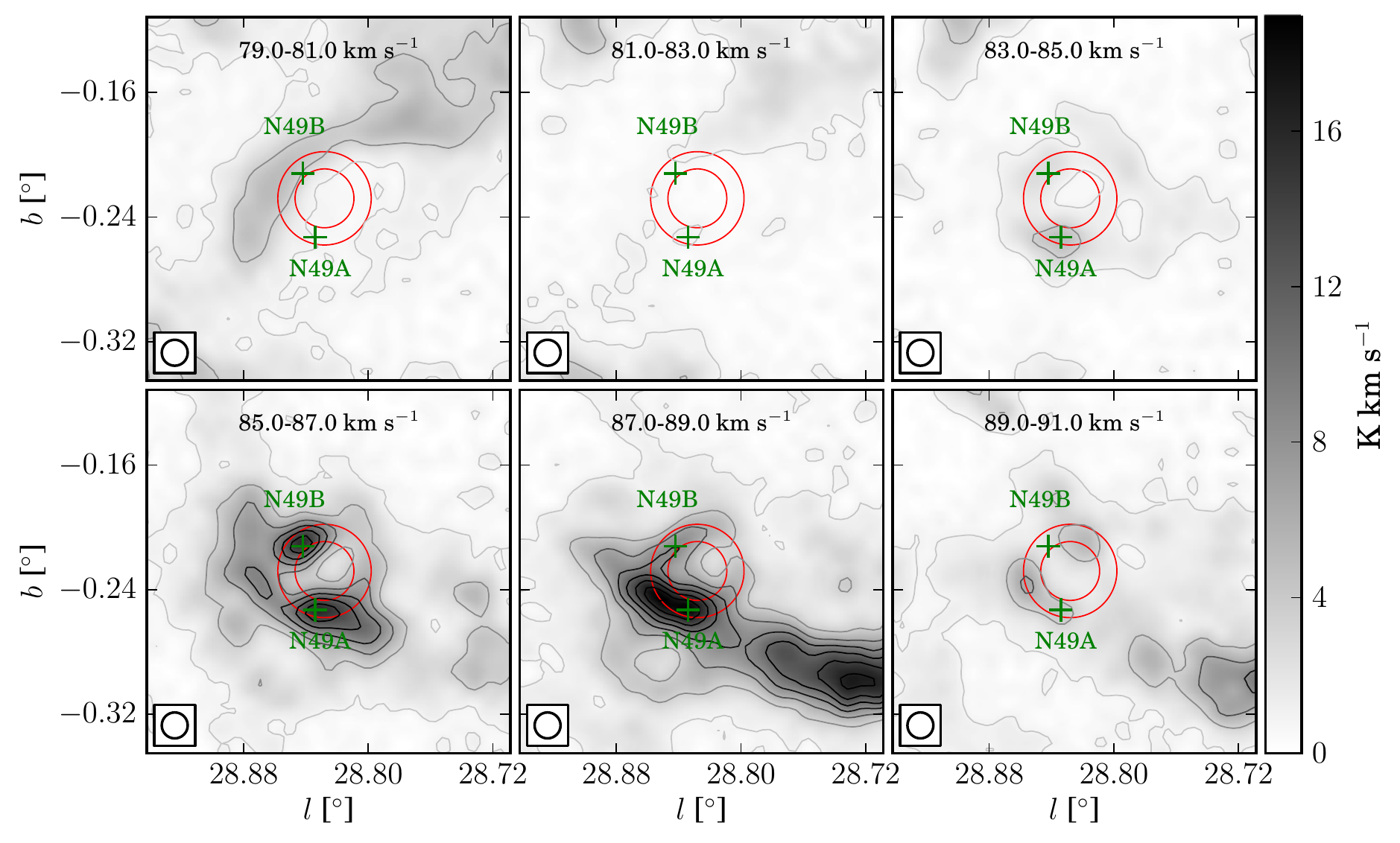}
\caption{Channel map of \coss\ for bubble N49 from 79.0 \kms\ to 91.0 \kms\ with 2.0 \kms\ intervals. The red circles mark the position of N49~\citep{2012MNRAS.424.2442S}, and the green crosses mark the peak positions of clumps. The rms ($\sigma$) of the image background is about 0.17 \intunit, and the contour levels space linearly from 5$\sigma$ to the peak with step 15.5$\sigma$.}
\label{Fig:N49chananelmap}
\end{figure}

 According to our observations, the CO emission agrees well with the  8 \mum\ band image, and all the five lines display arc structures along the edge of N49. The distribution of dense molecular gas traced by \hcos\ and \hcns\ suggests that they are probably accumulated by the expanding shell. We identified two clumps, N49A and N49B, at the rim of N49, as shown in Figure~\ref{Fig:N49chananelmap}. Interestingly, the channel map of \coss\ of N49 displays a clear shell structure at the velocity range of 85-87  \kms, which is most likely produced by the expansion of this \HII\ region. The profiles of CO and   \hcos\ lines are severely contaminated by other components with similar velocities, and therefore we cannot verify the outflow candidate offered by~\citet{2010ApJ...709..791B}. 


\subsubsection{N55}

\citet{2010ApJ...719.1104R} suggested that bubble N55 was associated with a star forming complex (SFC), which is SFCs 18 in their paper, and they classified it as a closed bubble. The distance and velocity of N55 provided by \citet{2010A&A...523A...6D} are 8.4 kpc and 93 \kms, respectively. 

\begin{figure} [H]
\center
\includegraphics[width=0.9\textwidth]{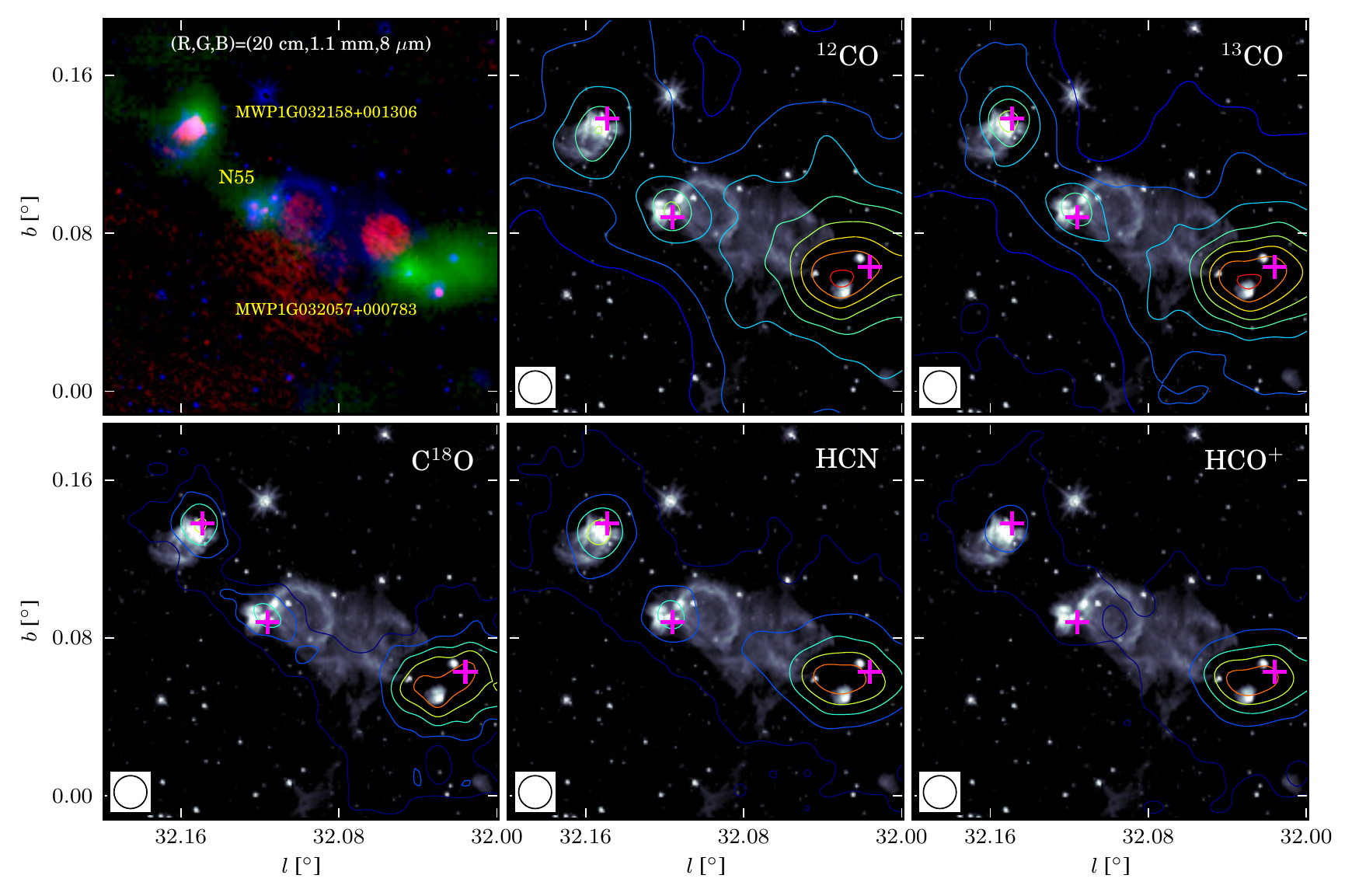}
\caption[]{Images of N55. The upper left panel is a three color image with 20 cm in red, 1.1 mm in green, and 8 \mum\ in blue. The other five panels are contours of five spectral line integrated intensity maps superimposed on the 8 \mum\ image with positions of clumps marked by purple crosses. The contour maps integrate the intensity from 89.0 \kms\ to 103.0 \kms. The rms ($\sigma$) of five line maps, \cofs, \coss, \cots, \hcns, and \hcos, are 0.49, 0.26, 0.27, 0.10, and 0.10 K \kms, respectively, and these contours all begin at 8$\sigma$, spacing with 47, 29, 7, 19, and 26$\sigma$, respectively.}
\label{Fig:N55Fiveline}
\end{figure}

\begin{figure} [H]
\center
\includegraphics[width=0.8\textwidth]{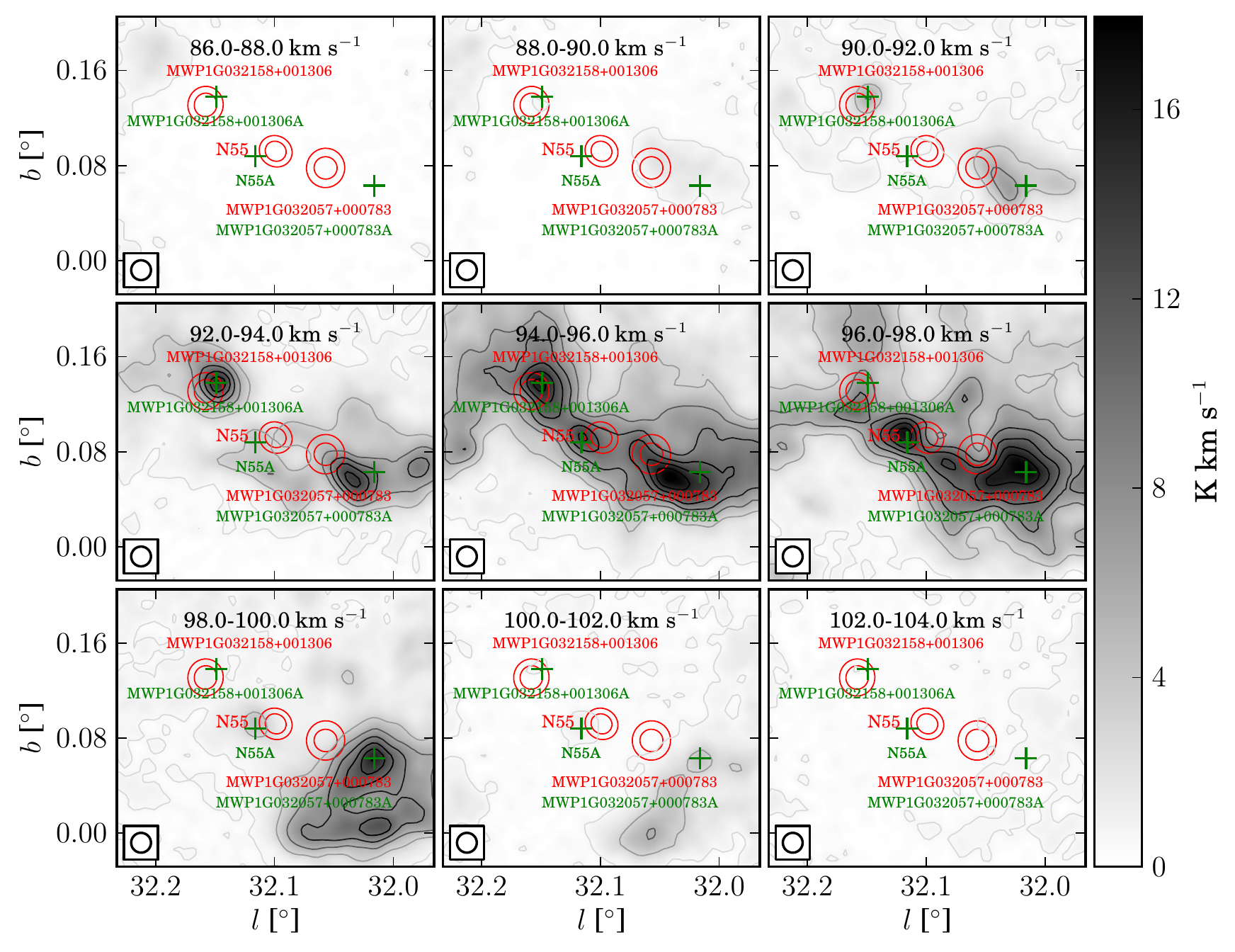}
\caption{Channel map of \coss\ for bubble N55 from 86.0 \kms\ to 104.0 \kms\ with 2.0 \kms\ intervals.  The red circles delineate the position of N55 from~\citet{2012MNRAS.424.2442S}, and the green crosses mark the peak position of clumps. The rms ($\sigma$) of the image background is about 0.10 \intunit, and the contour levels space linearly from 5$\sigma$ to the peak with step 25.3$\sigma$.
}
\label{Fig:N55chananelmap}
\end{figure}

This is a remarkable bubble region, the images of N55 are shown in Figure~\ref{Fig:N55Fiveline} ( the observation parameters are listed in Table~\ref{Tab:lineParameters}). There are  another two  prominent bubbles in this region, MWPIG032057+000783 and MWPIG032158+001306~\citep{2012MNRAS.424.2442S}. Diffuse and weak emissions of 20 cm continuum spread in and around N55, while each of the other two bubbles enclose a  20 cm continuum disk with relatively strong emissions. There are several UC \HII\ regions traced by the 20 cm continuum near N55, which indicates that  a cluster of  high-mass stars is emerging here. Interestingly, each of the three bubbles harbors a molecular clumps traced by the 1.1 mm continuum.  Those molecular clumps  are confirmed by our \cots\ lines, and the positions of these three clumps are marked with green crosses in Figure~\ref{Fig:N55chananelmap}.

We found a wide CO line wing at both red and blue ends of profiles of the molecular clump N55A.  Since there are at least three UC \HII\ regions near N55A, as shown in figure~\ref{Fig:N55Fiveline}, it is possible that more than one outflow is occurring here. Since they are not resolved, we simply regard it as one outflow when calculating the outflow parameters. However, we failed to estimate the age for this outflow  due to the overlap of peaks of the red and blue lobes, and the outflow parameters are given in Table~\ref{Tab:outflows}. The CO and \hcos\ line profiles of the other two molecular clumps  are contaminated with other components, and therefore, it is hard to perform outflow analysis.

\subsubsection{N75 and N74}

\begin{figure} [H]
\center
\includegraphics[width=0.9\textwidth]{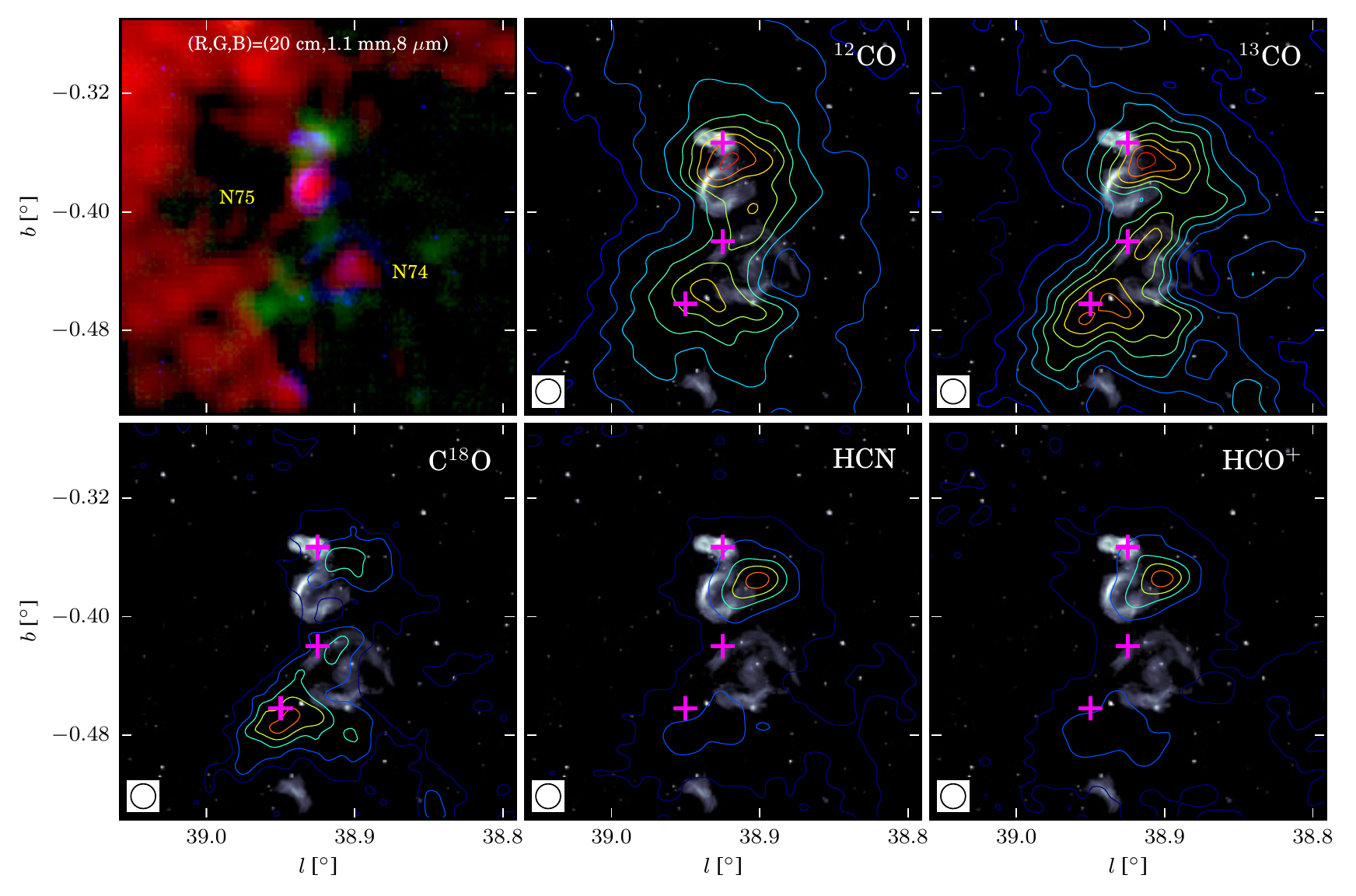}
\caption[]{Images of  N75 and N74. The upper left panel is a three color image with 20 cm in red, 1.1 mm in green, and 8 \mum\ in blue. The other five panels are contours of five spectral line integrated intensity maps superimposed on the 8 \mum\ image  with positions of clumps marked by purple crosses. The contour maps integrate the intensity from 37.0 \kms\ to 44.0 \kms. The rms ($\sigma$) of five line maps, \cofs, \coss, \cots, \hcns, and \hcos, are 0.32, 0.18, 0.18, 0.06, and 0.06 K \kms, respectively, and these contours  all begin at 8$\sigma$, spacing with 50, 26, 7, 21, and 24$\sigma$, respectively.}
\label{Fig:N75Fiveline}
\end{figure}

N74 and N75 are two prominent bubbles in  star formation region G38.9-0.40~\citep{2013ApJ...770....1A}. N75 is an open bubble with a bright rim and elliptical regular shape, while  N74 shows slightly weaker emission at 8 \mum\ band, as shown in Figure~\ref{Fig:N75Fiveline} (the observation parameters are shown in Table~\ref{Tab:lineParameters}).  Unexpectedly, no evident 20 cm emissions were detected within these two bubbles, which is abnormal for such  regular bubbles with strong 8\mum\ band emission.  The velocity of  Hydrogen recombination lines (RRLs) of this \HII\ region is about  42.1 \kms~\citep{2014ApJS..212....1A}, while the velocity of  carbon radio recombination lines is about  39.2 \kms~\citep{2013ApJ...764...34W}. Since the kinematic distance has not been resolved, we simply adopt a near kinematic distance of 2.8 kpc.

\citet{2012ApJ...760...58S} observed four molecular lines, HCN, HCO$^+$, N$_2$H$^+$, and CS,  as well as the 3.3 mm continuum emission,  using Combined Array for Research in Millimeter-wave Astronomy (CARMA)  towards N74.  Tentative evidence showed that the molecular clouds on bubble rims are more fragmented than dark clouds, and they suggest that triggered star formation may be occurring in this region, although their findings do not indicate a classic collect-and-collapse model.

\citet{2013ApJ...770....1A} performed a multi-wavelength study for this region. They claim that both bubbles  are powered by  O9.5V stars with star clusters surrounding them.    Although they identified 162 YSOs in this region,  they did  not find feedback-triggered star formation (SF), which may be due to the young ages of these bubbles. They found a strong correlation between areal YSO mass surface density and gas mass surface density, which suggests that gas density is a more important factor of star formation than stellar feedback.

\begin{figure} [H]
\center
\includegraphics[width=0.8\textwidth]{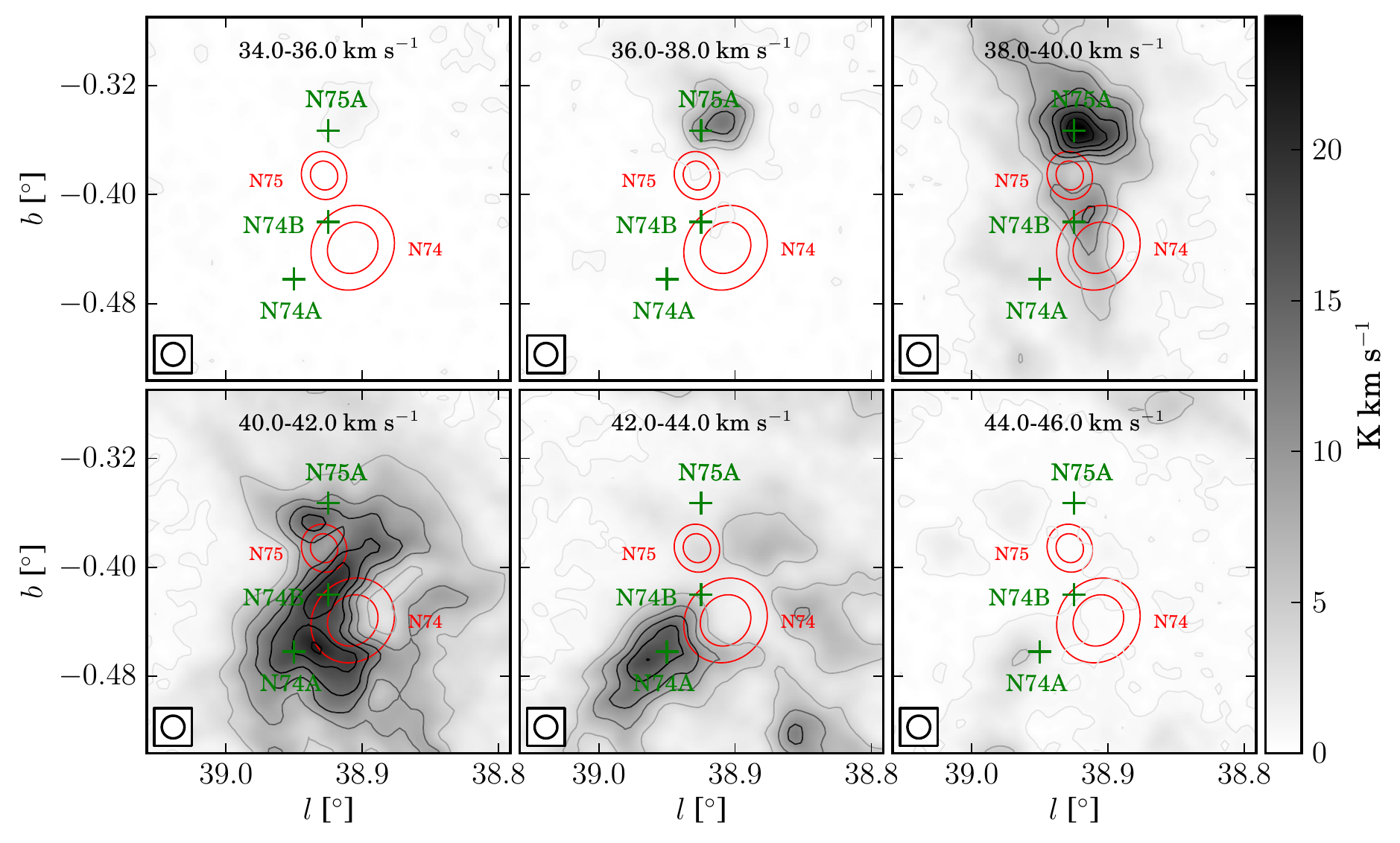}
\caption{Channel map of \coss\ for bubble N75 from 34.0 \kms\ to 46.0 \kms\ with 2.0 \kms\ intervals. The  two red circles delineate the position of N75~\citep{2012MNRAS.424.2442S}, and the green crosses mark the peak position of clumps. The rms ($\sigma$) of the image background is about 0.10 \intunit, and the contour levels space linearly from 5$\sigma$ to the peak with step 36$\sigma$.
}
\label{Fig:N75chananelmap}
\end{figure}

Our observations show CO emissions coincide well with these two bubbles, and three molecular cloud  clumps were revealed by \cots, which are marked by green crosses in Figure~\ref{Fig:N75chananelmap}.  However, at least two components are present at the border of N75 or N74, which hinders us from identifying outflows from the three clumps including the one identified by~\citet{2010ApJ...709..791B}. 

\subsubsection{N82}

\begin{figure} [H]
\center
\includegraphics[width=0.9\textwidth]{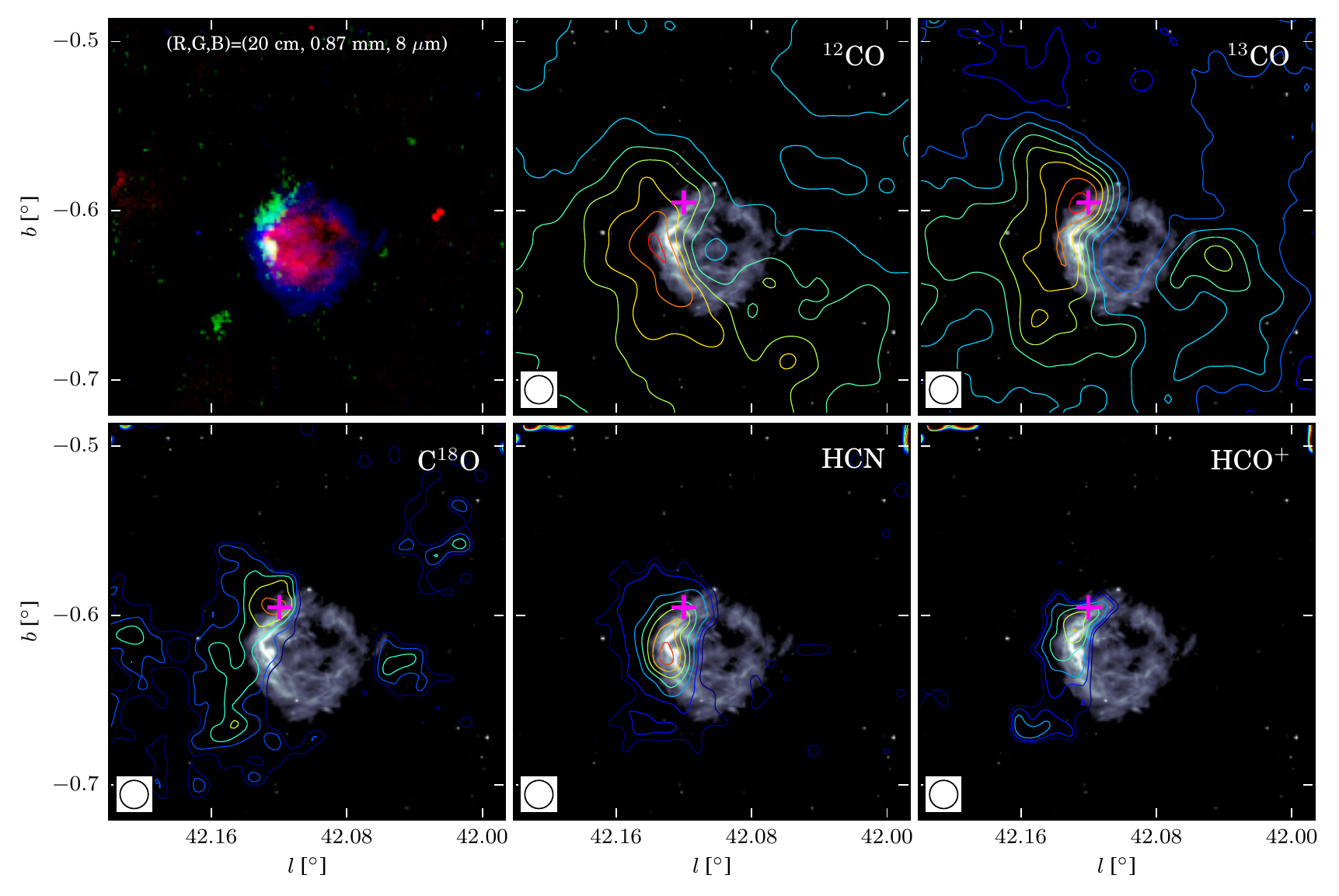}
\caption[]{Images of N82. The upper left panel is a three color image with 20 cm in red, 0.87 mm in green, and 8 \mum\ in blue. The other five panels are contours of five spectral line integrated intensity maps superimposed on the 8 \mum\ image with positions of clumps marked by purple crosses. The contour maps integrate  the intensity from 63.0 \kms\ to 72.0 \kms. The rms ($\sigma$) of five line maps, \cofs, \coss, \cots, \hcns, and \hcos, are 0.38, 0.21, 0.21, 0.08, and 0.08 K \kms, respectively, and these contours all begin at 8 $\sigma$, spacing with 28, 17, 3, 4, and 2$\sigma$, respectively.}
\label{Fig:N82Fiveline}
\end{figure}

N82 is a closed bubble containing a strong 20 cm continuum disk, which is shown in Figure~\ref{Fig:N82Fiveline} (see Table~\ref{Tab:lineParameters} for the molecular line parameters).   The velocity of H$\alpha$ is about 66.0 \kms~\citep{1989ApJS...71..469L}, which is in good agreement with  CO velocity, 66.5 \kms\ determined by~\citet{2010ApJ...709..791B} using the CO(J=3-2) line. The distance of this molecular clouds was resolved by~\citet{2010MNRAS.407..923S} using infrared extension, which is consistent with the result of~\citet{2009ApJ...699.1153R}. Consequently, we adopted a distance of 4.3 kpc for N82 following~\citet{2010MNRAS.407..923S}, and  the far distance suggested by~\citet{2010A&A...523A...6D} is rejected.


\citet{2010ApJ...716.1478W} identified  6 YSOs from the 8 \mum\ shell of N82, and they found a significantly increased YSO density at the 8 \mum\ emission peak.  \citet{2010ApJ...709..791B} observed the CO (J=3-2) line  using the JCMT, and they suggested the CO velocity   is    66.5 \kms. Using a new Monte Carlo method,~\citet{2011MNRAS.418.2219S} derived an age of approximately 1.8 Myr for  clusters in the vicinity of this bubble. 

\begin{figure} [H]
\center
\includegraphics[width=0.8\textwidth]{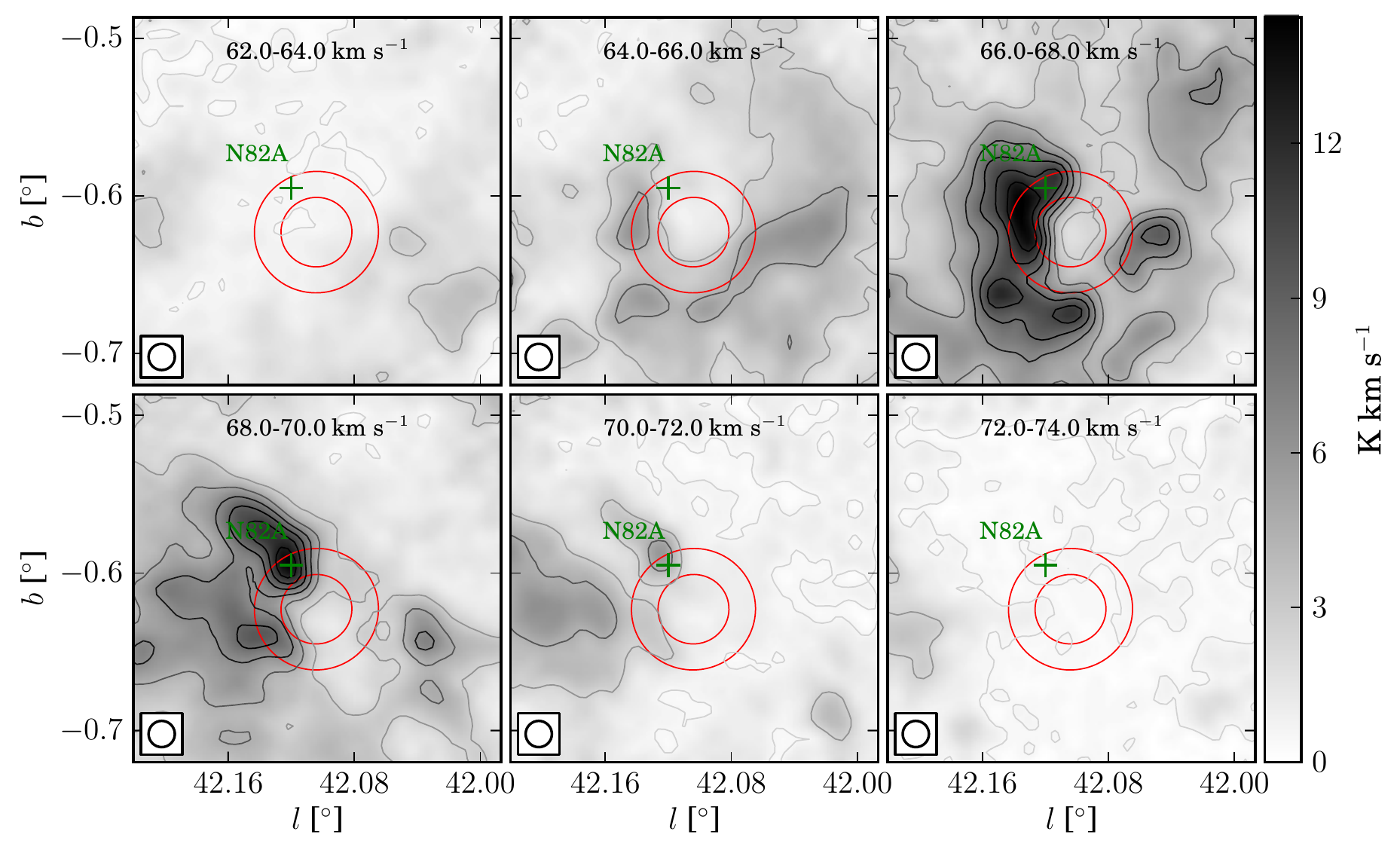}
\caption{Channel map of \coss\ for bubble N82 from 62.0 \kms\ to 74.0 \kms\ with 2.0 \kms\ intervals. The red circles delineate the position of N82 provided  by~\citet{2012MNRAS.424.2442S}, and the green cross marks the peak position of the clump. The rms ($\sigma$) of the image background is about 0.10 \intunit, and the contour levels space linearly from 3$\sigma$ to the peak with step 20.6$\sigma$.}
\label{Fig:N82chananelmap}
\end{figure}

According to  our  observations, the molecular clouds distribute alongside the bright rim of the bubble,  which indicates  the power source of the bubble is probably compressing the molecular clouds. We found a  molecular cloud clump near N82, N82A, which is marked in Figure~\ref{Fig:N82chananelmap}. However, the molecular line  profiles in this region show that multi-components are present here, which prevents us from search for outflows. 

\subsubsection{N89 and N90}

N89 and N90 are two adjacent bubbles  with  relatively large angle size and  weak 8 \mum\ band emissions.  N90 contains weak 20 cm continuum emission which is  absent  in bubble N89.   The  velocity of ionized gas from~\citet{2014ApJS..212....1A} is about 70.5 \kms. \citet{2012ApJ...759...96B} performed a survey of \HII\ regions using Arecibo radio telescope, and they suggest that  velocities for N89 and N90 are 73.1 \kms\ and 70.5 \kms, respectively. The distance of N90 given by~\citet{2010A&A...523A...6D} is 6.1 kpc. 

 Profiles of N89 and N90 shown in Figure~\ref{Fig:allspectral} demonstrate that a strong emission presents at $\sim$60 - 62 \kms, but this component is actually a part of Sagittarius Arm~\citep{2012APJ...752..118S}. Furthermore, no H-$\alpha$ emission is detected at this velocity, and  the integrated intensity map over this velocity range 	is not clearly related with N89 and N90. 

\begin{figure} [H]
\center
\includegraphics[width=0.9\textwidth]{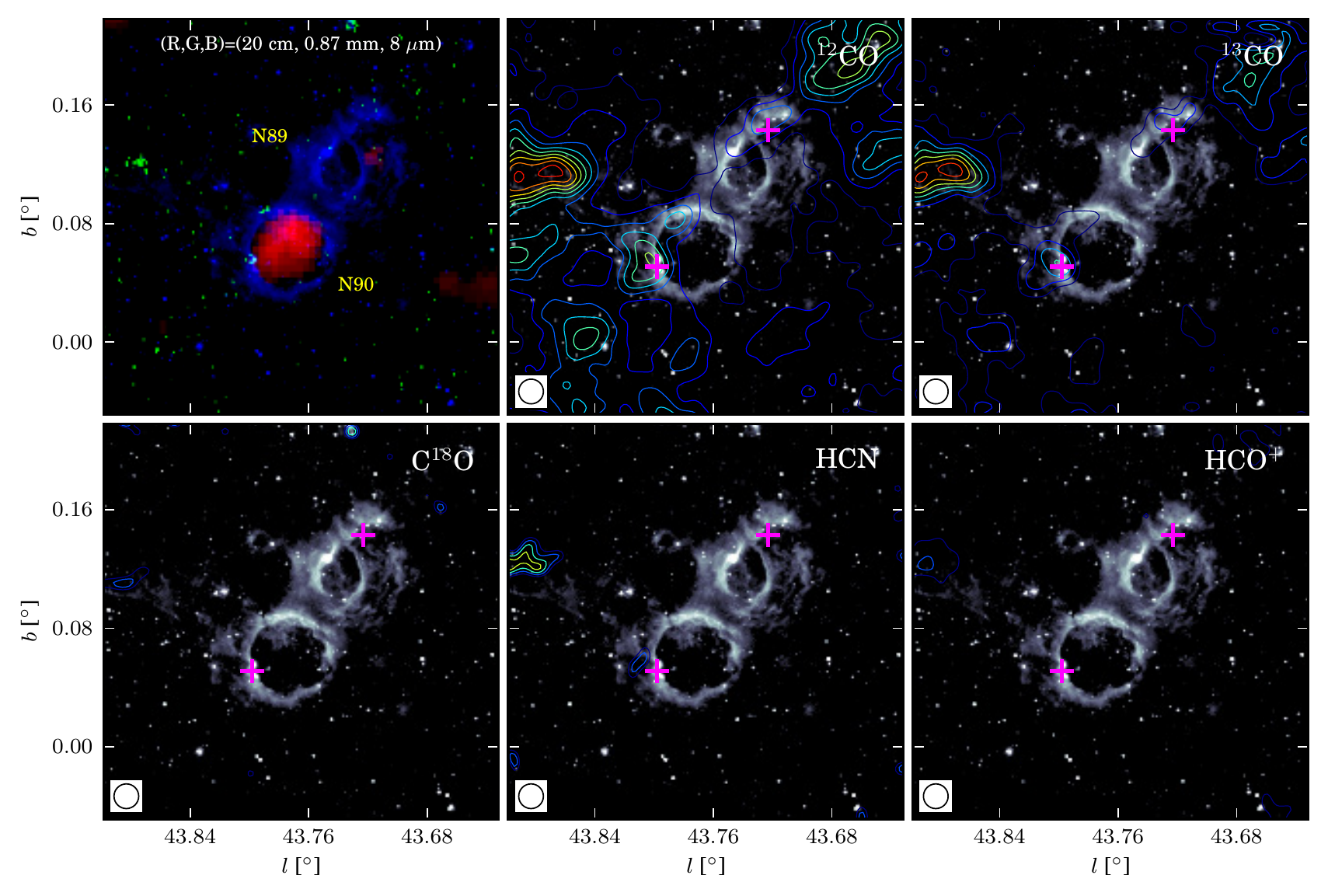}
\caption[]{Images of N89 and N90. The upper left panel is a three color image with 20 cm in red, 0.87 mm in green, and 8 \mum\ in blue. The other five panels are contours of five spectral line integrated intensity maps superimposed on the 8 \mum\ image with positions of clumps marked by purple crosses. The contour maps integrate the intensity from 64.0 \kms\ to 75.0 \kms. The rms  ($\sigma$) of five line maps, \cofs, \coss, \cots, \hcns, and \hcos, are 0.37, 0.18, 0.18, 0.06, and 0.07 K \kms, respectively, and these contours all begin at 8$\sigma$, spacing  with 15, 10, 3, 2, and 5$\sigma$, respectively.}
\label{Fig:N90Fiveline}
\end{figure}

\begin{figure} [H]
\center
\includegraphics[width=0.8\textwidth]{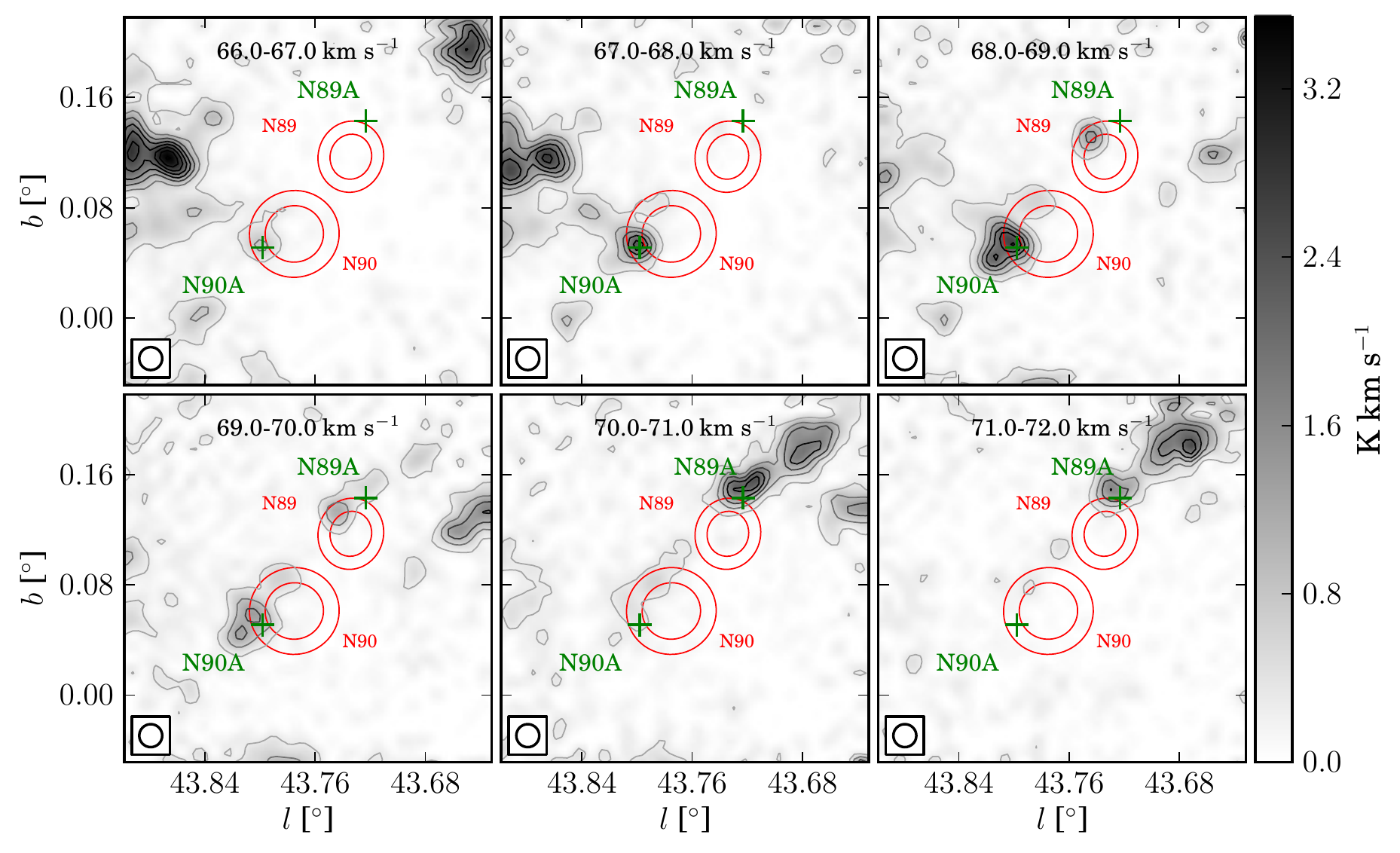}
\caption{Channel map of \coss\ for bubble N89 and N90 from 66.0 \kms\ to 72.0 \kms\ with 1.0 \kms\ intervals. The red circles delineate the position of N89 and N90~\citep{2012MNRAS.424.2442S}, and the green crosses mark the peak position of clumps. The rms ($\sigma$) of the image background is about 0.05 \intunit, and the contour levels space linearly from 5$\sigma$ to the peak with step 8.6$\sigma$.}
\label{Fig:N90chananelmap}
\end{figure}

According to our CO observation,  the molecular cloud at 70.3 and 68.4 \kms\ is most likely associated with bubble N89 and N90, respectively, as shown in Figure~\ref{Fig:N90Fiveline} (set Table~\ref{Tab:lineParameters} for the observation parameters). No \cots\ signal was detected in this region, and  the emission of \hcos\ and \hcns\ lines were rather weak. Since \cots\ emission is absent here, we used \coss\ line to identify clumps, and consequently, these two identified clumps are not concrete. These two clumps, N89A and N90A, are marked by green crosses in Figure~\ref{Fig:N90chananelmap}. We could not perform outflow identifications due to the contamination of CO profiles and  poor SNR of \hcos.

\subsubsection{N95}

\begin{figure} [H]
\center
\includegraphics[width=0.9\textwidth]{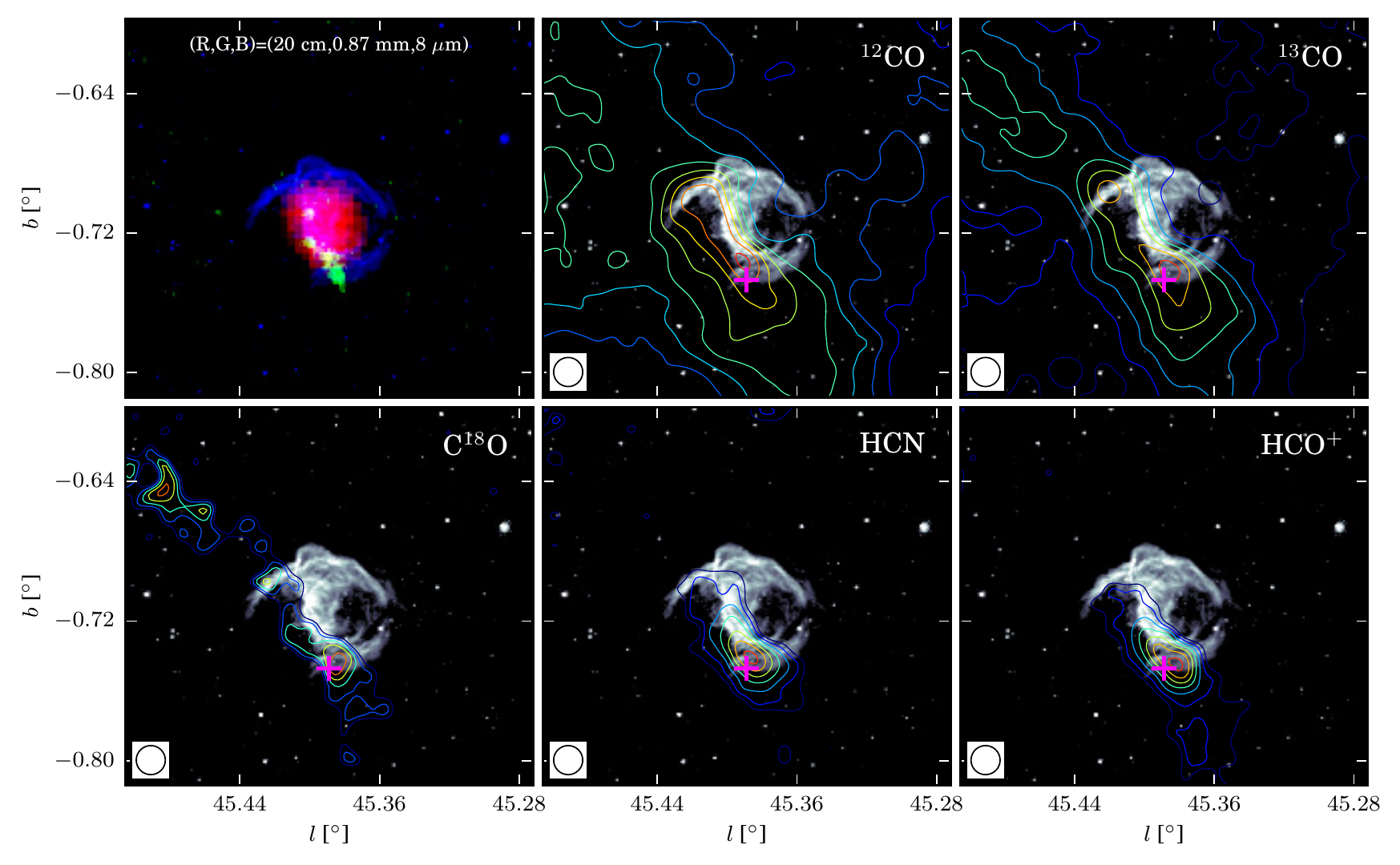}
\caption[]{Images of N95. The upper left panel is a three color image with 20 cm in red, 0.87 mm in green, and 8 \mum\ in blue. The other five panels are contours of five spectral line integrated intensity maps superimposed on the 8 \mum\ image with positions of clumps marked by purple crosses. The contour maps integrate the intensity from 57.0 \kms\ to 63.0 \kms. The rms ($\sigma$) of five line maps, \cofs, \coss, \cots, \hcns, and \hcos, are 0.32, 0.19, 0.19, 0.05, and 0.05 K \kms, respectively, and these contours all begin at 8$\sigma$,  spacing with 22, 18, 2, 5, and 4$\sigma$, respectively.}
\label{Fig:N95Fiveline}
\end{figure}

N95 is an open bubble with approximately round shape split by an arc, containing a strong 20 continuum disk, as demonstrated in Figure~\ref{Fig:N95Fiveline} (see Table~\ref{Tab:lineParameters} for the observations parameters). The velocity and distance offered by \citet{2010A&A...523A...6D} are 52.5 \kms\ and 8.0 kpc, respectively.

\begin{figure} [H]
\center
\includegraphics[width=0.8\textwidth]{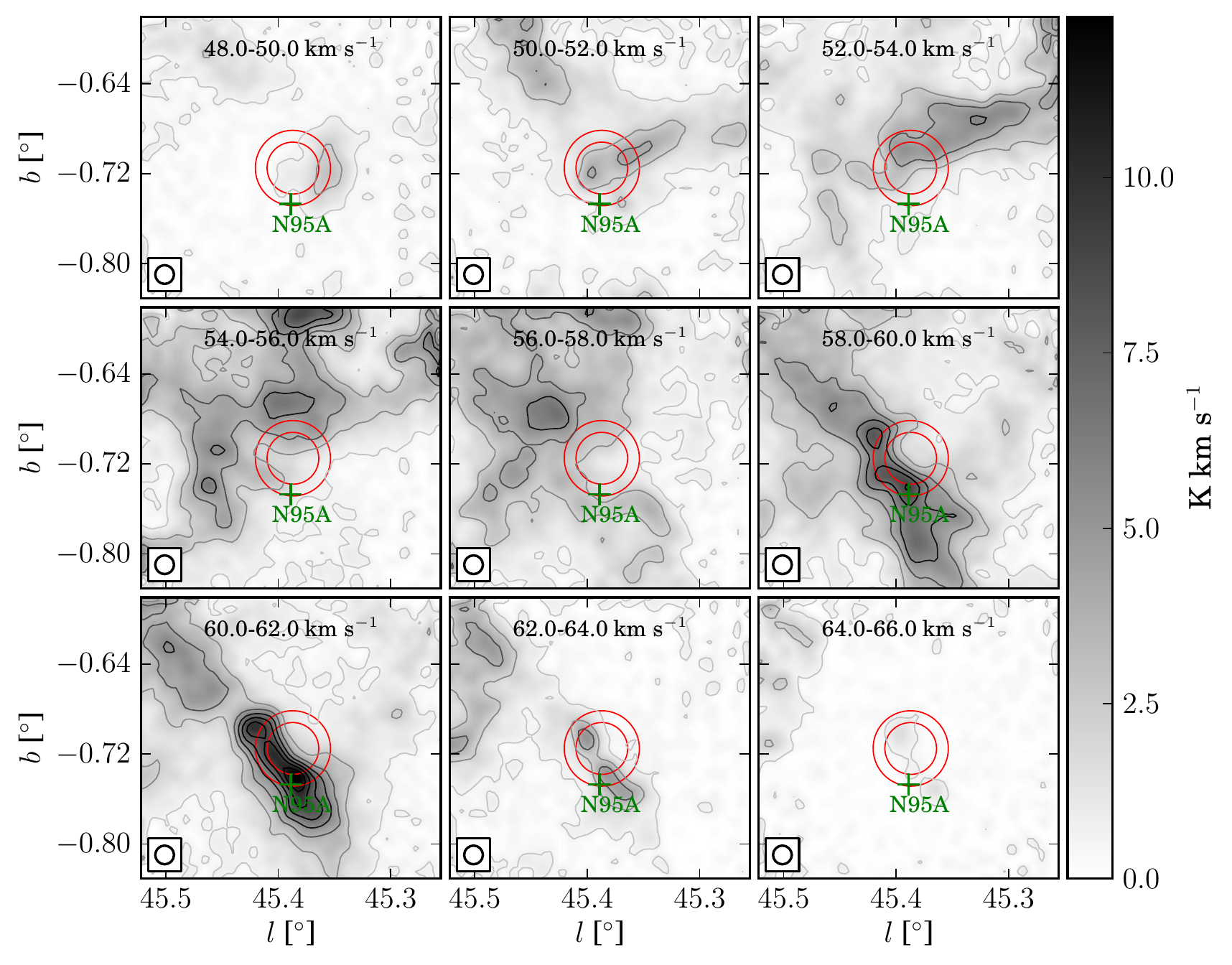}
\caption{Channel map of \coss\ for bubble N95 from 48.0 \kms\ to 66.0 \kms\ with 2.0 \kms\ intervals. The red circles delineate the the position of N95 given by~\citet{2012MNRAS.424.2442S}, and the green cross marks the peak position of the clump. The rms ($\sigma$) of the image background is about 0.11 \intunit, and the contour levels space linearly from 5$\sigma$ to the peak with step 14.9$\sigma$.}
\label{Fig:N95chananelmap}
\end{figure}

Our observations show that the molecular clouds distribute largely alongside the  8 \mum\ arc, and a molecular clump, N95A, is located at the border of N95, as shown in Figure~\ref{Fig:N95chananelmap}. The profiles of N95 is contaminated by other components with similar velocities, as shown in Figure~\ref{Fig:allspectral}, and consequently, we cannot perform outflow identifications in this region.

\subsubsection{N105}

\begin{figure} [H]
\center
\includegraphics[width=0.9\textwidth]{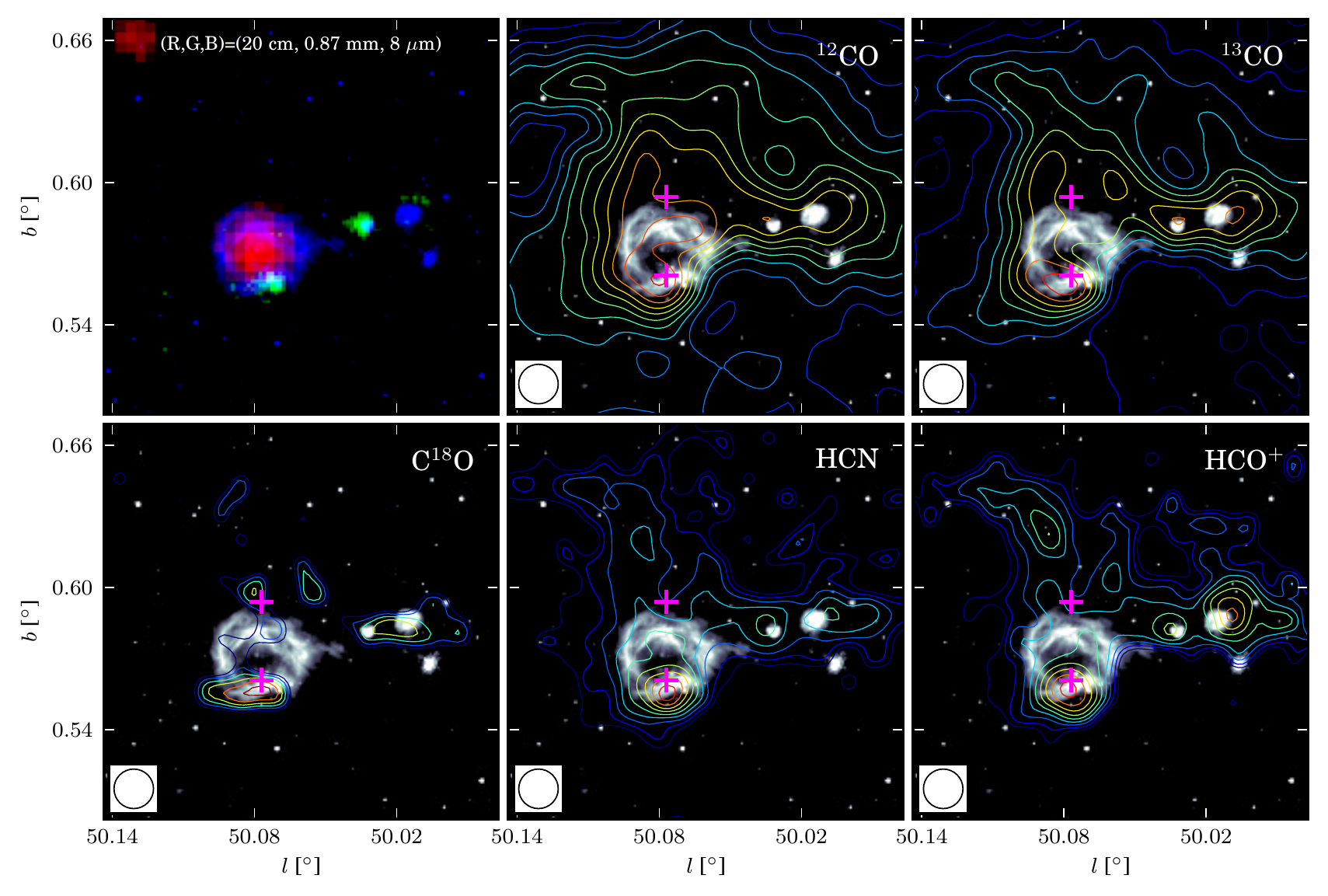}
\caption[]{Images of N105. The upper left panel is a three color image with 20 cm in red, 0.87 mm in green, and 8 \mum\ in blue. The other five panels are contours of five spectral line integrated intensity maps superimposed on the 8 \mum\ image with positions of clumps marked by purple crosses. The contour maps integrate the intensity from -6.0 \kms\ to 2.0 \kms. The rms ($\sigma$) of five line maps, \cofs, \coss, \cots, \hcns, and \hcos, are 0.40, 0.22, 0.22, 0.06, and 0.06 K \kms, respectively, and these contours all begin at 8$\sigma$ level, spacing with 21, 14, 2, 3, and 3$\sigma$, respectively.}
\label{Fig:N105Fiveline}
\end{figure}

N105 is a closed bubble with  small angular size  containing weak 20 cm continuum emissions, as shown in Figure~\ref{Fig:N105Fiveline} (see Table~\ref{Tab:lineParameters} for the observations parameters). \citet{2012ApJ...754...62A}  suggest a velocity of  -1.1 \kms\ and a kinematic distance of 11.2 kpc for this \HII\  region. This bubble is not included in the samples of~\citet{2010A&A...523A...6D}.

\begin{figure} [H]
\center
\includegraphics[width=0.8\textwidth]{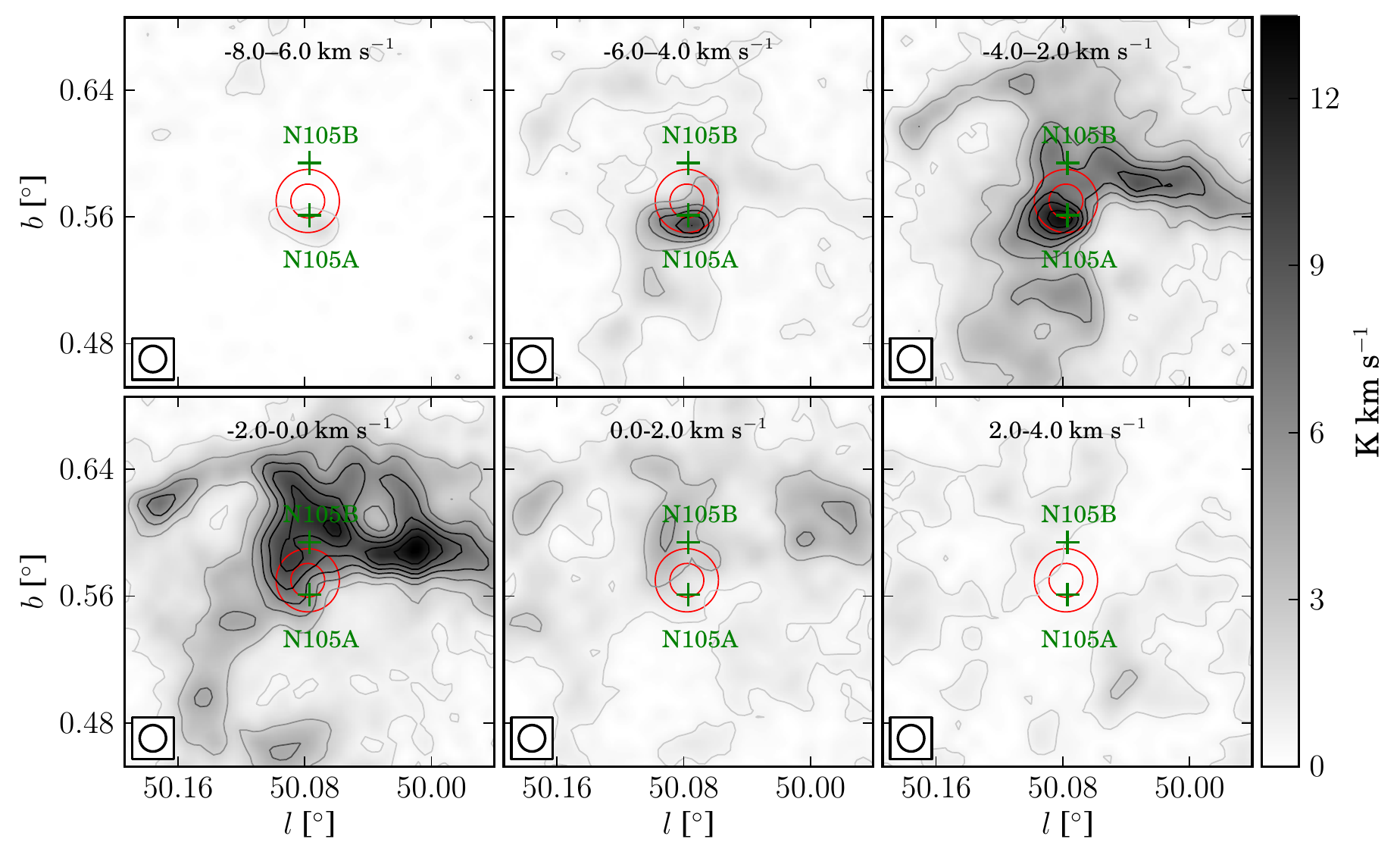}
\caption{Channel map of \coss\ for bubble N105 from -8.0 \kms\ to 4.0 \kms\ with 2.0 \kms\ intervals. The red circles delineate the position of N105 provided by~\citet{2012MNRAS.424.2442S}, and the green crosses mark the peak position of clumps. The rms ($\sigma$) of the image background is about 0.11 \intunit, and the contour levels space linearly from 5$\sigma$ to the peak with step 17$\sigma$.}
\label{Fig:N105chananelmap}
\end{figure}

 Our observations show that two molecular clumps located at the edge of N105, which are N105A and N105B, as shown in Figure~\ref{Fig:N105chananelmap}. No clear line wing is found toward 105A, and we found an outflow candidate in N105B. This outflow shows a clear blue wing, while the red end of the profiles is slightly contaminated by another molecular component. Details about this outflow is discussed in Section ~\ref{secoutflow}. 

\subsubsection{N123}

\begin{figure} [H]
\center
\includegraphics[width=0.9\textwidth]{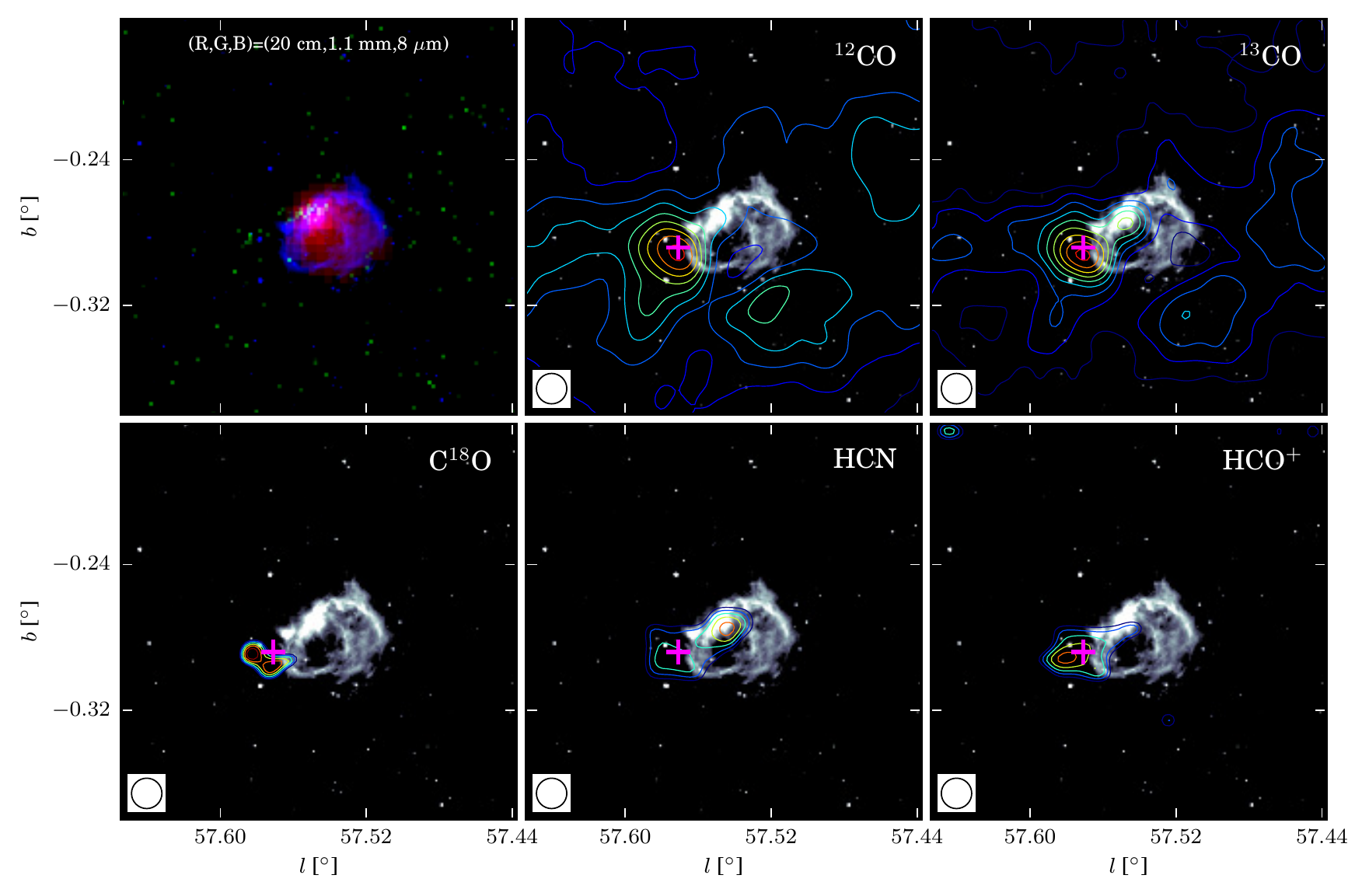}
\caption[]{Images of N123. The upper left panel is a three color image with 20 cm in red, 1.1 mm in green, and 8 \mum\ in blue. The other five panels are contours of five spectral line integrated intensity maps superimposed on 8 \mum\ image with positions of clumps marked by purple crosses. The contour maps integrate the intensity from -4.0 \kms\ to 6.0 \kms. The rms ($\sigma$) of five line maps, \cofs, \coss, \cots, \hcns, and \hcos, are 0.26, 0.17, 0.16, 0.05, and 0.06 K \kms, respectively, and these contours all begin at 8$\sigma$, spacing with 27, 13, 1, 3, and 3$\sigma$, respectively.}
\label{Fig:N123Fiveline}
\end{figure}

N123 is a noteworthy bubble with bright compact 20 and 6 cm continuum~\citep{2005AJ....130..586W}  sources in its rim, as shown in Figure~\ref{Fig:N123Fiveline} (see Table~\ref{Tab:lineParameters} for the observations parameters), and the flux at 24 \mum\ band~\citep{2009PASP..121...76C,2015AJ....149...64G} is saturated. Most likely, they are tracing an ultra-compact \HII\ region, which means a high-mass star might have been triggered by the expanding \HII\ shell of N123. Although  \citet{2010ApJ...716.1478W}  suggested a velocity of 25 \kms\ for N123, we did not find any consistency between CO and 8 \mum\ emission at this velocity.  However, the observation~\citep{2003ApJ...587..714W} of H110$\alpha$ and H$_2$CO show velocities of  6.3 and 2.0 \kms, respectively, which is more likely to be associated with N123. Following~\citet{2003ApJ...587..714W}, we adopted the distance of 8.6 kpc.

\begin{figure} [H]
\center
\includegraphics[width=0.8\textwidth]{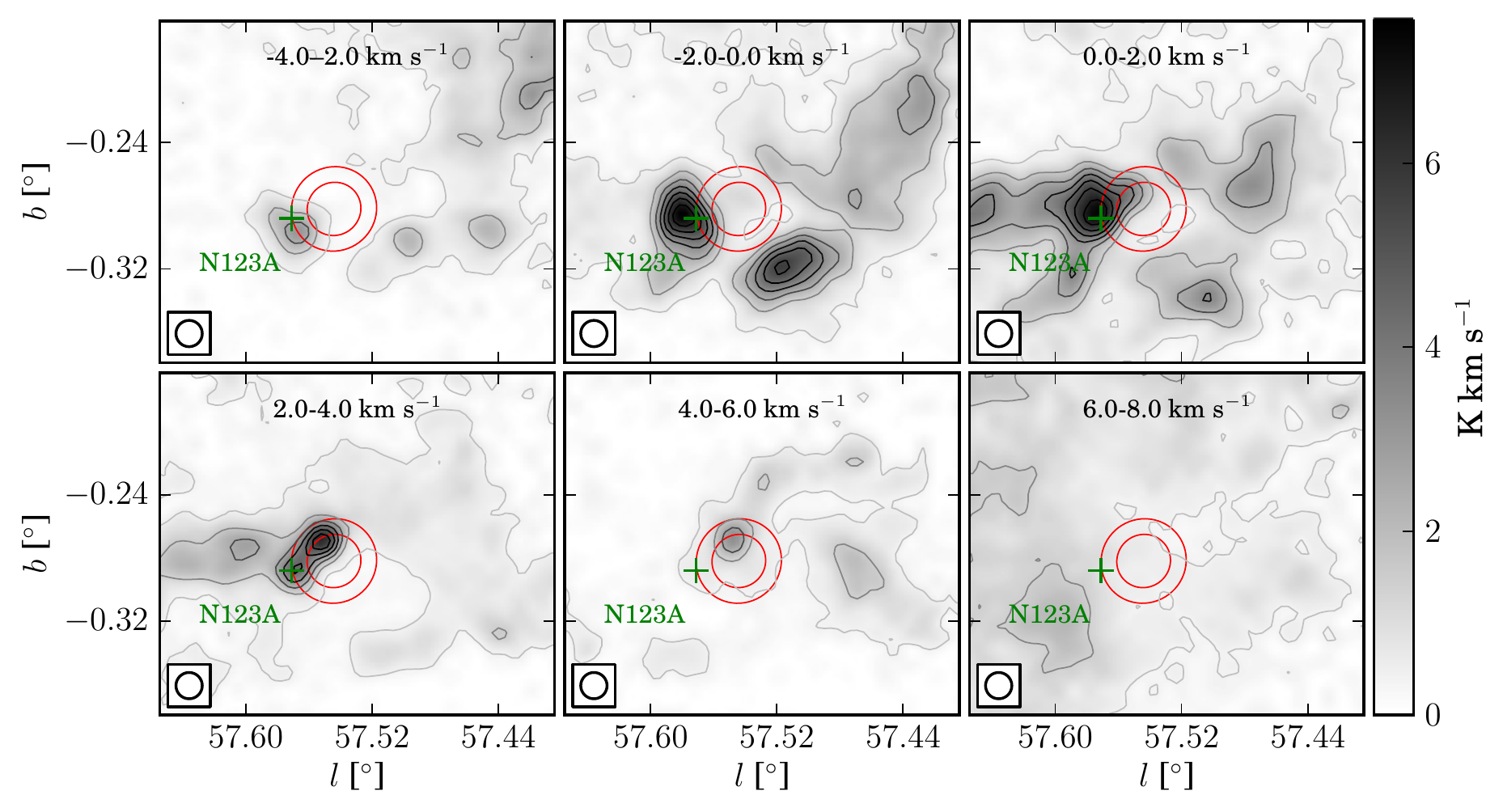}
\caption{Channel map of \coss\ for bubble N123 from -4.0 \kms\ to 8.0 \kms\ with 2.0 \kms\ intervals. The red circles delineate the position of N123 provided by~\citet{2012MNRAS.424.2442S}, and the green cross marks the peak position of the clump. The rms ($\sigma$) of the image background is about 0.11 \intunit, and the contour levels space linearly from 5$\sigma$ to the peak with step 13.7$\sigma$. }

\label{Fig:N123chananelmap}
\end{figure}

 According to our results, CO emissions at  2 \kms\ are  morphologically in good agreement with N123. The dense gas traced by \hcns\ and \hcos\ is present near the border of N123. We identified a molecular clump near N123, which is marked in Figure~\ref{Fig:N123chananelmap}. The profiles of N123 is severely contaminated by other components with adjacent velocities, and therefore we failed to perform outflow identification in this region.

\subsubsection{N133}

\begin{figure} [H]
\center
\includegraphics[width=0.9\textwidth]{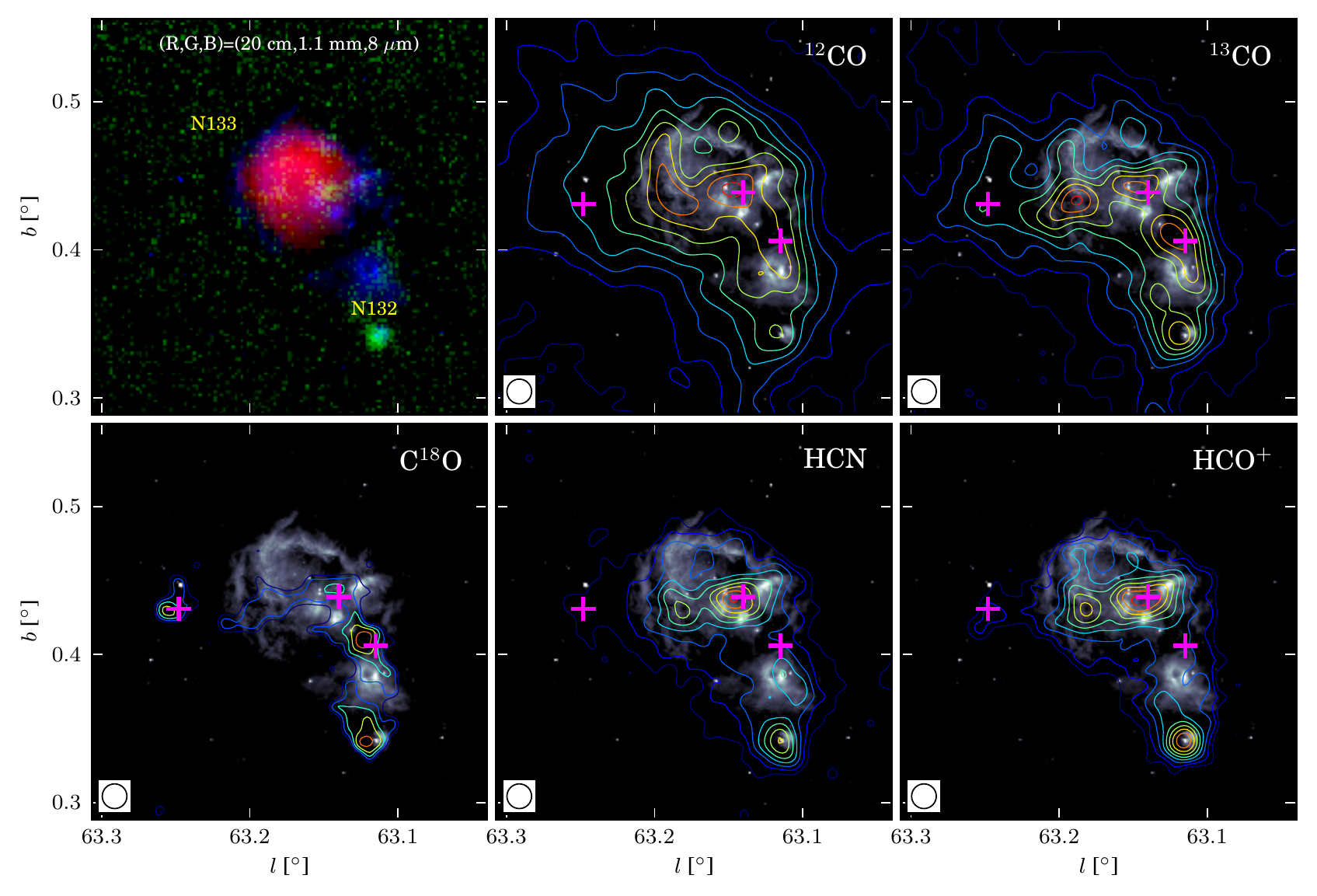}
\caption[]{Images of N133. The upper left panel is a three color image with 20 cm in red, 1.1 mm in green, and 8 \mum\ in blue. The other five panels are contours of five spectral line integrated intensity maps superimposed on the 8 \mum\ image with positions of clumps marked by purple crosses. The contour maps integrate from 15.0 \kms\ to 25.0 \kms. The rms ($\sigma$) of five line maps, \cofs, \coss, \cots, \hcns, and \hcos, are 0.46, 0.25, 0.24, 0.07, and 0.08 K \kms, respectively, and these contours all begin at 8$\sigma$, spacing with 40, 19, 3, 7, and 5$\sigma$, respectively.}
\label{Fig:N133Fiveline}
\end{figure}

N133 is the only optically visible \HII\ region in our samples. The 20 cm continuum emission is prominent, as shown in Figure~\ref{Fig:N133Fiveline} (see Table~\ref{Tab:lineParameters} for the line parameters). Its velocity and distance are approximately 21.4 \kms\ and 2.1 kpc, respectively~\citet{2010ApJ...716.1478W}, and there is another bubble, N132, with smaller angular size nearby. However, N132 is not resolved by the PMODLH 13.7 m telescope.

\citet{2010ApJ...716.1478W} identified 6 YSOs from this region, with positions clearly not coincident with the peaks of the 8 \mum\ emission. \citet{2014A&A...566A.122S} performed a multi-wavelength investigation focusing on star formation activity in the Sh2-90 \HII\ complex associated with N133. They found 129 low mass YSOs around this  bubble, and  confirmed the main ionization source as an O8-O9 V star suggested by~\citet{1983A&A...124....1L}. They proposed that multi-generation star formation is present in this complex, and argued that the expanding \HII\ region has triggered star formation at the borders of Sh2-90.

\begin{figure} [H]
\center
\includegraphics[width=0.8\textwidth]{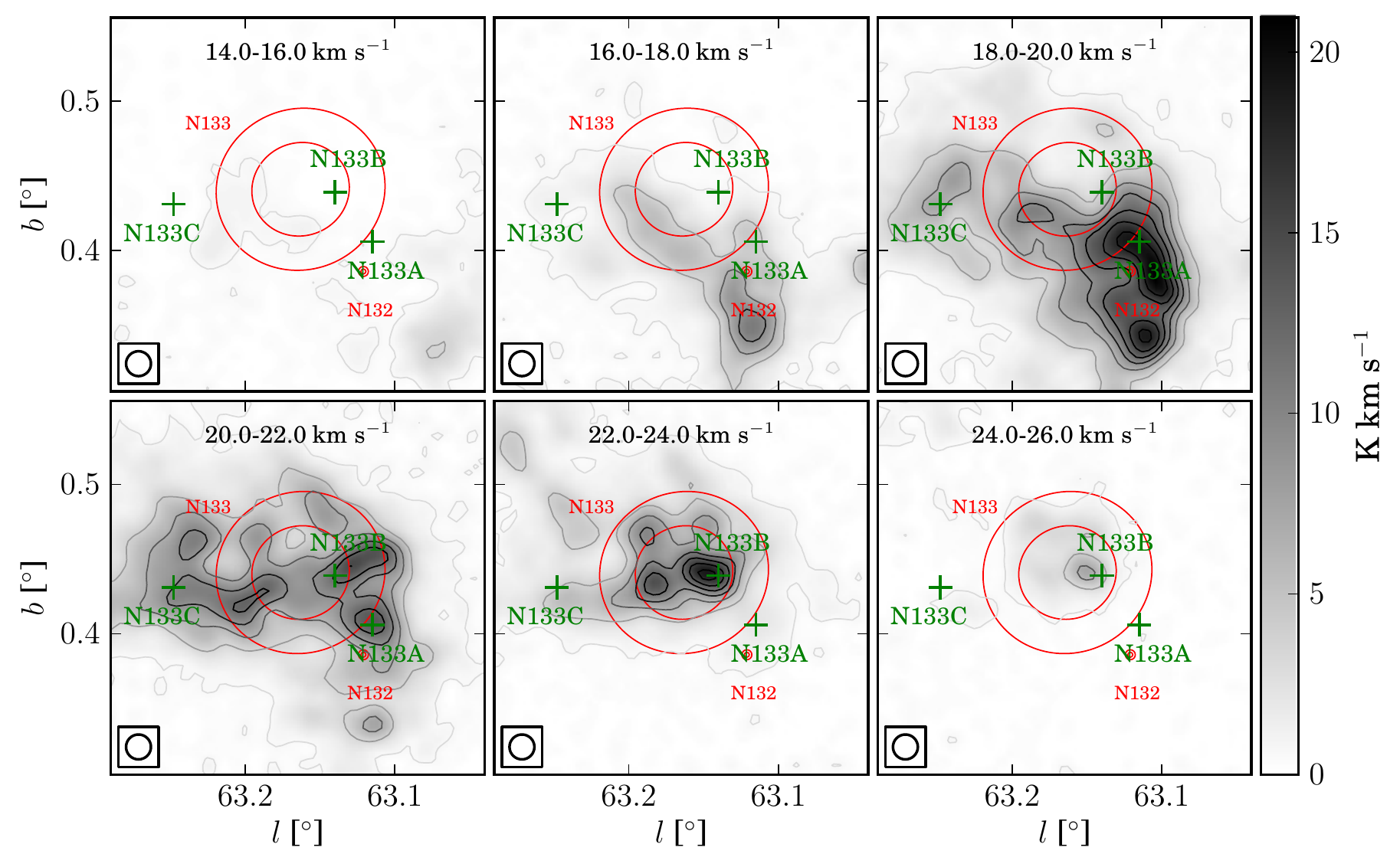}
\caption{Channel map of \coss\ for bubble N133 from 14.0 \kms\ to 26.0 \kms\ with 2.0 \kms\ intervals. The red circles delineate the position of N133 offered by~\citet{2012MNRAS.424.2442S},  and the green crosses mark the peak position of clumps. The rms ($\sigma$) of the image background is about 0.11 \intunit, and the contour levels space linearly from 5$\sigma$ to the peak with step 26.3$\sigma$.}

\label{Fig:N133chananelmap}
\end{figure}

From our observation, the five molecular line emissions agree well with 8 \mum emissions, and three molecular clumps were identified, which is displayed in Figure~\ref{Fig:N133chananelmap}. There is another adjacent component making the blue end of profiles unclean, while the red end is uncontaminated. We identified an outflow near N133, details of which are discussed in section~\ref{secoutflow}. 

\subsection{Outflows}
\label{secoutflow}

 Outflow is a direct signature of ongoing star formation.  Although it is hard for them to be detected, especially for bubble regions where high-mass stars are present, we still performed the identification for outflows around bubbles.

We identified outflows by checking the profiles, integrated intensity maps, and  particularly, the position velocity (P-V) diagrams,  in which outflows are characterized with protruding structures~\citep{1997A&A...323..223S,2004ApJ...608..330B}. However, we must be cautious about interpreting wing profiles near infrared bubbles as outflows, since the interaction between bubbles and molecular clouds can also produce similar structures. Therefore, intensity maps of  lobes  are also required to be clearly distinguished  from background emissions, and, most importantly, the line wings should be significantly wide.
 
The specific steps to identify  an outflow were:
\begin{enumerate}

\item Estimate the center velocity and position  of the drive source using \cots\ line. For a molecular clump in Table~\ref{Tab:clumps}, we integrated \cots\ line over the entire velocity range, and regard the peak of integrated intensity map of \cots\ as the possible position of the driven source.  Subsequently, we determined the center velocity using the averaged \cots\ profile over the peak position and its adjacent pixels (within 1 arcmin). 

\item   In order to improve SNRs of lines, we smooth  the CO and \hcos\ lines to a  velocity resolution of 0.8 \kms. This is slightly smaller than the value adopted  by~\citet{2005ApJ...625..864Z}, about 1 \kms, which is sufficient to identify outflows with different kinds of masses.  

\item   Determine the velocity ranges of red and blue wings, using \cofs\ or \hcos line. We keep the velocity interval  between the center velocity and the inner edge of red and blue lobes equal, and extend the outer end of their velocity ranges to the first zero point of their line profiles.  We gradually increased this velocity interval, until at least one integrated intensity map of these two lobes are clearly distinguished  from the background.

\item Draw the P-V diagram to confirm outflow wings. 
\end{enumerate}

We collected samples for searching for outflows by investigating the \cofs\ or \hcos\ profiles of bubble regions visually. For a bubble, if the red or blue or both ends of the profiles are uncontaminated by other components,  it is appended to the sample list. Totally, there are seven bubbles in this list, N14, N37, N55, N74, N105, and N133. Although the CO profiles of N37 and N14 are complicated as shown in Figure~\ref{Fig:allspectral}, their \hcos\ emissions are relatively strong and uncontaminated, and therefore they are included in our samples. 24 \mum~\citep{2009PASP..121...76C, 2015AJ....149...64G} emissions,  which can trace the warmed dust emissions probably heated by protostellars, were also involved for confirmation.

Four of these six samples, N14, N55, N105, and N133, are found harboring outflow candidate. However, only N55 shows evident  bipolar structure, while the other three bubbles display single wings, mainly due to the contamination of adjacent components with similar velocities.
  
We checked the outflow candidates identified by \citet{2010ApJ...709..791B} using an visual method based on CO (J=3-2). Three of their candidates, N37, N49, and N74, were included in our bubble observations.  For the clumps near N37,  the \cofs\ emissions are quite complex in the velocity range 30-40 \kms, while no clear line wing is shown by \hcos. Furthermore, no 24 \mum\ source is found in this clumps,   indicating no outflow is present here.   For N49,  both of the CO and \hcos\ profiles are severely contaminated by adjacent components with approximate velocities. Consequently, we cannot perform outflow identification for this region.  For the outflow candidate  The outflow candidate near N74 is actually associated with N75, and however, there are two components present here, which made it difficult to identify outflows. 
 
Following~\citet{1984ApJ...284..176S}, we calculated the mass, momentum, and energy entrained  in the outflowing gas without any projection effect calibrated.  The LTE assumption was applied, and we also assumed the optical depth of the \cofs\ were thin at line wings. \cofs\ column densities were estimated by 
\begin{equation} 
N_{\mathrm{^{12}CO}} = \frac{4.2\times10^{13}T_{\mathrm{ex}}\int T_R(\mathrm{^{12}CO})\mathrm{d}V}{\mathrm{exp}(-5.5/T_{\mathrm{ex}})} \  \rm cm^{-2},
\end{equation}
 where $T_{\mathrm{ex}}$  is the excitation temperature of $^{12}$CO, calculated using the \cof\ bright temperature at  line centers; and $\int T_R(\mathrm{^{12}CO})\mathrm{d}V$  is the integrated intensity over the velocity range of lobes~\citep{1984ApJ...284..176S}. The spectrum was averaged over the area which is determined by the  half  peak value contour of the  integrated intensity over the corresponding velocity range. We adopted a value of $\sim1\times10^4$ for  the ratio of $^{12}$CO to H$_2$ column density, $N{(\mathrm{H}_2)}/N( \rm ^{12}CO)$, following~\citet{1984ApJ...284..176S}.

The momentum, $P$, is proportional  to $\Sigma \int T_R(\mathrm{^{12}CO})V\mathrm{d}V$, and  the energy, $E$, is proportional to $\Sigma \int T_R(\mathrm{^{12}CO})V^2\mathrm{d}V$, where $V$ is the velocity of gas with respect to the driven source. The scale of a single lobe is defined by the the separation between the lobe peaks and the central driven source. The dynamical age, $t_{\rm dyn}$, was estimated  simply through dividing the scale by the mean outflow velocity that is defined as $P/M$~\citep{1990ApJ...348..530C}. The rate of outflow mass, $\dot M = M/t_{\rm dyn}$, the mechanical force, $F_{\mathrm{outflow}} = P/t_{\rm dyn}$, and outflow luminosity $L_{\mathrm{outflow}} = E/t_{\rm dyn}$, were also calculated. Details regarding the calculation of  outflow parameters can be found in~\citet{1996ApJ...472..225S, 2005ApJ...625..864Z}. The estimated parameters are summarized in Table~\ref{Tab:outflows}.

\subsubsection{Outflow N14}

\begin{figure} [H]
\center
\includegraphics[width=0.8\textwidth]{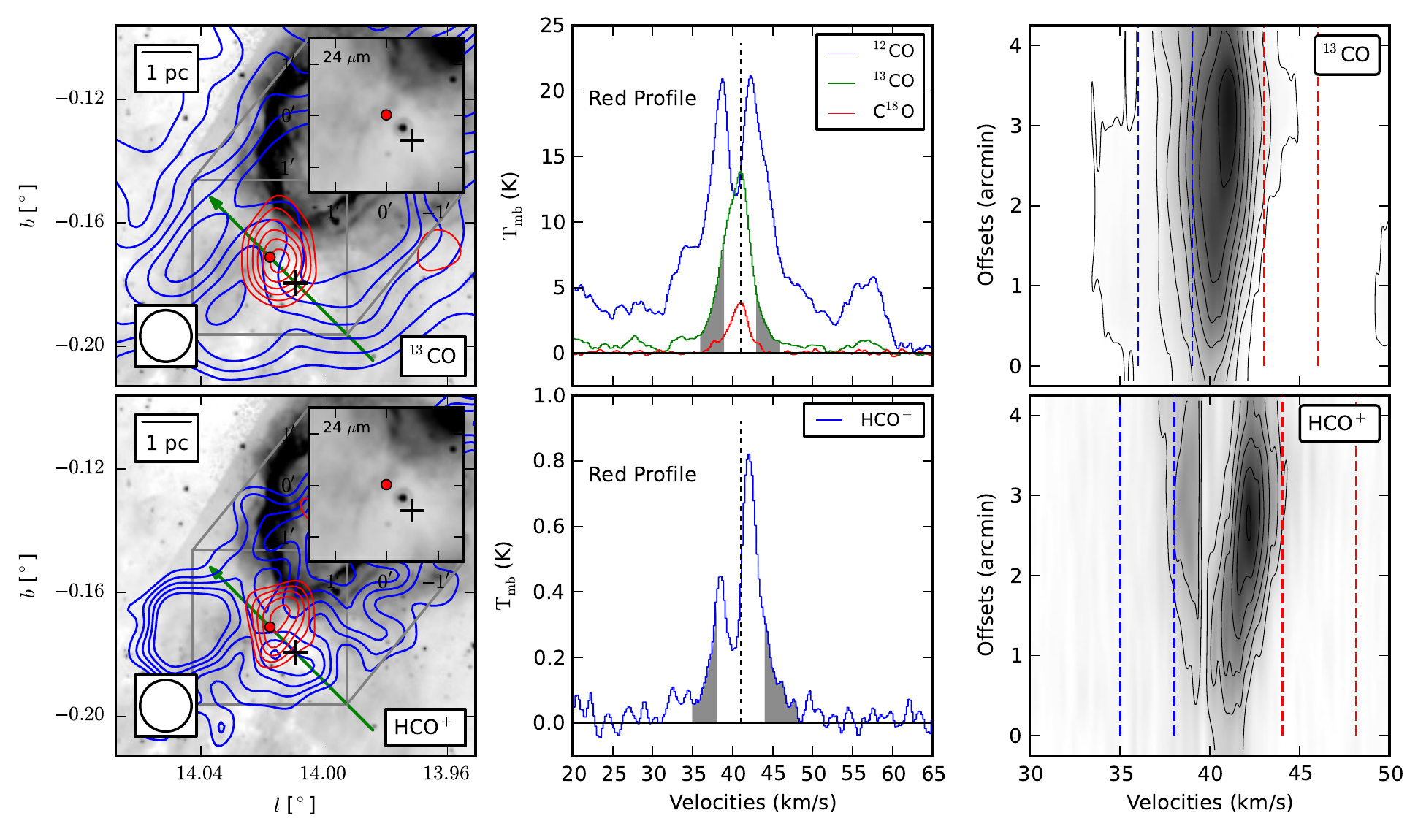}
\caption[]{Outflow map of N14. The left map shows the 8 \mum\ background image with the \cos\ integrated intensity contour map superimposed on it. The inset shows a close section of the MIPSGAL 24 \mum\ image. The integrating ranges for the blue and red lobes are shown in the line profiles and  position velocity (P-V) slice maps.  The black cross marks the peak position of \cots\ integrated intensity map. The green arrow delineates the origin and direction of the P-V diagram. The \cofs\ contours start at the 50\% level and increase by 10\% of the peak value, while the  \hcos\ contours start at the 60\% level and increase by 10\% of the peak value.} 
\label{Fig:N14outflow}
\end{figure}

Although  CO profiles provided limited information about this outflow due to their complexities,  \hcos\ reveals an outflow with a clear red wing near N14, as shown in Figure ~\ref{Fig:N14outflow}.

The drive source of this outflow is  probably located in the clump N14A, towards which we found a 24 \mum\ point source which probably marks a more precise position of the driven source.  A clearly protruded structure can be seen in the  P-V diagram of the \coss\ lines, as demonstrated in Figure~\ref{Fig:N14outflow}. However, we cannot  calculate the outflow parameters because the CO profiles of this region is exceedingly complicated.

\subsubsection{Outflow N55}

\begin{figure} [H]
\center
\includegraphics[width=0.9\textwidth]{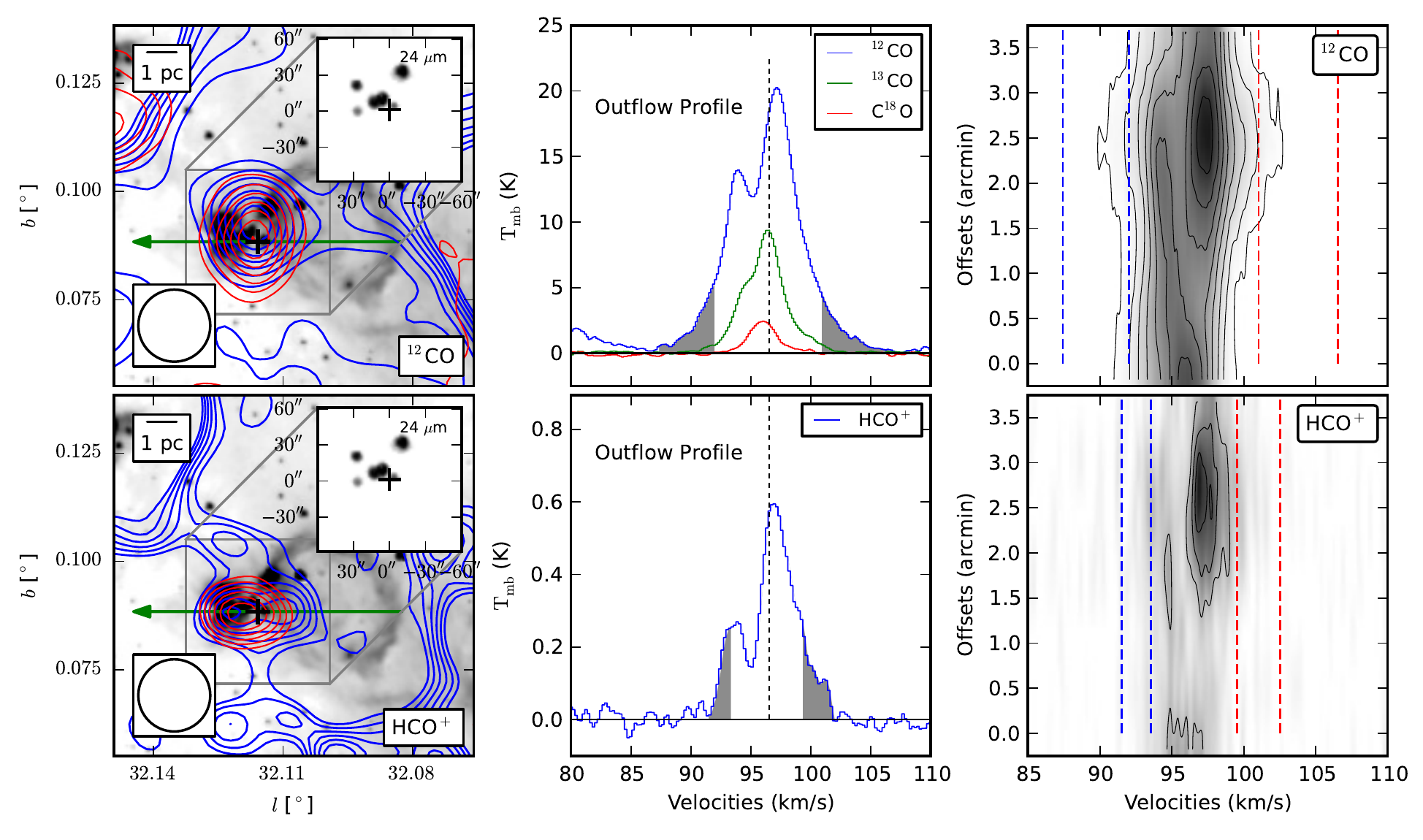}
\caption[]{Outflow maps of N55. The left map shows the 8 \mum\ background image with \cof\ integrated intensity contour map superimposed on it. The inset shows a close section of the  MIPSGAL 24 \mum\ emissions.  The integrating ranges for the blue and red lobes are shown in the line profiles and position velocity (P-V) slice maps.  The black cross marks the peak position of \cots\ integrated intensity map. The green arrow delineates the origin and direction of the P-V diagram. The \cofs\ contours start at  the 40\% level and increase by 10\%  of the peak, while  the  \hcos\ contours start at the 50\% and increase by 10\%  of the peak value.} 
\label{Fig:N55outflow}
\end{figure}

This is a bipolar outflow, and as shown in Figure~\ref{Fig:N55outflow}, the red and blue wings of the CO profiles are wide. There are several ultra-compact \HII\ regions traced by the 20 cm continuum in the clump, N55A, associated with this outflow candidate. Consequently, there are probably more that one driven source here, and they are most likely high-mass.  Since they are not resolved by our observation, and we simply treat this candidate as a single outflow when we calculating its parameters. However, we cannot estimate the outflow scale for this outflow, because the blue lobe totally overlaps with the red one.  Consequently, the time related parameters cannot be calculated either.

\subsubsection{Outflow N105}

\begin{figure} [H]
\center
\includegraphics[width=0.8\textwidth]{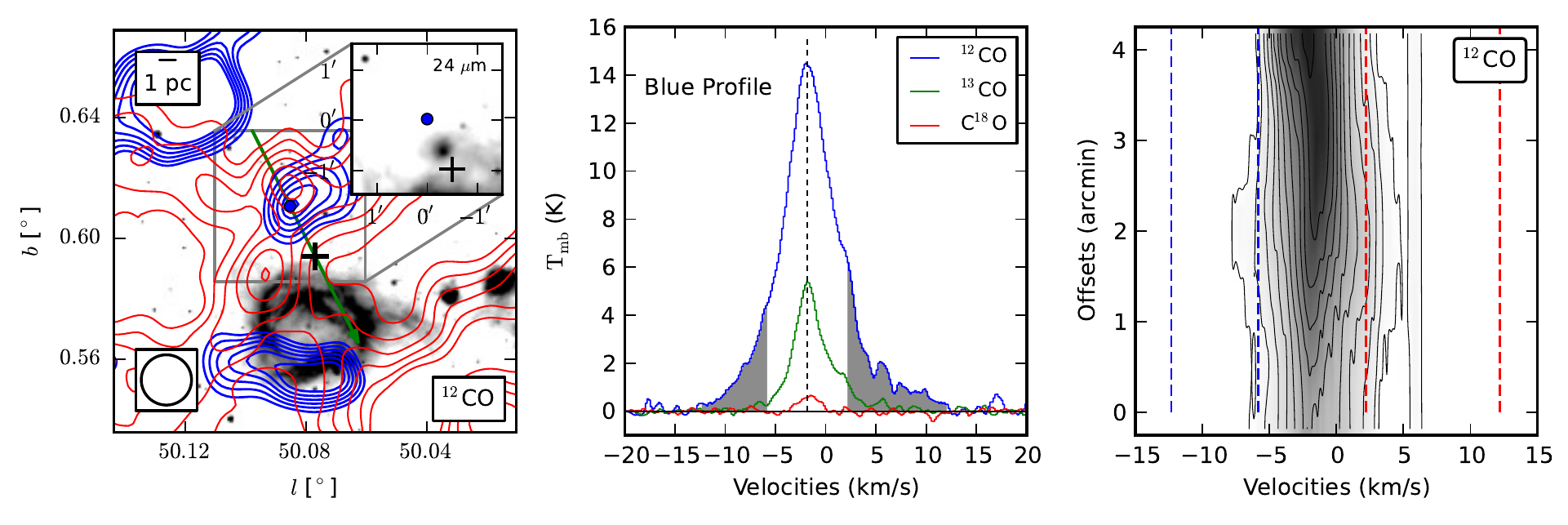}
\caption[]{Outflow maps of N105. The left map shows the 8 \mum\ background image with \cof\ integrated intensity contour map superimposed on it. The inset shows a close section of the MIPSGAL 24 \mum\ background. The integrating ranges for the blue and red  lobes are shown in the line profiles and position velocity (P-V) slice map.  The black cross marks the peak position of \cots\ integrated intensity map. The green arrow delineates the origin and direction of the P-V diagram. The \cofs\ contour levels start at 50\%, and increase by 10\%  of the peakvalue.} 
\label{Fig:N105outflow}
\end{figure}

This single pole outflow candidate was located in the clump N105B, and its \cofs\ profile  displayed a wide clear blue wing, as shown in Figure~\ref{Fig:N105outflow}. Unfortunately, the red end of the  profile is slightly contaminated by another component. Near the peak of the \cots\ integrated  intensity map, we found a 24 \mum\ continuum source which is probably tracing the driven source. We did not draw  \hcos\ line maps due to its low SNR.  The P-V diagram of \cofs\ displays a  clearly protruded structure, confirming the existence of the outflow.

\subsubsection{Outflow N133}

\begin{figure} [H]
\center
\includegraphics[width=0.8\textwidth]{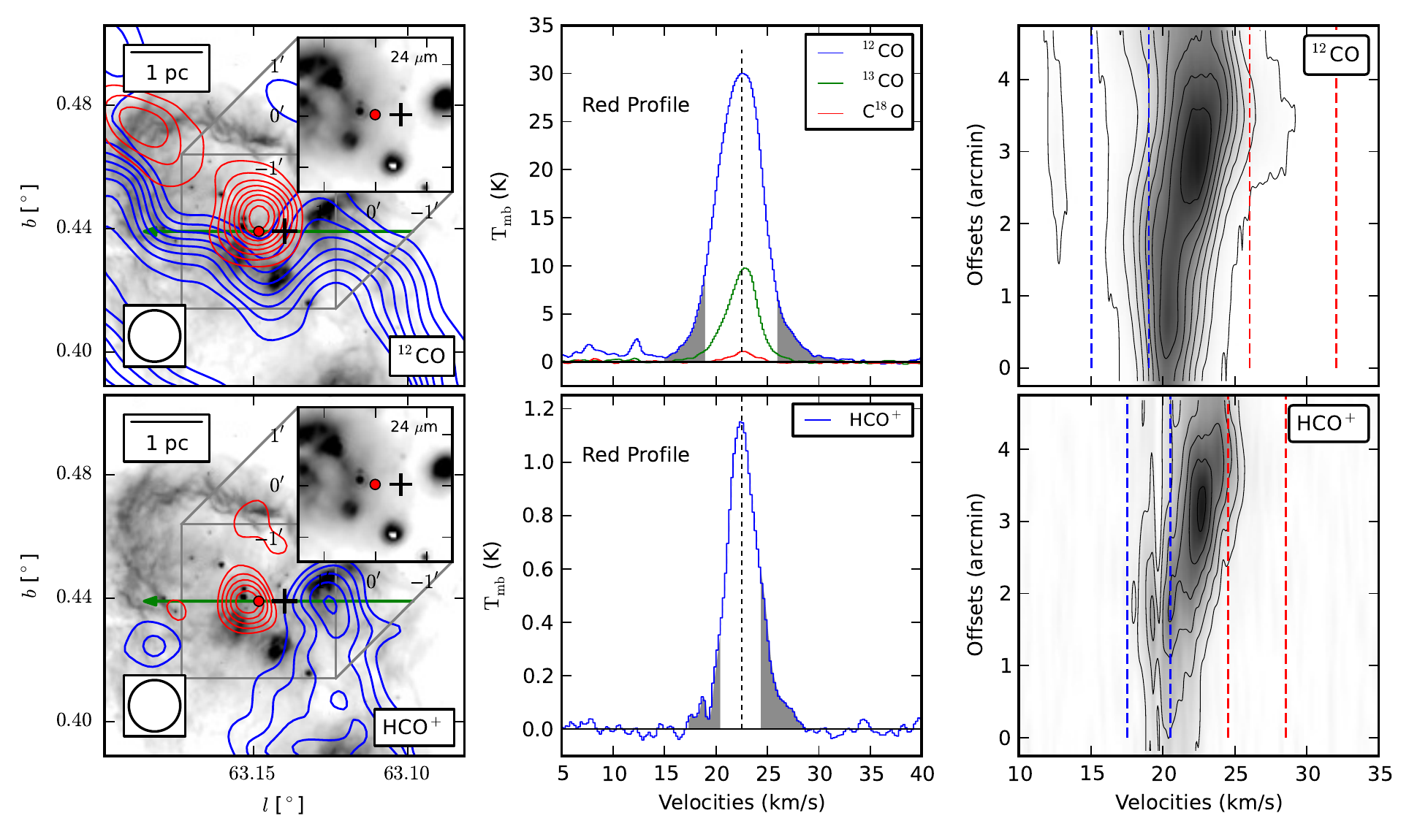}
\caption[]{Outflow maps of N133. The left map shows the 8 \mum\ background image with \cof\ integrated intensity contour map superimposed on it. The inset shows a close section of the  MIPSGAL 24 \mum\ background. The integrating ranges for the blue and red  lobes are shown in the line profiles and position velocity (P-V) slice maps. The black cross marks the peak position of \cots\ integrated intensity map. The green arrow delineates the origin and direction of the P-V diagram. The \cofs\ contours start at 30\% and increase by 10\%  of the peak while the \hcos\ contours start at 50\% and increase by 10\%  of the peak.} 
\label{Fig:N133outflow}
\end{figure}

The blue profiles in this region are mixed with another component located near N133. However, the \cofs, \coss, and \hcos\ line profiles shows clear red wings, which indicates an outflow may be present here. As shown in Figure~\ref{Fig:N133outflow}, there are several 24 \mum\ sources around the \cots\ clump. Consequently, we could not distinguish which is responsible for this outflow due to the limited resolution.

\begin{table}[H]
 \tiny
\setlength{\tabcolsep}{2pt}
  \centering
  \begin{center}
\caption{Outflow parameters.\label{Tab:outflows}}
\begin{tabular}{lcccccccccccccccc}
   \tableline\tableline

 Name& Distance&  Lobe& Peak &Velocity Range   & Mass&Momentum&Energy&$\bar{v}$ & Scale & t$_{\rm dyn}$  & $\dot{M}$& $F_{\rm outflow}$  & $ L_{\rm outflow}$  \\

&  &&  &   &  &   & (10$^{45}$ )  & && (10$^{4}$) & (10$^{-4}$) & (10$^{-3}$)& \\

&(kpc) &&  (\deg, \deg)& (\kms)   & ($M_\odot$)& ($M_\odot$ km s$^{-1}$)  & (ergs)&(\kms)&(pc)& (yr)  & ($M_\odot/yr$) & ($M_\odot$ km s$^{-1} yr^{-1}$) &$L_\odot$\\
   \tableline

 N14 & 3.6 & Red &(14.018, -0.171) &44.0~-~47.0  & ... & ...& ...& ... & ...&... &... &... &...\\
\tableline
 
N55 & 8.4 & Red &(32.116, 0.088) &101.0~-~106.5 & 17 & 103 & 6.7 &... &... &... &...&...&... \\
 
N55 & 8.4 & Blue &(32.116, 0.088) &87.4~-~92.0 & 24 & 144 & 8.8  &...&...&... &... &... &...\\ 
 
 \tableline
 
  N105 & 11.2 & Blue &(50.085, 0.611) &-12.3~-~-5.8 &  40.4  & 231.4 &  14.0  &5.7 &  2.4  &  40.6  & 1.00  & 0.57  & 0.28 \\
 
 \tableline

 N133 & 2.5 & Red &(63.148, 0.439) &26.0~-~32.0 & 1.4 & 6.8 & 0.3 &4.9 & 0.4 & 7.3 &0.19 &0.09 &0.04 \\

\tableline
\end{tabular}
 \vspace{0.4cm}
\flushleft
 
\end{center}
  
\end{table}

\section{DISCUSSION}

\subsection{Molecular Clouds and Clumps Around Bubbles}

Our observations show that all of the 18  bubbles are associated with molecular clouds, and  the majority of them are giant and  characterized with complex profiles.  All five molecular lines were detected for all bubbles except N89 and N90, where \cots\ emissions are absent and \hcos\ emissions are relatively weak. Clumps present around all 18 bubbles, and molecular clouds near four of these bubbles, N4, N14, N49, and N133, display shell morphologies. As shown in Figure~\ref{Fig:allspectral}, some molecular clouds exhibited broad shifted profiles, e.g. MWP1G032057+000783, N82, and N95. These profiles deviate  significantly from the Gaussian curve, which indicates that their environments are greatly affected by the expansion of \HII\ regions. Since these profiles are similar to those displayed by molecular clouds interacting with supernova remnants (SNRs)~\citep{2014ApJ...791..109Z,2014ApJ...796..122S}, we speculate that these bubbles are probably sweeping the molecular clouds.  However, we should be aware that these broad line features could also be caused by adjacent components and cloud-wide rotation. Since these broad line features are situated near bubbles, we consider that the interaction between bubbles and their surrounding molecular clouds is the more likely explanation. 

We estimated physical properties for the molecular clouds and clumps using three isotopic CO  lines and based on LTE assumption. Average spectra were used when calculating the mass for each bubble, and the optical depths calculated from \coss\  were generally below 0.5, supporting the assumption of optically thin  for the \coss\ line. Molecular cloud masses within bubble regions are of the order of 10$^3$-10$^4$ $M_\sun$,  close to the value of $\sim10^4M_\sun$ calculated by \citet{2005ApJ...623..917H} for a high-mass star with an age of 1 Myr. According to our calculations, most molecular clumps possess a typical mass of $10^3 M_\sun$, and their line with of \cots\ are generally 2 \kms. However, these masses inevitability include the envelopes around the molecular clumps, and they should be treated as upper limits.  Nonetheless,    12 of the 23 identified molecular clumps possess virial masses which are significantly smaller than LTE masses, indicating  these clumps are mostly gravitationally bound.

 \subsection{Outflows and Star Formation Processes}

Most of the bubbles display complex profiles due to the complicated environment associated with high-mass stars.  We found that 6  of the 18 bubble regions satisfy the requirements for searching for outflows, which means at least one end  of their profiles are uncontaminated. We identified 4 outflow candidates from four bubble regions,  N14, N55, N105,  and N133. However, only N55 displays a bipolar structure, and the other three outflows show single lobes.

The detection ratio of outflows is approximately 67\% based on our biased samples. The presence of outflows indicates that star formation processes are indeed occurring around the bubbles which are essentially expanding \HII\ regions. Previous studies of star formations around bubbles mainly focused on YSOs.  However,  it is hard to determine  the masses and ages of those YSOs,  and furthermore,  they are not  necessarily precursors of stars. In contrast, outflows are reliable tracers of ongoing star formation, and are more powerful tools to confirm the existence of star formation activity.  The estimated dynamical ages for outflow near N105 and N133 are about 0.4 and 0.07 Myr respectively, while the age of bubble N105 and N133 are approximately 1.6 and 0.4 Myr which are both significantly larger than the ages of outflows.  Limited by the spatial resolution of PMODLH, we speculate that the star formation processes around these bubbles might be triggered or accelerated by expanding \HII\  regions. However, higher spatial resolution observations are needed to confirm the relationship between bubbles and star formation processes around them.

 \citet{2005ApJ...625..864Z} carried out an CO (J=2-1) survey towards 69 high-mass IRAS sources, with a resolution of approximately 29$''$, and they found a typical energy of $1\times10^{46}$ ergs for the identified outflows. The energy of blue lobe of N105 outflow is approximately $1.3\times10^{46}$ ergs, and the sum  energy of red and blue lobe of N55  is about  $1.6\times10^{46}$ ergs.  Therefore, the drive sources of these two outflows are probably high-mass. We performed a search for EGOs for the clumps which are associated with these four outflows, using the method mentioned by~\citet{2008AJ....136.2391C}. However, we did not identify any EGO from these clumps, and no corresponding source was found from the catalog provided by them. Two possibilities can be responsible for this result, one is that these two outflows are not high-mass, the other is that they are not evolved enough to exhibit excessive 4.5 \mum\ emission which are mainly tracing shocked molecular gas.

Interestingly, there are three bubbles, N55, N49, and N123 display ultra-compact (UC) \HII\ regions at their borders, which are traced by 20 or 6 cm continuum emission.  These UC \HII\ regions are mostly tracing high-mass stars which could be triggered by bubbles.

\subsection{Compare with Previous Molecular Line Studies on Bubbles}

 Some investigations on single bubbles using GRS \cos\ data show that clumps are also present in the vicinity of bubbles beyond our samples. For instance, \citet{2010A&A...513A..44P} found some clumps near N65 are probably produced by the fragmentation of their natal molecular clouds, and they speculate that the  collect-and-collapse process might be occurring here, which is consistent with what we have found.  Furthermore, they identified an EGO in the molecular shell, the mass of which is the same order of magnitude with the clumps identified by us. They identified 22 YSOs candidates around N65, and according to the result of SED fitting for those YSOs, they concluded that the formation of a second generation of stars had occurred.  Similar results are found around N107~\citep{2014A&A...565A...6S} and N115~\citep{2014A&A...569A..36X}. The age of N107 is about 2 Myr, while the age of N115 is about 1.5 Myr. These two values are approximate with the estimated ages of our samples.  The molecular mass along the border of N107 is about $4.0\times10^4 M_{\odot}$, which is comparable with our results.

 \citet{2014MNRAS.438..426H} performed a statistical study of infrared bubbles using GRS \cos\ data. They found 60 percent of 309 MWP bubbles are associated with molecular clumps. The correlation between bubble morphology and molecular gas distribution make them believe some of  these clumps near bubbles are produced by the expansion of bubbles.  
 
 All these studies show that bubbles are likely associated with clumps. According to single bubble studies, the typical age of bubbles is 1Myr, and the masses of molecular clouds along their borders are approximately $1\times10^4 M_{\odot}$.

\section{CONCLUSIONS}

 We have presented an investigation of 13 regions containing 18 infrared bubbles, using three CO isotopic lines and two high density molecular lines, \hcos\ and \hcn. Some profiles of bubble regions show broad redshifted \cofs\ profiles, arc structures, and \cofs\ velocity gradients, indicating they are probably interacting with the molecular clouds around them. Most of the bubbles are associated with dense molecular gas traced by \hcos\ and \hcns, and using \cots, we identified 24 molecular clumps near 18 infrared bubbles, most of which are gravitationally bound. These facts are generally in agreement with either the collect-and-collapse or the RDI model.

 A search of outflow was carried out in six bubble regions. Four bubble regions, N14, N37, N55, and N133, show outflow activities, while only N55 shows a bipolar structure. The energy of outflows indicates the driven source of this outflow is probably high-mass, and however, no EGO is found towards these outflow candidates.  These outflows are convincing evidence that the star formation processes are occurring around bubbles. Besides outflows, ultra-compact \HII\ regions are found on the border of N55, N49, and N123. Among the 18 selected bubbles, six bubbles possess star formation activities nearby, and the detection ratio of outflows and UC \HII\ regions near bubbles is about 0.3.  We speculate that star formation processes might have been triggered around these bubbles. However, higher spatial resolution observations are needed to confirm these star formation activities.

\begin{acknowledgements}
 We are grateful to two anonymous referees for careful readings of the manuscript and constructive comments that make this study complete. Thanks to P. Tremblin  for providing us a machine-readable dataset result of numerical simulation of the expansion of \HII\ regions in turbulent environments.  We would also like to thank Sam McSweeney for his helpful report.  This work was supported by the National Science Foundation of China (Grants No. 11233007 and 11133008),  the National Science Foundation of Shanghai (Grant No. 15ZR1446900),  and the Key Laboratory for Radio Astronomy, CAS.
\end{acknowledgements}


\begin{thebibliography}{88}
\providecommand{\natexlab}[1]{#1}
\providecommand{\selectlanguage}[1]{\relax}

\bibitem[{{Aguirre} et~al.(2011){Aguirre}, {Ginsburg}, {Dunham}
  et~al.}]{2011ApJS..192....4A}
{Aguirre}, J.~E., {Ginsburg}, A.~G., {Dunham}, M.~K., et~al. 2011, \apjs, 192,
  4

\bibitem[{{Alexander} et~al.(2013){Alexander}, {Kobulnicky}, {Kerton}, \&
  {Arvidsson}}]{2013ApJ...770....1A}
{Alexander}, M.~J., {Kobulnicky}, H.~A., {Kerton}, C.~R., \& {Arvidsson}, K.
  2013, \apj, 770, 1

\bibitem[{{Anderson} \& {Bania}(2009)}]{2009ApJ...690..706A}
{Anderson}, L.~D., \& {Bania}, T.~M. 2009, \apj, 690, 706

\bibitem[{{Anderson} et~al.(2011){Anderson}, {Bania}, {Balser}, \&
  {Rood}}]{2011ApJS..194...32A}
{Anderson}, L.~D., {Bania}, T.~M., {Balser}, D.~S., \& {Rood}, R.~T. 2011,
  \apjs, 194, 32

\bibitem[{{Anderson} et~al.(2012{\natexlab{a}}){Anderson}, {Bania}, {Balser},
  \& {Rood}}]{2012ApJ...754...62A}
{Anderson}, L.~D., {Bania}, T.~M., {Balser}, D.~S., \& {Rood}, R.~T.
  2012{\natexlab{a}}, \apj, 754, 62

\bibitem[{{Anderson} et~al.(2014){Anderson}, {Bania}, {Balser}
  et~al.}]{2014ApJS..212....1A}
{Anderson}, L.~D., {Bania}, T.~M., {Balser}, D.~S., et~al. 2014, \apjs, 212, 1

\bibitem[{{Anderson} et~al.(2012{\natexlab{b}}){Anderson}, {Zavagno},
  {Deharveng} et~al.}]{2012A&A...542A..10A}
{Anderson}, L.~D., {Zavagno}, A., {Deharveng}, L., et~al. 2012{\natexlab{b}},
  \aap, 542, A10

\bibitem[{{Bania} et~al.(2012){Bania}, {Anderson}, \&
  {Balser}}]{2012ApJ...759...96B}
{Bania}, T.~M., {Anderson}, L.~D., \& {Balser}, D.~S. 2012, \apj, 759, 96

\bibitem[{{Beaumont} \& {Williams}(2010)}]{2010ApJ...709..791B}
{Beaumont}, C.~N., \& {Williams}, J.~P. 2010, \apj, 709, 791

\bibitem[{{Benjamin} et~al.(2003){Benjamin}, {Churchwell}, {Babler}
  et~al.}]{2003PASP..115..953B}
{Benjamin}, R.~A., {Churchwell}, E., {Babler}, B.~L., et~al. 2003, \pasp, 115,
  953

\bibitem[{{Bergin} \& {Tafalla}(2007)}]{2007ARA&A..45..339B}
{Bergin}, E.~A., \& {Tafalla}, M. 2007, \araa, 45, 339

\bibitem[{{Bertoldi}(1989)}]{1989ApJ...346..735B}
{Bertoldi}, F. 1989, \apj, 346, 735

\bibitem[{{Beuther} et~al.(2004){Beuther}, {Schilke}, \&
  {Gueth}}]{2004ApJ...608..330B}
{Beuther}, H., {Schilke}, P., \& {Gueth}, F. 2004, \apj, 608, 330

\bibitem[{{Cabrit} \& {Bertout}(1990)}]{1990ApJ...348..530C}
{Cabrit}, S., \& {Bertout}, C. 1990, \apj, 348, 530

\bibitem[{{Carey} et~al.(2009){Carey}, {Noriega-Crespo}, {Mizuno}
  et~al.}]{2009PASP..121...76C}
{Carey}, S.~J., {Noriega-Crespo}, A., {Mizuno}, D.~R., et~al. 2009, \pasp, 121,
  76

\bibitem[{{Caswell} et~al.(1995){Caswell}, {Vaile}, {Ellingsen}, {Whiteoak}, \&
  {Norris}}]{1995MNRAS.272...96C}
{Caswell}, J.~L., {Vaile}, R.~A., {Ellingsen}, S.~P., {Whiteoak}, J.~B., \&
  {Norris}, R.~P. 1995, \mnras, 272, 96

\bibitem[{{Christopher} et~al.(2005){Christopher}, {Scoville}, {Stolovy}, \&
  {Yun}}]{2005ApJ...622..346C}
{Christopher}, M.~H., {Scoville}, N.~Z., {Stolovy}, S.~R., \& {Yun}, M.~S.
  2005, \apj, 622, 346

\bibitem[{{Churchwell} et~al.(2009){Churchwell}, {Babler}, {Meade}
  et~al.}]{2009PASP..121..213C}
{Churchwell}, E., {Babler}, B.~L., {Meade}, M.~R., et~al. 2009, \pasp, 121, 213

\bibitem[{{Churchwell} et~al.(2006){Churchwell}, {Povich}, {Allen}
  et~al.}]{2006ApJ...649..759C}
{Churchwell}, E., {Povich}, M.~S., {Allen}, D., et~al. 2006, \apj, 649, 759

\bibitem[{{Churchwell} et~al.(2007){Churchwell}, {Watson}, {Povich}
  et~al.}]{2007ApJ...670..428C}
{Churchwell}, E., {Watson}, D.~F., {Povich}, M.~S., et~al. 2007, \apj, 670, 428

\bibitem[{{Condon} et~al.(1998){Condon}, {Cotton}, {Greisen}
  et~al.}]{1998AJ....115.1693C}
{Condon}, J.~J., {Cotton}, W.~D., {Greisen}, E.~W., et~al. 1998, \aj, 115, 1693

\bibitem[{{Cyganowski} et~al.(2009){Cyganowski}, {Brogan}, {Hunter}, \&
  {Churchwell}}]{2009ApJ...702.1615C}
{Cyganowski}, C.~J., {Brogan}, C.~L., {Hunter}, T.~R., \& {Churchwell}, E.
  2009, \apj, 702, 1615

\bibitem[{{Cyganowski} et~al.(2013){Cyganowski}, {Koda}, {Rosolowsky}
  et~al.}]{2013ApJ...764...61C}
{Cyganowski}, C.~J., {Koda}, J., {Rosolowsky}, E., et~al. 2013, \apj, 764, 61

\bibitem[{{Cyganowski} et~al.(2008){Cyganowski}, {Whitney}, {Holden}
  et~al.}]{2008AJ....136.2391C}
{Cyganowski}, C.~J., {Whitney}, B.~A., {Holden}, E., et~al. 2008, \aj, 136,
  2391-2412

\bibitem[{{Dale} et~al.(2015){Dale}, {Haworth}, \&
  {Bressert}}]{2015MNRAS.450.1199D}
{Dale}, J.~E., {Haworth}, T.~J., \& {Bressert}, E. 2015, \mnras, 450, 1199

\bibitem[{{Dame} et~al.(2001){Dame}, {Hartmann}, \&
  {Thaddeus}}]{2001ApJ...547..792D}
{Dame}, T.~M., {Hartmann}, D., \& {Thaddeus}, P. 2001, \apj, 547, 792

\bibitem[{{Deharveng} et~al.(2010){Deharveng}, {Schuller}, {Anderson}
  et~al.}]{2010A&A...523A...6D}
{Deharveng}, L., {Schuller}, F., {Anderson}, L.~D., et~al. 2010, \aap, 523, A6

\bibitem[{{Deharveng} et~al.(2005){Deharveng}, {Zavagno}, \&
  {Caplan}}]{2005A&A...433..565D}
{Deharveng}, L., {Zavagno}, A., \& {Caplan}, J. 2005, \aap, 433, 565

\bibitem[{{Dewangan} \& {Ojha}(2013)}]{2013MNRAS.429.1386D}
{Dewangan}, L.~K., \& {Ojha}, D.~K. 2013, \mnras, 429, 1386

\bibitem[{{Elmegreen} \& {Lada}(1977)}]{1977ApJ...214..725E}
{Elmegreen}, B.~G., \& {Lada}, C.~J. 1977, \apj, 214, 725

\bibitem[{{Everett} \& {Churchwell}(2010)}]{2010ApJ...713..592E}
{Everett}, J.~E., \& {Churchwell}, E. 2010, \apj, 713, 592

\bibitem[{{Frerking} et~al.(1982){Frerking}, {Langer}, \&
  {Wilson}}]{1982ApJ...262..590F}
{Frerking}, M.~A., {Langer}, W.~D., \& {Wilson}, R.~W. 1982, \apj, 262, 590

\bibitem[{{Gutermuth} \& {Heyer}(2015)}]{2015AJ....149...64G}
{Gutermuth}, R.~A., \& {Heyer}, M. 2015, \aj, 149, 64

\bibitem[{{Helfand} et~al.(2006){Helfand}, {Becker}, {White}, {Fallon}, \&
  {Tuttle}}]{2006AJ....131.2525H}
{Helfand}, D.~J., {Becker}, R.~H., {White}, R.~L., {Fallon}, A., \& {Tuttle},
  S. 2006, \aj, 131, 2525

\bibitem[{{Heyer} \& {Dame}(2015)}]{2015ARA&A..53..583H}
{Heyer}, M., \& {Dame}, T.~M. 2015, \araa, 53, 583

\bibitem[{{Hildebrand}(1983)}]{1983QJRAS..24..267H}
{Hildebrand}, R.~H. 1983, \qjras, 24, 267

\bibitem[{{Hollenbach} \& {Tielens}(1997)}]{1997ARA&A..35..179H}
{Hollenbach}, D.~J., \& {Tielens}, A.~G.~G.~M. 1997, \araa, 35, 179

\bibitem[{{Hosokawa} \& {Inutsuka}(2005)}]{2005ApJ...623..917H}
{Hosokawa}, T., \& {Inutsuka}, S.-i. 2005, \apj, 623, 917

\bibitem[{{Hou} \& {Gao}(2014)}]{2014MNRAS.438..426H}
{Hou}, L.~G., \& {Gao}, X.~Y. 2014, \mnras, 438, 426

\bibitem[{{Jackson} et~al.(2006){Jackson}, {Rathborne}, {Shah}
  et~al.}]{2006ApJS..163..145J}
{Jackson}, J.~M., {Rathborne}, J.~M., {Shah}, R.~Y., et~al. 2006, \apjs, 163,
  145

\bibitem[{{Ji} et~al.(2012){Ji}, {Zhou}, {Esimbek}
  et~al.}]{2012A&A...544A..39J}
{Ji}, W.-G., {Zhou}, J.-J., {Esimbek}, J., et~al. 2012, \aap, 544, A39

\bibitem[{{Kaufman} et~al.(1999){Kaufman}, {Wolfire}, {Hollenbach}, \&
  {Luhman}}]{1999ApJ...527..795K}
{Kaufman}, M.~J., {Wolfire}, M.~G., {Hollenbach}, D.~J., \& {Luhman}, M.~L.
  1999, \apj, 527, 795

\bibitem[{{Kendrew} et~al.(2012){Kendrew}, {Simpson}, {Bressert}
  et~al.}]{2012ApJ...755...71K}
{Kendrew}, S., {Simpson}, R., {Bressert}, E., et~al. 2012, \apj, 755, 71

\bibitem[{{Lafon} et~al.(1983){Lafon}, {Baudry}, {de La Noe}, \&
  {Deharveng}}]{1983A&A...124....1L}
{Lafon}, G., {Baudry}, A., {de La Noe}, J., \& {Deharveng}, L. 1983, \aap, 124,
  1

\bibitem[{{Leger} \& {Puget}(1984)}]{1984A&A...137L...5L}
{Leger}, A., \& {Puget}, J.~L. 1984, \aap, 137, L5

\bibitem[{{Li} et~al.(2013){Li}, {Jiang}, {Liu}, \&
  {Wang}}]{2013RAA....13..921L}
{Li}, J.-Y., {Jiang}, Z.-B., {Liu}, Y., \& {Wang}, Y. 2013, Research in
  Astronomy and Astrophysics, 13, 921

\bibitem[{{Liu} et~al.(2014){Liu}, {Wang}, \& {Xu}}]{2014MNRAS.443.2264L}
{Liu}, X.-L., {Wang}, J.-J., \& {Xu}, J.-L. 2014, \mnras, 443, 2264

\bibitem[{{Lockman}(1989)}]{1989ApJS...71..469L}
{Lockman}, F.~J. 1989, \apjs, 71, 469

\bibitem[{{MacLaren} et~al.(1988){MacLaren}, {Richardson}, \&
  {Wolfendale}}]{1988ApJ...333..821M}
{MacLaren}, I., {Richardson}, K.~M., \& {Wolfendale}, A.~W. 1988, \apj, 333,
  821

\bibitem[{{Nagahama} et~al.(1998){Nagahama}, {Mizuno}, {Ogawa}, \&
  {Fukui}}]{1998AJ....116..336N}
{Nagahama}, T., {Mizuno}, A., {Ogawa}, H., \& {Fukui}, Y. 1998, \aj, 116, 336

\bibitem[{{Petriella} et~al.(2010){Petriella}, {Paron}, \&
  {Giacani}}]{2010A&A...513A..44P}
{Petriella}, A., {Paron}, S., \& {Giacani}, E. 2010, \aap, 513, A44

\bibitem[{{Rahman} \& {Murray}(2010)}]{2010ApJ...719.1104R}
{Rahman}, M., \& {Murray}, N. 2010, \apj, 719, 1104

\bibitem[{{Rawlings} et~al.(2004){Rawlings}, {Redman}, {Keto}, \&
  {Williams}}]{2004MNRAS.351.1054R}
{Rawlings}, J.~M.~C., {Redman}, M.~P., {Keto}, E., \& {Williams}, D.~A. 2004,
  \mnras, 351, 1054

\bibitem[{{Roman-Duval} et~al.(2009){Roman-Duval}, {Jackson}, {Heyer}
  et~al.}]{2009ApJ...699.1153R}
{Roman-Duval}, J., {Jackson}, J.~M., {Heyer}, M., et~al. 2009, \apj, 699, 1153

\bibitem[{{Rosolowsky} et~al.(2010){Rosolowsky}, {Dunham}, {Ginsburg}
  et~al.}]{2010ApJS..188..123R}
{Rosolowsky}, E., {Dunham}, M.~K., {Ginsburg}, A., et~al. 2010, \apjs, 188, 123

\bibitem[{{Samal} et~al.(2014){Samal}, {Zavagno}, {Deharveng}
  et~al.}]{2014A&A...566A.122S}
{Samal}, M.~R., {Zavagno}, A., {Deharveng}, L., et~al. 2014, \aap, 566, A122

\bibitem[{{Sawada} et~al.(2012){Sawada}, {Hasegawa}, {Sugimoto}, {Koda}, \&
  {Handa}}]{2012APJ...752..118S}
{Sawada}, T., {Hasegawa}, T., {Sugimoto}, M., {Koda}, J., \& {Handa}, T. 2012,
  \apj, 752, 118

\bibitem[{{Schuller} et~al.(2009){Schuller}, {Menten}, {Contreras}
  et~al.}]{2009AA...504..415S}
{Schuller}, F., {Menten}, K.~M., {Contreras}, Y., et~al. 2009, \aap, 504, 415

\bibitem[{{Scoville} et~al.(1986){Scoville}, {Sargent}, {Sanders}
  et~al.}]{1986ApJ...303..416S}
{Scoville}, N.~Z., {Sargent}, A.~I., {Sanders}, D.~B., et~al. 1986, \apj, 303,
  416

\bibitem[{Shan et~al.(2012)Shan, Yang, Shi et~al.}]{shan2012development}
Shan, W., Yang, J., Shi, S., et~al. 2012, IEEE, 2, 593

\bibitem[{{Shepherd} \& {Churchwell}(1996)}]{1996ApJ...472..225S}
{Shepherd}, D.~S., \& {Churchwell}, E. 1996, \apj, 472, 225

\bibitem[{{Sherman}(2012)}]{2012ApJ...760...58S}
{Sherman}, R.~A. 2012, \apj, 760, 58

\bibitem[{{Shirley}(2015)}]{2015PASP..127..299S}
{Shirley}, Y.~L. 2015, \pasp, 127, 299

\bibitem[{{Sidorin} et~al.(2014){Sidorin}, {Douglas}, {Palou{\v s}},
  {W{\"u}nsch}, \& {Ehlerov{\'a}}}]{2014A&A...565A...6S}
{Sidorin}, V., {Douglas}, K.~A., {Palou{\v s}}, J., {W{\"u}nsch}, R., \&
  {Ehlerov{\'a}}, S. 2014, \aap, 565, A6

\bibitem[{{Simpson} et~al.(2012){Simpson}, {Povich}, {Kendrew}
  et~al.}]{2012MNRAS.424.2442S}
{Simpson}, R.~J., {Povich}, M.~S., {Kendrew}, S., et~al. 2012, \mnras, 424,
  2442

\bibitem[{{Smith} et~al.(1997){Smith}, {Suttner}, \&
  {Yorke}}]{1997A&A...323..223S}
{Smith}, M.~D., {Suttner}, G., \& {Yorke}, H.~W. 1997, \aap, 323, 223

\bibitem[{{Snell} et~al.(1984){Snell}, {Scoville}, {Sanders}, \&
  {Erickson}}]{1984ApJ...284..176S}
{Snell}, R.~L., {Scoville}, N.~Z., {Sanders}, D.~B., \& {Erickson}, N.~R. 1984,
  \apj, 284, 176

\bibitem[{{Sobolev} et~al.(2005){Sobolev}, {Ostrovskii}, {Kirsanova}
  et~al.}]{2005IAUS..227..174S}
{Sobolev}, A.~M., {Ostrovskii}, A.~B., {Kirsanova}, M.~S., et~al. 2005, in
  Massive Star Birth: A Crossroads of Astrophysics, \emph{IAU Symposium}, vol.
  227, edited by R.~{Cesaroni}, M.~{Felli}, E.~{Churchwell}, \& M.~{Walmsley},
  174--179

\bibitem[{{Stead} \& {Hoare}(2010)}]{2010MNRAS.407..923S}
{Stead}, J.~J., \& {Hoare}, M.~G. 2010, \mnras, 407, 923

\bibitem[{{Stead} \& {Hoare}(2011)}]{2011MNRAS.418.2219S}
{Stead}, J.~J., \& {Hoare}, M.~G. 2011, \mnras, 418, 2219

\bibitem[{{Stil} et~al.(2006){Stil}, {Taylor}, {Dickey}
  et~al.}]{2006AJ....132.1158S}
{Stil}, J.~M., {Taylor}, A.~R., {Dickey}, J.~M., et~al. 2006, \aj, 132, 1158

\bibitem[{{Su} et~al.(2014){Su}, {Yang}, {Zhou}, {Zhou}, \&
  {Chen}}]{2014ApJ...796..122S}
{Su}, Y., {Yang}, J., {Zhou}, X., {Zhou}, P., \& {Chen}, Y. 2014, \apj, 796,
  122

\bibitem[{{Thompson} et~al.(2012){Thompson}, {Urquhart}, {Moore}, \&
  {Morgan}}]{2012MNRAS.421..408T}
{Thompson}, M.~A., {Urquhart}, J.~S., {Moore}, T.~J.~T., \& {Morgan}, L.~K.
  2012, \mnras, 421, 408

\bibitem[{{Tremblin} et~al.(2014){Tremblin}, {Anderson}, {Didelon}
  et~al.}]{2014A&A...568A...4T}
{Tremblin}, P., {Anderson}, L.~D., {Didelon}, P., et~al. 2014, \aap, 568, A4

\bibitem[{{Watson} et~al.(2003){Watson}, {Araya}, {Sewilo}
  et~al.}]{2003ApJ...587..714W}
{Watson}, C., {Araya}, E., {Sewilo}, M., et~al. 2003, \apj, 587, 714

\bibitem[{{Watson} et~al.(2009){Watson}, {Corn}, {Churchwell}
  et~al.}]{2009ApJ...694..546W}
{Watson}, C., {Corn}, T., {Churchwell}, E.~B., et~al. 2009, \apj, 694, 546

\bibitem[{{Watson} et~al.(2010){Watson}, {Hanspal}, \&
  {Mengistu}}]{2010ApJ...716.1478W}
{Watson}, C., {Hanspal}, U., \& {Mengistu}, A. 2010, \apj, 716, 1478

\bibitem[{{Watson} et~al.(2008){Watson}, {Povich}, {Churchwell}
  et~al.}]{2008ApJ...681.1341W}
{Watson}, C., {Povich}, M.~S., {Churchwell}, E.~B., et~al. 2008, \apj, 681,
  1341

\bibitem[{{Wenger} et~al.(2013){Wenger}, {Bania}, {Balser}, \&
  {Anderson}}]{2013ApJ...764...34W}
{Wenger}, T.~V., {Bania}, T.~M., {Balser}, D.~S., \& {Anderson}, L.~D. 2013,
  \apj, 764, 34

\bibitem[{{White} et~al.(2005){White}, {Becker}, \&
  {Helfand}}]{2005AJ....130..586W}
{White}, R.~L., {Becker}, R.~H., \& {Helfand}, D.~J. 2005, \aj, 130, 586

\bibitem[{{Wu} et~al.(2010){Wu}, {Evans}, {Shirley}, \&
  {Knez}}]{2010ApJS..188..313W}
{Wu}, J., {Evans}, N.~J., II, {Shirley}, Y.~L., \& {Knez}, C. 2010, \apjs, 188,
  313

\bibitem[{{Xu} \& {Ju}(2014)}]{2014A&A...569A..36X}
{Xu}, J.-L., \& {Ju}, B.-G. 2014, \aap, 569, A36

\bibitem[{{Yoshida} et~al.(2010){Yoshida}, {Kitamura}, {Shimajiri}, \&
  {Kawabe}}]{2010ApJ...718.1019Y}
{Yoshida}, A., {Kitamura}, Y., {Shimajiri}, Y., \& {Kawabe}, R. 2010, \apj,
  718, 1019

\bibitem[{{Yuan} et~al.(2014){Yuan}, {Wu}, {Li}, \&
  {Liu}}]{2014ApJ...797...40Y}
{Yuan}, J.-H., {Wu}, Y., {Li}, J.~Z., \& {Liu}, H. 2014, \apj, 797, 40

\bibitem[{{Zavagno} et~al.(2010){Zavagno}, {Anderson}, {Russeil}
  et~al.}]{2010A&A...518L.101Z}
{Zavagno}, A., {Anderson}, L.~D., {Russeil}, D., et~al. 2010, \aap, 518, L101

\bibitem[{{Zhang} et~al.(2001){Zhang}, {Hunter}, {Brand}
  et~al.}]{2001ApJ...552L.167Z}
{Zhang}, Q., {Hunter}, T.~R., {Brand}, J., et~al. 2001, \apjl, 552, L167

\bibitem[{{Zhang} et~al.(2005){Zhang}, {Hunter}, {Brand}
  et~al.}]{2005ApJ...625..864Z}
{Zhang}, Q., {Hunter}, T.~R., {Brand}, J., et~al. 2005, \apj, 625, 864

\bibitem[{{Zhou} et~al.(2014){Zhou}, {Yang}, {Fang}, \&
  {Su}}]{2014ApJ...791..109Z}
{Zhou}, X., {Yang}, J., {Fang}, M., \& {Su}, Y. 2014, \apj, 791, 109

\end{thebibliography}
\end{document}